\documentclass{ar2e}
\usepackage{ulem}  

\input psfig.sty

\usepackage{ARAstroBib}
\usepackage{aas_macros} 

\newcommand{\sun}{\ensuremath{\odot}}

\begin{document}
\title{Teraelectronvolt Astronomy}

\markboth{J.A. Hinton and W. Hofmann}{Teraelectronvolt Astronomy}
\author{J.A. Hinton and W. Hofmann}


\begin{abstract}
Ground-based $\gamma$-ray astronomy, which provides access to the TeV
energy range, is a young and rapidly developing discipline.
Recent discoveries in this waveband have important consequences
for a wide range of topics in astrophysics and astroparticle physics.
This article is an attempt to review the experimental status of this 
field and to provide the basic formulae and concepts required to begin the 
interpretation of TeV observations.
\end{abstract}

\maketitle
\
\section{Introduction}

Ground-based $\gamma$-ray astronomy effectively began in 1989, with the
first robust detection of a TeV $\gamma$-ray source, the Crab Nebula, using
the 10\,m diameter air-Cherenkov telescope of the Whipple Observatory 
\citep{weekes89}. Nineteen years on, Cherenkov telescopes have been used 
to detect $\sim$80 very high energy (VHE; E $>$ 100 GeV) sources, firmly 
establishing a new astronomical domain, and the first discoveries of sources 
using air-shower particle detectors have been made \citep{abdo07}. 
The key
advantage of ground-based instrumentation over satellite-based GeV
instruments such as EGRET \citep{hartman99} and the recently launched 
GLAST (now Fermi) large area telescope (LAT) \citep{thompson04} 
is collection area. The typical effective collection
area of a single Cherenkov telescope is $10^{5}$\,m$^2$, almost five
orders of magnitude larger than can realistically be achieved via
direct detection in space. So far, the Imaging Atmospheric-Cherenkov
Telescope (IACT) technique has proven to be the most powerful approach in
this energy regime, with sensitivity of $\sim$1\% of the flux
of the Crab Nebula, angular resolution for single $\gamma$-rays of around
5$'$, energy resolution of $\approx$15\%, and the ability to locate sources with
precision down to 10$''$. The limitations of the IACT technique in comparison
to air-shower particle detectors include a limited field of view ($\sim5^{\circ}$)
and a relatively poor duty cycle,
with about 1000 hours of useful observations obtainable per year.

The current status of VHE $\gamma$-ray astronomy is best summarized by
Figure~\ref{fig_tevsky}, showing the distribution of different types
of VHE $\gamma$-ray source on the sky. In addition to a collection of
extragalactic sources at high latitudes --- identified as active
galactic nuclei, a number of clearly Galactic sources line the
Galactic equator.  Their low latitudes imply kpc distances and
most are extended with respect to the $\sim$5$'$ resolution
of the instruments, implying emission region sizes of $\cal
O$(10\,pc). These sources include shell-type supernova remnants,
pulsar wind nebulae such as the Crab Nebula, binary systems, the
Galactic Center and a number of unidentified sources.

This review aims to summarize the basic emission mechanisms suspected
to be active in these objects, to provide a {\it cookbook} of useful
formula for calculations in this field and to give an overview of the
principle astrophysical results of TeV astronomy in the
last few years. This article does not attempt to provide a complete
list of VHE $\gamma$-ray sources and their properties --- given the rapid
progress in the field this task is better served by online databases
such as {\it TeVCat} --- but rather aims to illustrate key properties and
mechanisms using selected objects.  The reader is also referred to
recent reviews covering related subjects, such as the review of
high-energy astrophysics with ground-based detectors by
\citet{aharonian_buckley08} including a more detailed discussion of
air-shower physics, the review of supernova remnants at high energy by
\citet{reynolds08}, the discussion of whether acceleration in supernova
remnants can account for galactic cosmic rays by \citet{hillas05}, the
review of cosmic-ray propagation and interactions by \citet{strong07}
and the review of pulsar wind nebulae by \citet{gaensler06}.

\section{Generation and propagation of VHE $\gamma$-rays}
\label{sec:CB}

In this section some of the basic characteristics of acceleration and
radiation processes relevant to VHE astronomy are summarized, providing
simple expressions, approximations, and references to detailed work.

\subsection{Particle acceleration and propagation}
\label{sec:acc}

Except for possible production by top-down processes such as the
decay of heavy particles, VHE $\gamma$-rays are produced only in the
interactions of accelerated charged particles - nuclei or electrons\footnote{Here and
in the following, `electrons' stands for `electrons and positrons'} -
with ambient matter or radiation fields. The $\gamma$-ray production rate
reflects the product of the densities of cosmic ray (CR) particles and `targets',
tracking particles as they propagate away from the acceleration
site due to diffusion in
magnetic fields or convection. $\gamma$-ray sources will therefore be
extended objects, with sizes and shapes governed by particle flow
speeds, distribution of targets, and possibly by the finite lifetime of
particles due to interactions or radiative cooling. 
$\gamma$-ray sources
which appear point-like, given the $\sim$5$'$ resolution of current
instruments, are either very distant (extragalactic), or associated with
compact targets such as dense molecular clouds or stellar radiation fields.

Unless convective flows or bulk motion dominate, particle transport is
governed by scattering off inhomogeneous ambient magnetic fields,
resulting in diffusive propagation, $\langle r^2 \rangle = 2D t$. The diffusion
coefficient $D$ is determined by the average strength $B$ of the magnetic field
and its degree of turbulence $\delta B$ on length-scales comparable
to the gyration radius $R_g$. The slowest possible diffusion in an isotropic medium
is achieved in the Bohm limit $\eta =1$,
where the mean free path of particles 
is given by the gyration radius, resulting in $D \approx \eta R_g c/3$ or, with
$R_{g,pc} \approx 0.0011 E_{\rm TeV}/B_{\perp,\mu {\rm G}}$,
and particles of unit charge
\begin{equation}
\langle r_{\rm pc}^2 \rangle^{1/2} \approx 0.01 (\eta E_{\rm TeV} t_{\rm yr} / B_{\mu {\rm G}})^{1/2}
\label{eq:diffusion}
\end{equation}
 Typical interstellar fields are of the order of 3\,$\mu G$.
The coefficient $\eta$ can be
estimated as $\eta \approx (\delta B/B)^{-2}$ \citep{strong07}. 
Bohm diffusion is very slow, compared for example to the speeds of
young supernova shocks. Large-scale propagation in the Milky Way \citep[e.g.][]{strong07}
is governed
by larger diffusion coefficients; in leaky box models one obtains $D
\sim k 10^{28} E_{\rm GeV}^\alpha$~cm$^2$/s
where $k$ is a coefficient of order unity and $\alpha = 0.3-0.6$, or
$\langle r_{\rm pc}^2 \rangle^{1/2} \sim (E_{\rm TeV}^\alpha t_{\rm yr})^{1/2}$, with the
energy dependence reflecting the scale distribution of magnetic
turbulence. 

Diffusive shock acceleration is most likely the principle mechanism
behind the production of high energy particles, with supernova
remnants (SNRs) believed to be one of the main sources of CRs. For a
recent overview of shock acceleration in SNRs see \citet{reynolds08}.
A supersonic flow, for example due to ejecta from a supernova
explosion or a pulsar wind, terminates in a shock, balancing the
pressure of the ambient medium. Best viewed in the rest frame of the
shock, material is streaming into the shock from upstream at velocity
$u$, is compressed by a compression factor $r$ and flows away from the
shock with a speed reduced by the shock compression ratio. High-energy
particles scatter off turbulent magnetic fields on both sides on the
shock and may diffusively cross the shock many times. Each time they
cross the shock and are isotropized by scattering in the medium on the
other side, they gain an energy of order $\Delta E/E \approx u/c$.  In
each cycle, there is a finite probability for escape with the
downstream flow, naturally creating a power-law spectrum of
accelerated particles, $N(E) \sim E^{-\Gamma}$ with $\Gamma \approx
(r+2)/(r-1)$, with $r=4$ and $\Gamma \approx 2$ for shocks with high
Mach number. Neglecting radiative cooling and adiabatic losses, the
distribution of accelerated particles is uniform in the region
downstream of the shock, where particles are swept away, and extends
over a scale $D/u$ into the upstream region. The acceleration rate
$(1/E)(\mbox{d}E/\mbox{d}t) \approx u^2/D$ is governed by the rate of
shock crossings, determined by the diffusion coefficient. A small
diffusion coefficient will keep particles near the shock and ensure
rapid return across the shock front. The maximum energy is governed
either by the finite lifetime of the shock, by synchrotron losses in
case of accelerated electrons, or by the gyro-radius exceeding the
shock size. Assuming Bohm diffusion, maximum proton energies
achievable in supernova shocks of age $1000\,t_{3}$\,yr are of order
\begin{equation}
E_{\rm max} \sim u_8^2 t_{3} B_{\mu G} \mathrm{TeV} 
\end{equation}
where $u_8$ is the shock speed in units
of 1000\,km s$^{-1}$ (e.g. \citet{reynolds08}); for electrons, synchrotron
losses limit the peak energy to 
\begin{equation}
E_{\rm max} \sim 100 u_8 B_{\mu G}^{-1/2} \mathrm{TeV} 
\end{equation}
See \citet{zirakashvili07}.

In shock acceleration in SNRs, a significant fraction of kinetic energy of the
flow can be transferred to high-energy particles. Once the energy
density in particles is comparable to that in the shock, nonlinear
effects start to play a role \citep[see][and references therein]{caprioli08}. 
The overall compression ratio $r$ is
increased beyond 4, but particles scattered upstream of the shock
decelerate the inflowing material and generate a precursor, reducing
the compression ratio at the main shock. Particles with gyro-radii
that are small compared to the size of the precursor experience only the
reduced compression ratio $r$, resulting in a steeper spectral index, 
whereas for the highest-energy particles
$r>4$.  Nonlinear shock acceleration hence produces concave spectra
with $\Gamma \ge 2$ at low energy and $\Gamma$ somewhat below 2 at
high energy.

Both acceleration speed and maximum energy increase for small diffusion
coefficients, i.e. from high magnetic fields and maximal turbulence of
the field on all scales, resulting in $\eta \approx 1$. In this
context, turbulent field amplification by streaming CRs has
received increasing attention. Both resonant and non-resonant
instabilities of the magnetic field are driven by
CR currents (\citet{lucek00},
see \citet{caprioli08} for further
references). In SNRs, magnetic fields can be estimated
from the cooling length-scale of electrons produced at the shock front
\citep[see for example][]{vink03, berezhko03}, with results consistent with the predicted
$B^2 \propto n u^3$ dependence, where $n$ is the ambient density and
$u$ the shock velocity. Magnetic field amplification boosts the maximum energy of
accelerated protons, but reduces the maximum electron energy due to increased losses.

\subsection{Electronic origin of $\gamma$-rays}
\label{sec:elec}

Electrons produce high-energy radiation primarily via the inverse
Compton (IC) process \citep{blumenthal70}, by up-scattering ambient
cosmic microwave background (CMB), infrared, optical or, in special cases, X-ray, photons. 

{\it Energy loss rates:}
in the Thomson limit, $b = 4E_{e}E_T/m^2c^4 \ll 1$ or
$b \approx\, 15 E_{e,\mathrm{TeV}}E_{T,\mathrm{eV}} \ll 1$, 
with $E_e$ and $E_T$ denoting electron
and target photon energies, respectively, the energy loss rate
is given by 
\begin{equation}
\mbox{d}E/\mbox{d}t = (4/3)\sigma_{T}c\gamma^{2}U_{\rm rad}
\end{equation} 
A black-body target radiation field can be approximated by setting
$E_T$ to $2.8kT$. At higher electron energies, around 300\,TeV,
10\,TeV and 30\,GeV for scattering off CMB, IR from dust and visible
light, respectively, the Klein--Nishina (KN) regime begins and the
energy loss rate is reduced, with a $\log{E_e}$ dependence in the deep KN
regime. In general the {\it cooling time}, $\tau = E_{e}/
(\mbox{d}E_{e}/\mbox{d}t)$, for IC scattering is given by

\begin{equation}
\tau_{\rm yr} = E_e/(\mbox{d}E_e/\mbox{d}t) \approx 3.1 \cdot 10^{5} U_{{\rm rad,eV\,cm}^{-3}}^{-1} E_{e,{\rm TeV}}^{-1} f_{\rm KN}^{-1} 
\label{eq:ic_cooltime}
\end{equation}

where the KN suppression factor $f_{\rm KN}$ can be parametrized as 
\begin{equation}
f_{\rm KN} \approx (1+b)^{-1.5} \approx (1 + 40 E_{\rm e,TeV} kT_{\rm eV})^{-1.5}
\end{equation}
for $b < 10^4$ \citep[][ and Erratum]{moderski05a}.
The synchrotron cooling time is given by the very similar expression:

\begin{equation}
\tau_{\rm yr} = E_e/(\mbox{d}E_e/\mbox{d}t) \approx 1.3 \cdot 10^7 B_{\mu G}^{-2} E_{e,{\rm TeV}}^{-1}
\label{eq:sync_cooltime}
\end{equation}

\begin{figure}
\centerline{\psfig{figure=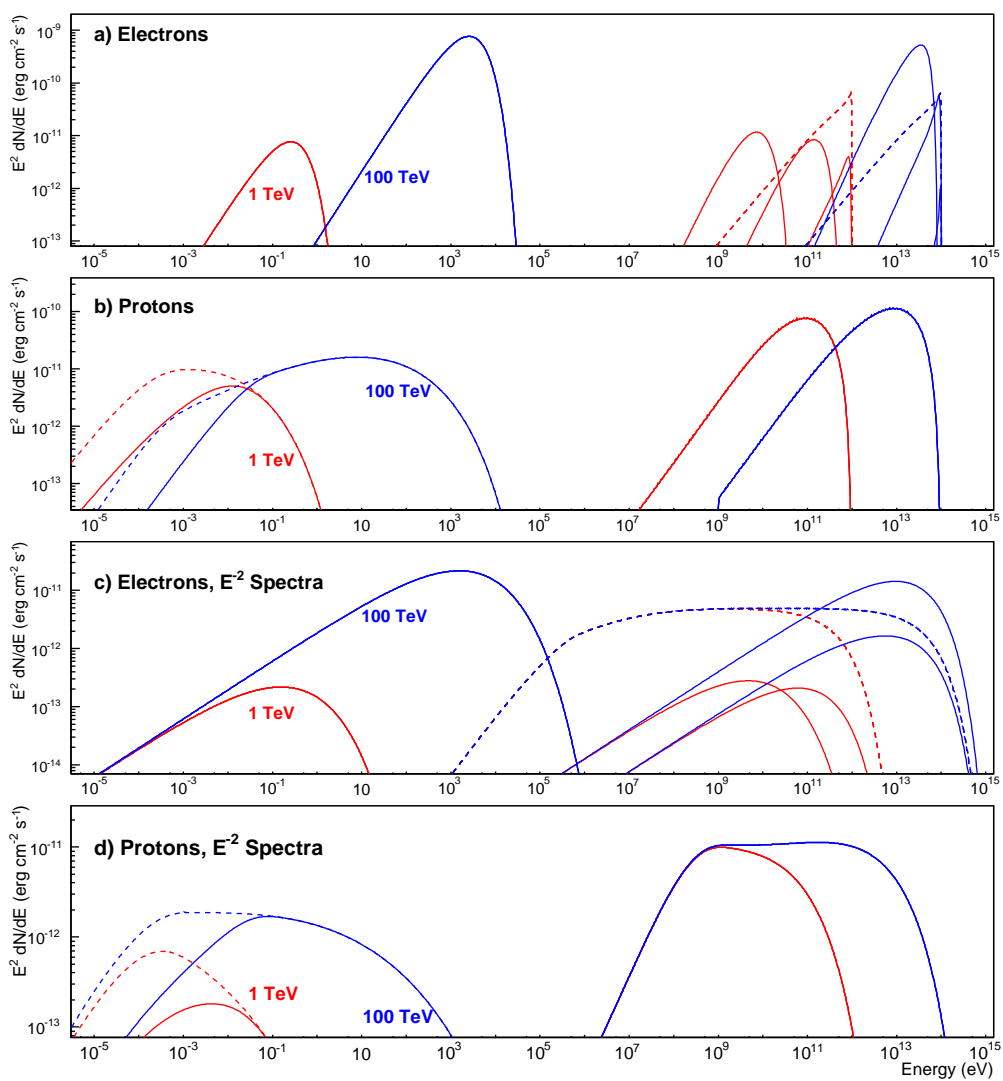,height=34pc}} 
\caption{{\bf a)} SEDs for radiation of 
	mono-energetic 1/100 TeV electrons (red/blue
	curves): Synchrotron and IC (solid curves) and Bremsstrahlung
	(dashed curves). Three IC curves are shown for each primary energy:
	(from low to high) on the CMB ($kT=2.35\times10^{-4}$ eV, 
	$b \approx  4\times10^{-3}/0.4$), on dust-emitted FIR ($kT=0.02$ eV, $b \approx 0.3/30$), and on visible
	(star) light ($kT=1.5$ eV, $b \approx 20/2000$).  
	Note that for 100 TeV electrons scattering on optical photons the IC energy distribution
	is effectively a delta-function at 100 TeV.
	The curve normalizations
	are appropriate for a total particle energy of $10^{47}$ erg at 1 kpc
	distance in a magnetic field of 3 $\mu$G, a matter density of 100
	hydrogen atoms cm$^{-3}$ and radiation fields of density 0.26 eV
	cm$^{-3}$ (CMB and FIR) and 1 eV cm$^{-3}$ (starlight).
	{\bf b)} SEDs for $\gamma$-rays and synchrotron radiation of secondary electrons from
	strong interactions of mono-energetic protons. The magnetic field is increased to 30 $\mu$G to illustrate 
	the effects of cooling and steady injection over $10^{4}$ yr (dashed curves $10^{5}$ yr)
	is assumed. The input energy is $10^{48}$ erg.
	{\bf c)} and {\bf d)} -- as for a) and b) but for cut-off power-law distributions of particles:
	$dN/dE \propto E^{-2} \exp{-E/E_{c}}$ with $E_{c}$ = 1 TeV (red) and 100 TeV (blue).
}
\label{fig_IC} 
\end{figure}

{\it Radiation spectra:} scattering of mono-energetic electrons on a blackbody distribution of
target photons of temperature $T$ results in a broad spectral energy distribution (SED) 
of the resulting $\gamma$-rays (Figure~\ref{fig_IC}). In the Thomson regime,
energy losses are gradual and the $\gamma$-ray SED peaks at 
\begin{equation}
E_{\gamma,\mathrm{TeV}} \approx 33 E_{e,\mathrm{TeV}}^{2} kT_{\rm eV}
\label{eq:egammaIC}
\end{equation}
resulting in 
\begin{equation}
E_{e,\mathrm{TeV}} \approx 11 E_{\gamma,\mathrm{TeV}}^{1/2}
\label{eq:egammaIC_CMB}
\end{equation}
for scattering of CMB photons. In the KN regime,
electrons tend to lose a large fraction of their energy in a single IC
event and the corresponding SED peak energy is shifted to:
\begin{equation}
E_{\gamma,{\rm TeV}} \approx E_{\rm e,TeV} {2.1 b \over (1 + (2.1 b)^{0.8} )^{1/0.8}}
\label{eq:egammaIC_KN}
\end{equation}

The equivalent expression for the typical energy of synchrotron
photons is 
\begin{equation}
E_{\rm s,eV} = 0.087 E_{e,\mathrm{TeV}}^2 B_{\mu \mathrm{G}}
\label{eq:esync}
\end{equation}
Whilst the energy loss rate of electrons depends (in the Thompson regime) only on the
energy density in target radiation fields, the spectrum of $\gamma$-rays
is strongly influenced by the spectral distribution of target photons
(Figure~\ref{fig_IC}). Since in the Thomson limit the $\gamma$-ray energy
scales with the square of the electron energy, the IC $\gamma$-ray
spectrum is harder than the spectrum of parent electrons: assuming
isotropic distributions of electrons and target photons, a power-law distribution of
electrons $N_e(E) \sim E_e^{-\Gamma_e}$ will generate IC (and
synchrotron) spectra of index $\Gamma_\gamma = (\Gamma_e+1)/2$,
$\Phi_\gamma \propto E_\gamma^{-(\Gamma_e+1)/2}$. Deep in the KN
regime, $\gamma$-ray spectra steepen by $\Delta \Gamma_\gamma \approx
(\Gamma_e+1)/2$, $\Phi_\gamma \sim E_\gamma^{-(\Gamma_e+1)}$, up to
$\log{E_\gamma}$ terms \citep{blumenthal70}. Therefore, even for a pure
power-law distribution of electrons, $\gamma$-ray spectra exhibit a break
corresponding to the transition to the KN regime. 

IC $\gamma$-rays of energy $E_\gamma$ (in the Thomson regime) and
synchrotron photons of energy $E_{\rm sync}$ probe the identical
electron population provided that $E_{\gamma,{\rm TeV}} = 380 E_{\rm sync,
eV} (kT_{\rm eV}/B_{\mu{\rm G}})$.  For example, 11 TeV electrons
produce both 1 TeV IC photons (on the CMB) and 1 keV synchrotron
photons, if $B$=100\,$\mu$G. In this case, the ratio of energy flux in
the synchrotron and IC bands is just $U_{\rm mag}/U_{\rm rad}$, allowing
the magnetic field strength in the source region to be determined. For
the smaller (several $\mu$G) fields found typically in the ISM, X-ray synchrotron
emission traces electrons of considerably higher energies than the
TeV IC $\gamma$-rays and a model for the shape of the electron spectrum
is required to deduce the magnetic field. 

Which target photon fields are most relevant for $\gamma$-ray production
depends on the electron spectrum and on the $\gamma$-ray energy
considered. The relative yields of CMB target photons, IR target
photons and visible target photons vary depending on the location of
sources, e.g. with Galacto-centric radius, and
with the proximity of strong radiation sources.
At the location of the sun, typical radiation energy densities 
are 0.3\,eV\,cm$^{-3}$ for
IR from dust, and 0.3\,eV\,cm$^{-3}$ for visible light, 
see \citet{porter06} (c.f. 0.26\,eV\,cm$^{-3}$ for the CMB).
For electron spectra
without a cutoff, CMB photons will often dominate at sufficiently high
energies, as the contributions of IR and visible photons are suppressed 
in the KN regime. In this case, $\gamma$-rays directly map the distribution of VHE
electrons. For electron spectra with a cutoff, the situation can,
however, be reversed, with the efficient transfer of energy from an
electron to a $\gamma$-ray over-compensating for the drop in scattering
cross-section. 

Electrons passing through a medium containing atoms or plasma will
also create $\gamma$-rays by bremsstrahlung \citep{blumenthal70}.  The
loss timescale is energy-independent, with loss rates and spectra
depending on the state of the medium (mainly the shielding length for
ions). For neutral hydrogen atoms of density $n$ per cm$^3$, $\tau =
E_e/(\mbox{d}E_e/\mbox{d}t) \approx 3.9 \cdot 10^{7} n^{-1}$ years.
IC scattering on the
CMB dominates in the 1~TeV emission of electrons in (neutral) media
as long as $n\,<\,240$ cm$^{-3}$.  Bremsstrahlung $\gamma$-ray spectra
produced by a power-law distribution of electrons are also 
power-laws, with $\Gamma_\gamma = \Gamma_e$.

\subsection{Modelling radiation spectra from electron populations}

\begin{figure}[h]
\centerline{\psfig{figure=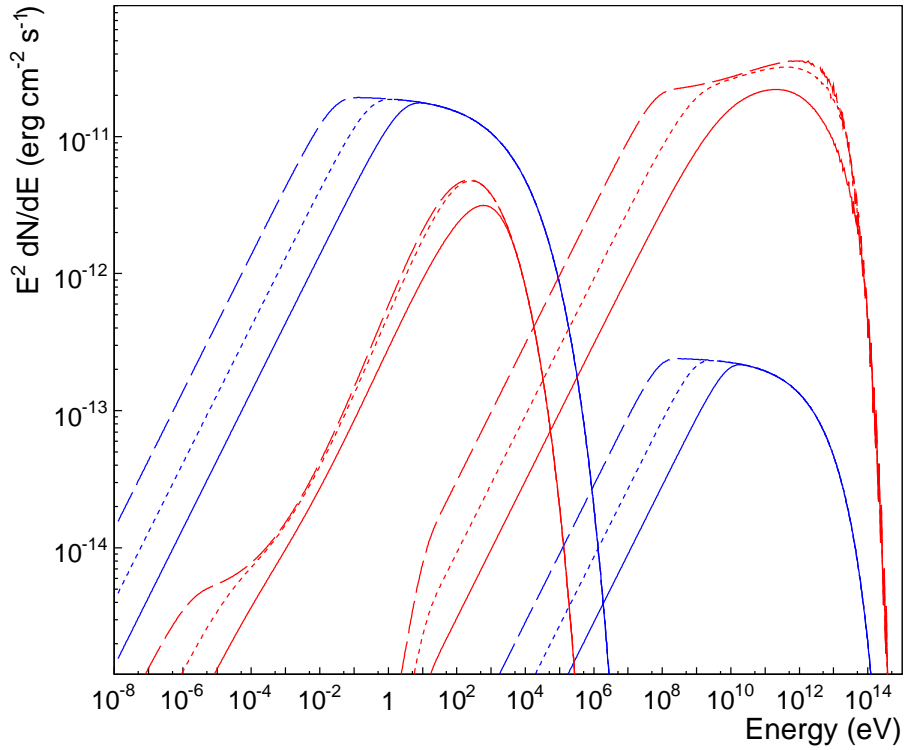,height=25pc}} 
\caption{SEDs for continuous injection and cooling 
of a population of electrons with an $E^{-2}$ injection spectrum and an exponential
cut-off at 100 TeV.
Solid, short-dashed and
long-dashed curves show injection timescales of $10^{4}$ yr, $3 \times 10^{4}$ yr
and $10^{5}$ yr, respectively. The blue curves show synchrotron and IC emission in
the case of synchrotron-dominated cooling, with $B=30\mu$G and the CMBR as target for IC.
The red curves illustrate the effects of IC-dominated cooling with 
a lower magnetic field ($B=3\mu$G) and a higher energy radiation 
field (black-body photons with $kT$=1.5~eV, with density 1000 eV cm$^{-3}$) where 
KN effects become important.
}
\label{fig_sedcool} 
\end{figure}

Energy losses by synchrotron radiation or IC will modify electron
spectra compared to the index at injection, $\Gamma_{\rm i}$, above
energies where the energy loss time (Eq. \ref{eq:sync_cooltime}) becomes
comparable to the age $T$ of the electron source. If synchrotron
losses or IC losses in the Thomson regime dominate, $(dE/dt)_{\rm sync+IC} = -\kappa E^2$, the electron
energy distribution will be cut off (for burst injection
in a source of age $T_{s}$) at $E = 1/(\kappa T_{s})$, the energy where the radiative
lifetime $(1/E)(dE/dt)$ equals the age $T_{s}$ , or will exhibit a spectral break at the same
energy, with index $\Gamma_e$ increasing by 1 (in case of continuous
injection over time $T_s$) \citep{kardashev62}. This will cause a cutoff or
break with index change $\Delta \Gamma_\gamma = 1/2$ in IC and synchrotron
spectra at the corresponding energies (equations~\ref{eq:egammaIC} and \ref{eq:esync}).  
The situation is more
complicated if IC losses dominate over synchrotron losses, as might be
the case near strong radiation sources, as discussed by \citet{moderski05a}. 
Entering the KN regime, electron energy losses are reduced,
resulting in a harder ``cooled'' spectrum $\Gamma_e \approx \Gamma_{\rm i}+1+\Delta
\Gamma$ with $\Delta \Gamma \approx -1...-1.5$, and $\Gamma_\gamma
\approx \Gamma_{\rm i}$. With IC losses scaling slower than $E_e^2$,
synchrotron losses will, however, always dominate above a
certain energy, causing $\Gamma_\gamma$ to change to $\approx
\Gamma_{\rm i}+2$. Figure~\ref{fig_sedcool} shows the effects of both
IC and synchrotron dominated cooling on the SED from an injected 
power-law of relativistic electrons.

The above discussion assumes particle injection with a power-law
spectrum, followed by radiative losses. If acceleration timescales and
loss timescales are comparable, acceleration and losses cannot be
factored and complex spectral patterns can arise, as discussed, 
for example, by \citet{zirakashvili07}.

In modelling radiation spectra of astrophysical sources, several approaches
are followed:

\begin{itemize}

\item One can determine the spectrum (and composition) of particles which,
interacting with suitable target fields and radiation fields, gives rise
to the observed wide-band SED \citep[e.g.][]{HESS:g09}.

\item One can determine the energy spectrum of injected particles which,
after accounting for the modification of the spectrum over time due to
radiation losses and interactions etc., creates the observed SED. 
\citep[e.g.][and in Figure~\ref{fig_sedcool}]{hinton07}.

\item  One can model the full dynamics of the source, including particle
acceleration mechanisms, losses and interactions \citep[e.g.][]{berezhko06}.

\end{itemize}

The first approach is self-consistent but has the potential problem
that, due to the many degrees of freedom in the choice of spectral
parametrisation and targets, it may be difficult to
arrive at a unique solution. Frequently, spectral parametrisation
assumes a broken power law, with the fit sometimes resulting in a large increase
in spectral index at the break 
\citep[see e.g.][]{HESS:g09}. It is non-trivial to see which mechanisms would
produce such spectra; at synchrotron cooling breaks in electron spectra, the index
increases only by one.

The second intermediate approach attempts to cure this deficit by
including (some of) the mechanisms which cause spectral breaks and
cutoffs. However, simplifying assumptions are often  made concerning the
history of particle injection. In sources where particles are confined
within an expanding envelope, such as SNRs or pulsar
wind nebulae, adiabatic losses usually need to be taken into account, but
frequently are not. Cooling during acceleration may modify
injection spectra compared to the often assumed power law. 
This approach usually gives reliable answers only if cooling times
are short compared to the evolution/expansion timescales of the
system, but long compared to acceleration timescales.

The third approach is ``best'' in that it attempts to account for all
relevant effects, but is most demanding and has the disadvantage that
solutions are usually numerical and that it is non-trivial to understand
the systems in terms of a clear one-to-one connection between input
assumptions and predicted radiation SED.

\subsection{Hadronic origin of $\gamma$-rays}
\label{sec:hadrons}

An alternative source of VHE $\gamma$-rays are interactions of
high-energy protons and nuclei with interstellar material. Through this
mechanism, $\gamma$-ray emission traces CR acceleration sites and
CR propagation, as demonstrated in the GeV energy range by
the study of diffuse $\gamma$-ray emission from the Milky Way \citep{hunter97}.
The interaction cross-section of VHE protons on hydrogen nuclei of density
$n$ per cm$^{3}$ is only weakly energy-dependent, $\sigma_{pp,inel}
\approx$35\,mb, resulting in a lifetime of $\tau \approx 3 \cdot
10^7 n^{-1}\,\mbox{yr}$ at multi-TeV energies. The number of secondary
particles produced per interaction increases with energy. Typically,
half of the energy $E_p$ of the primary is carried away by a leading
nucleon, the other half is split between charged and neutral pions and
a small fraction of heavier hadrons. 
This implies that about 1/6 of the
primary energy is carried by a number of $\gamma$-rays produced in
$\pi^0$ decays. To a good approximation, and for $E_\gamma \gg m_\pi$, the
distribution in energy of $\gamma$-rays produced per interaction is scale
invariant, $\mbox{d}n_\gamma/\mbox{d}E_\gamma(E_\gamma,E_p) =
D_\gamma(x = E_\gamma/E_p)$. Parameterisations for the fragmentation
function $D$ are given for example by \citet{kelner06}.
Figure~\ref{fig_IC} illustrates the SED of $\gamma$-rays produced by
mono-energetic protons. The SED of secondary $\gamma$-rays peaks at about
1/10 of the energy of the primary. For a power-law proton distribution
$\mbox{d}n_p/\mbox{d}E_p = k_p E_p^{-\Gamma}$ the $\gamma$-ray distribution is again a
power law with the same index:
\begin{equation}
{\mbox{d}n_\gamma \over \mbox{d}E_\gamma \mbox{d}t} = c n \sigma_{pp,inel} \int_{E_\gamma} {\mbox{d}n_p \over \mbox{d}E_p}\,D\left({E_\gamma \over E_p}\right) {\mbox{d}E_p \over E_p} \\
= c n \sigma_{pp,inel} k_p E_\gamma^{\Gamma} \int_0^1 D(x) x^{\Gamma-1} \mbox{d}x
\end{equation}
It is convenient to write this as
\begin{equation}
{\mbox{d}n_\gamma \over \mbox{d}E_\gamma \mbox{d}t} = c n \sigma_{pp,inel} \kappa k_p \left( {E_\gamma \over f} \right)^{-\Gamma}
\end{equation}
 since for $f \approx 0.15$ the coefficient $\kappa \approx 6$ is
approximately constant for $\Gamma$ between 2 and
4. This property should not be misinterpreted in the sense of a
delta-function approximation, namely that in general the $\gamma$-ray
spectrum at energy $E_\gamma$ traces the proton spectrum at energy
$E_p/f$. In particular, \citet{kelner06} show that
an exponential cut-off in the proton spectrum, $\exp(-E_p/E_{cut})$ is
transformed into a more gradual cut-off, $\exp(-(16
E_\gamma/E_{cut})^{1/2})$ in the $\gamma$-ray spectrum. Obviously, given
the width of the $\gamma$-ray SED illustrated in Figure~\ref{fig_IC}, any
feature in the proton spectrum will re-appear smoothed in the
$\gamma$-ray spectrum.

One application is the estimation of the $\gamma$-ray flux from molecular
clouds illuminated by CRs, possibly from a nearby
accelerator. For a cloud of mass $M$ and distance $d$ illuminated by 
a proton flux $\Phi(E) = k_p E_p^{-\Gamma}$ (per area, solid angle and energy),
one obtains a $\gamma$-ray flux
\begin{equation}
\Phi_\gamma (E) = {\sigma_{pp,inel} \over d^2} {M \over m_H} \kappa k_p \left( {E_\gamma \over f} \right)^{-\Gamma}
\end{equation}
with the hydrogen mass $m_H$. With a small correction for heavier nuclei and assuming a CR proton flux
as measured on Earth, this translates, for example, into a flux
\begin{equation}
  \label{eq:cloudflux}
  \Phi_{\gamma}(> 1 \mathrm{TeV}) \approx 1.6 \cdot 10^{-12} M_6/d^2_{\rm kpc}\,\mathrm{cm}^{-2} \mathrm{s}^{-1}
\end{equation}
where $M_6$ is the cloud mass in units
of $10^6$ solar masses \citep[see e.g.][]{aharonian91}.
These estimates assume that the CRs permeate the cloud; the
increased magnetic fields inside dense clouds may however influence
the diffusion coefficient and hence the spectrum and flux of 
CRs. In the case of clouds illuminated by nearby CR accelerators
which were active for a relatively short time, energy-dependent
diffusion will modify the spectrum at the cloud as compared to the
spectrum generated at the accelerator; high-energy particles reach the
cloud first, resulting in a harder $\gamma$-ray spectrum at the
cloud  \citep[see][and references therein]{gabici07a}.

Dense environments will also result in production of bremsstrahlung
$\gamma$-rays by electrons; assuming that electron and proton spectra
are identical and that the ratio of electrons to protons is $k_{ep}$,
one obtains a flux ratio $\Phi_{\gamma,{\rm brems}}/\Phi_{\gamma,{\rm
hadr}} \approx 3.3\,k_{ep}$ at 1~TeV, with only weak dependence on
photon energy due to the changing $\sigma_{pp}$ (in the region far
from cut-offs).

\subsection{Absorption during propagation and opacity of sources}
\label{sec:abs}

\begin{figure}
\centerline{\psfig{figure=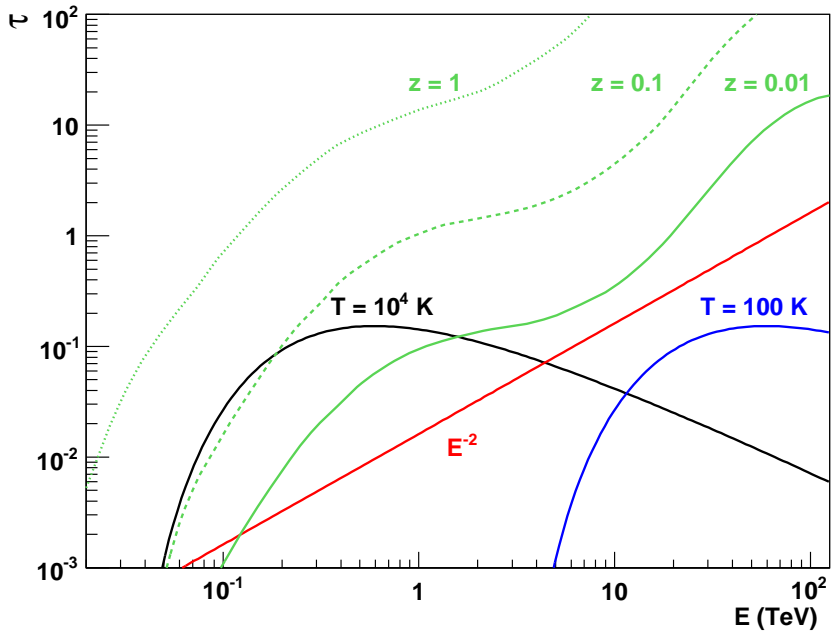,height=20pc}} 
\caption{
  Optical depth ($\tau$) of $\gamma$-rays as a function of energy, for FIR
  ($kT = 0.008$\,eV) and visible ($kT = 0.8$\,eV) target photon fields, of
  column density $3 \cdot 10^{22}$\,eV cm$^{-2}$ for FIR (1\,eV cm$^{-3}$ over
  10\,kpc) and $3 \cdot 10^{24}$\,eV cm$^{-2}$ for visible (1\,eV cm$^{-3}$ over
  1\,Mpc). The effect of absorption on a non-thermal photon distribution
  (with $dN/dE \propto E^{-2}$) is shown for comparison. Note that the 
  absorption is constant for an $E^{-1}$ photon field. The green curves
  show the optical depth for pair production on the EBL for redshifts 
  of 0.01, 0.1 and 1 for the model of \citet{2008A&A...487..837F}.
}
\label{fig_abs} 
\end{figure}

The mean free path of VHE $\gamma$-rays in hydrogen gas is governed by the
electron-positron pair production cross-section and has a value of 80\,g\,cm$^{-2}$ or 
equivalently $5\cdot 10^{25}$\,hydrogen atoms cm$^{-2}$; for all practical purposes, the
universe is transparent to this process.
The more relevant process is the absorption by pair production
on ambient (CMB, IR, visible or X-ray) photons of energy $E_{T}$. The
process acts above the threshold $E_\gamma E_T = m^2c^4$ or
$E_{\gamma,{\rm TeV}} = 0.26/E_{T,{\rm eV}}$ and the absorption cross section for
an isotropic photon field $\sigma_{\gamma\gamma}$ 
\citep[see e.g.][]{aharonian08} peaks close to threshold at
$E_{\gamma,{\rm TeV}} = 0.9/E_{T,{\rm eV}}$. Figure~\ref{fig_abs} illustrates
absorption by different (blackbody and power-law) photon spectra.

Two situations where absorption is important are (a) the observation of
extragalactic sources and (b) TeV emission near intense sources of radiation. 
Figure~\ref{fig_abs} also shows the optical depth for VHE $\gamma$-rays
interacting with intergalactic radiation fields taking into
account cosmological evolution of the background fields; the
range of $\gamma$-rays is about $ z=0.03, \sim0.1, \sim1$ for $E_\gamma = 10, \sim1,
\sim0.1$\,TeV. At PeV energies, the range is reduced to galactic distance
scales. At energies up to tens of TeV, on the other hand, absorption
of galactic sources is almost negligible (Figure~\ref{fig_abs}), even
for sources near the Galactic Center with its increased radiation
fields \citep{moskalenko06}.

Absorption is relevant for $\gamma$-rays produced in radiation-intensive
environments, for example in systems of compact objects in close orbit 
around massive stars. Radiation produced, for example, in a massive X-ray binary at
0.1\,AU from the massive star traverses ${\cal O}$($10^{27}$)\,eV\,cm$^{-2}$, 
resulting in a large optical depth at
TeV energies if the $\gamma$-ray source is behind the star, allowing
head-on collisions between $\gamma$-rays and stellar photons. Because of
the peaked absorption cross section, narrow (black-body) background
photon spectra cause selective absorption in part of the VHE energy
range and lead to significantly modified spectra (see Figure~\ref{fig_abs} and
\citet{aharonian08}). In such dense absorbers,
the electrons produced may undergo IC scattering, creating a pair
cascade which proceeds until $\gamma$-ray energies are low enough 
that the medium becomes transparent. The absorption dip in the $\gamma$-ray spectrum
is then accompanied by a corresponding enhancement at lower energies
\citep[see for example][]{protheroe93}. Absorption within the
source is also a key factor for $\gamma$-ray
emission from compact regions (knots, blobs...) within AGN jets, due to the 
high radiation density in these regions; see Section~\ref{sec:AGN}.

\section{Instruments for TeV astronomy and their characteristics}

\begin{figure}
\label{fig_sketch} 
\centerline{\psfig{figure=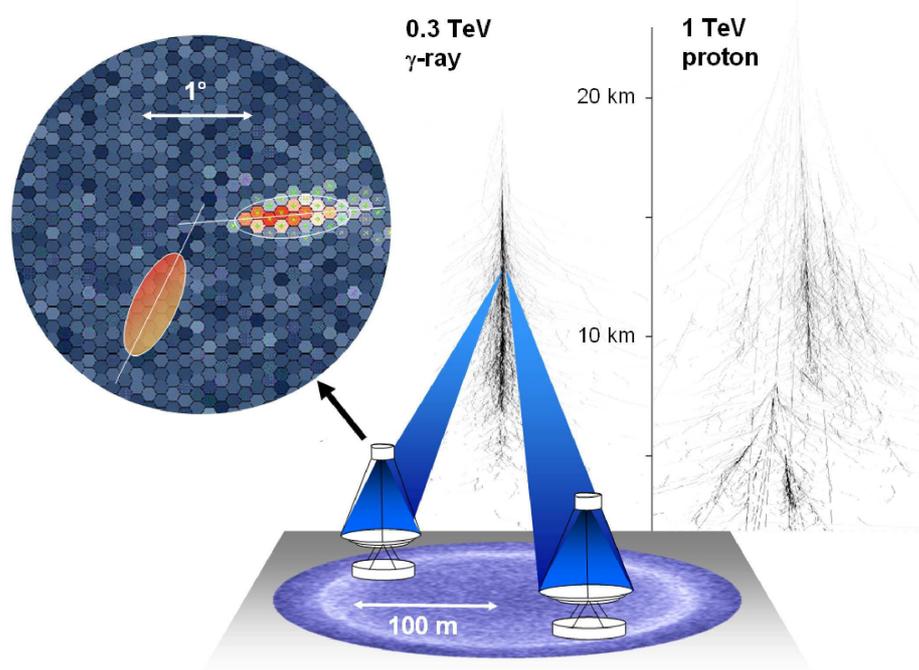,height=22pc}} 
\caption{A sketch of the imaging atmospheric Cherenkov technique showing the
formation of an electromagnetic cascade for a 300~GeV photon primary, 
the production of Cherenkov light,
and the formation of an image in the camera of a Cherenkov telescope.
Cherenkov light production for a proton initiated cascade is shown
for comparison. Shower images produced by Konrad Bernl\"ohr.}
\end{figure}

Given the very low fluxes of $\gamma$-rays in the VHE regime --- ${\cal
O}(10^{-11})$\,photons per cm$^2$-second (a few photons per m$^2$-year) above
1 TeV for strong sources, direct detection by space-based instruments is
excluded.  Ground-based instruments detect secondary products
resulting from the development of $\gamma$-ray initiated air-showers;
either particles reaching the ground or Cherenkov light emitted by
shower particles in the atmosphere. In contrast to the well-collimated
electromagnetic air-showers induced by $\gamma$-rays (or electrons),
air-showers initiated by CR nucleons 
typically feature a number of electromagnetic sub-showers induced by
$\pi^0$ decays and contain muons from charged pion decays
(see Figure~\ref{fig_sketch}). Rejection of the background of showers
initiated by charged CRs is a key performance criterion for $\gamma$-ray detection
systems, and is usually achieved on the basis of shower shape or muon
content. A more detailed discussion of air-shower characteristics and
the detection systems used can be found for example in \citet{aharonian_buckley08}.

\subsection{Instrument characteristics}

For ground-based instruments detecting $\gamma$-rays via their shower
development in the atmosphere, effective detection areas, $A(E)$
(defined such that the differential detection rate $R_{\gamma}(E) =
\Phi_{\gamma}(E) A(E)$), have a sub-threshold region where they
exhibit a steep rise with energy, and a high-energy region where
$A(E)$ varies only weakly with energy. In the sub-threshold region,
the detector triggers only because of favorable fluctuations in the
development of an air-shower. In the high-energy region, every
air-shower within a certain fiducial region is recorded. The `energy
threshold' of a detection system is usually quoted as the energy at
which the peak detection rate $R(E)$ occurs for typical power-law
$\gamma$-ray spectra. The threshold thus determined obviously depends
on the assumed spectral index, but is always located in the transition
region between the steeply rising part and the nearly constant region
of $A(E)$.  Individual events may be detected at energies well below
this nominal threshold.

Two criteria govern detectability of a source during an exposure time
$T$: (a) a minimum number $n_0$ (usually 5...10) of $\gamma$-rays must
be detected, $T R_\gamma > n_0$, and (b) the $\gamma$-ray signal must
be be significant above fluctuations in the background, approximately
$(T R_\gamma)/(T \eta_{\rm CR} R_{\rm CR} \Omega)^{1/2} > \sigma_0$.
Here, $R_{\rm CR}$ is the detection rate of background CRs per solid
angle, $\eta_{\rm CR}$ the efficiency for CRs passing analysis cuts
relative to the corresponding efficiency for $\gamma$-rays, and
$\Omega$, the solid angle over which the signal from a source has to
be integrated, given either by the point spread function (PSF) of the
instrument or the source size. For point sources and a Gaussian PSF,
$\Omega \approx \pi \theta_{68}^2$ where $\theta_{68}$ is the 68\%
containment radius of the PSF.  Current instruments typically operate
in the background dominated regime, implying that minimal detectable
fluxes scale as $T^{-1/2} \eta_{\rm CR}^{1/2} \theta_{68}$. For
sources which are large compared to the PSF, $\Omega \approx \pi
\theta_{s}^2$, and sensitivity degrades linearly with source radius
$\theta_{s}$.

\subsection{Cherenkov imaging of air-showers}

In the past decade, Imaging Atmospheric Cherenkov Telescopes (IACTs)
have emerged as the most powerful instrument for pointed observations
of VHE $\gamma$-ray sources.  IACTs use focusing mirrors to image the
Cherenkov light emitted by shower particles onto a pixelated photon
detection system (see Figure~\ref{fig_sketch}); a summary of the
characteristics of current and selected previous instruments is given
in Table~\ref{tab_iacts}.

At the maximum of the shower development, around 10\,km a.s.l. for TeV
energies, the Cherenkov threshold for electrons is around 40\,MeV and
the Cherenkov angle is $0.7^\circ$ or less.  Light emitted at the
Cherenkov angle reaches the ground within a circle of 100 to 150\,m
depending on the height above sea level of the detection system.
Multiple-scattering angles of shower particles near the Cherenkov
threshold are comparable to the Cherenkov angle, resulting in a more
or less uniformly filled light pool, with typically 10 detected
Cherenkov photons per TeV shower energy and m$^2$ mirror area for
photomultiplier sensors. With increasing energy, the central density
in the light pool is enhanced due to deeper penetration of
showers. Triggering and image reconstruction usually requires 50 to
100 detected photons and sets the scale for the dish size.  The pixel
size of the detection system should be matched to the size of features
in air-shower images; simulation studies show saturation of
performance for pixels much below $0.1^\circ$ diameter --- close to
the typical rms width of a $\gamma$-ray image at TeV energies.  The
asymptotic collection area for IACTs is determined by the maximum
impact distance for which shower images still fall within a camera and
hence by the camera field of view (FOV). At 2000~m a.s.l. the impact
distance limitation is approximatey 100~m per degree of the opening
angle of the camera field of view (for showers close to zenith).

Most modern instruments use multiple telescopes (a) to image the
air-shower from different viewing angles for improved reconstruction
of $\gamma$-ray direction and rejection of CR background and (b) to
apply a coincidence requirement rejecting single-telescope triggers
caused by CR muons with impact points close to a telescope mirror, or
by night sky background. Telescope spacing needs to be large enough to
provide a sufficient baseline for stereoscopic measurements, but small
enough that multiple telescopes fit within the Cherenkov light pool;
the exact spacing tends to be uncritical within a range of $\sim$
70\,m to 150\,m. Depending on selection cuts, telescope systems such
as H.E.S.S. (Table \ref{tab_iacts}) provide an angular resolution for
single $\gamma$-rays of 3$'$ to 6$'$, a $\gamma$-ray energy resolution
of around 15\% and a CR rejection factor of ${\cal O}$($10^{-2}$) or
better. Combined with the energy threshold of about 100\,GeV and a
high-energy effective area of some $10^5$\,m$^2$, this allows
detection of sources of 1\% of the strength of the Crab Nebula ($\nu
F_{\nu} \sim 3 \times 10^{-13}$ erg cm$^{-2}$ s$^{-1}$ around 1~TeV)
within 25~h of observations close to Zenith.  Performance of IACTs is
at some point limited by fluctuations in air-shower development; for
example, at energies below 10~GeV an air-shower is in principle still
detectable, but because of the small number of Cherenkov-emitting
particles, the energy determination is quite unreliable. Current
instruments, however, are still relatively far from reaching
fundamental limits; for example, shower fluctuations allow an angular
resolution of a fraction of an arc-minute at TeV energies, provided
that the number of Cherenkov photons detected is sufficient, and a CR
rejection factor of order $10^{-4}$ at TeV energies
\citep{2006astro.ph..3076H}.
A non-trivial issue in the analysis of data is the
absolute energy calibration. Cherenkov light from local muons is often
used to calibrate the response of instruments, but variations in
atmospheric profile and transmission and of the orientation of the
shower axis relative to the geomagnetic field can 
influence the shower development and the light yield
\citep[e.g.][]{2000APh....12..255B} and cause systematic calibration
uncertainties at the 10--20\% level.

\begin{table}
\small 
  \begin{tabular}{|l|c|c|c|c|c|c|c|c|c|c|} \hline
    Instrument & Lat. & Long. & Alt. & Tels. & Area & Pixels & FoV & Thresh. & Sens. \\
    & ($^{\circ}$) & ($^{\circ}$) & (m) &            & (m$^{2}$)  &            & ($^{\circ}$) & (TeV) & (\% Crab) \\\hline 
    H.E.S.S. & -23 & 16 & 1800 & 4 & 428 & 960 & 5 & 0.1 & 0.7 \\ 
    VERITAS & 32 & -111 & 1275 & 4 & 424 & 499 & 3.5 & 0.1 & 1 \\ 
    MAGIC & 29 & 18 & 2225 & 1 & 234 & 574 &  3.5$^{\dagger}$ & 0.06 & 2 \\ 
    CANGAROO & -31 & 137 & 160 & 3 & 172 & 427 & 4 & 0.4 & 15 \\
    Whipple & 32 & -111 & 2300 & 1 & 75 & 379 & 2.3 & 0.3 & 15 \\
    \it{HEGRA} & 29 & 18 & 2200 & 5 & 43 & 271 & 4.3 & 0.5 & 5 \\
    \it{CAT} & 42 & 2 & 1650 & 1 & 17.8 & 600 & 4.8$^{\dagger}$ & 0.25 & 15 \\
    \hline
  \end{tabular}
  \label{tab:inst}
  \caption{Properties of selected air-Cherenkov instruments, including two of historical interest (HEGRA and CAT).
  $^{\dagger}$ These instruments have pixels of two different sizes.   
  Adapted from \citet{hinton08}. 
}
\label{tab_iacts}
\end{table}

\subsection{Ground-based particle detectors}

The direct detection of air-shower particles offers a method of
$\gamma$-ray detection with close to 100\% duty cycle and very wide
($\sim$1 steradian) field of view. Because of these advantages this
method offers an interesting complementary approach, despite the fact
that the point-source sensitivity of such detectors is currently
almost two orders of magnitude poorer than the most sensitive IACTs
(at a few TeV).  Indeed the extensive air-shower sampling (EAS)
technique has recently produced its first contributions to the
catalogue of TeV sources~\citep{abdo07} and proved its usefulness for
all sky surveys (at modest sensitivity) and the detection of diffuse
emission~\citep{abdo08}.

The main challenges of the EAS approach are the rejection of the
CR background and directional and energy reconstruction using
the exponentially decreasing tail of particles detected well beyond
shower maximum.  High altitudes are therefore critical to achieve low
($<1$ TeV) thresholds with such instruments. In addition, whilst in
principle all-sky detectors, the field of view obtained is limited in
practice by the rapid increase in energy threshold with zenith angle;
typically a factor two increase between 0$^{\circ}$ and 30$^{\circ}$
zenith \citep[see e.g][]{abdo07}.  The reconstruction of the primary
$\gamma$-ray direction is based on shower front timing. The arrival
time of the shower front can be determined with an accuracy of a few
nanoseconds over $\sim$ 100~m baselines leading to typical resolutions
of 0.3$^{\circ}$ -- 1$^{\circ}$ \citep[see e.g.][]{atkins03}. 
The much larger fluctuations present
in the particle number at ground level with respect to the essentially
calorimetric air-Cherenkov approach, make primary energy determination
extremely difficult for EAS detectors.  The rejection of the hadronic
background is based on the muon content of showers and/or the
distribution of shower particles on the ground.

The Water Cherenkov approach pioneered by the MILAGRO collaboration,
appears to represent the most cost effective method of achieving
complete ground coverage. MILAGRO is a $80\times60\times8$ m pond 
instrumented with PMTs and surrounded by 175 water tanks, located at 
an altitude of 2630~m \citep{atkins03}. MILAGRO achieves its best
background rejection power and sensitivity in the regime above 10~TeV. 
Widely spaced ($\sim$1 m$^{2}$ detectors $>5$ m apart) 
scintillation-based detectors have been used for CR measurements 
and in the search for UHE $\gamma$-ray sources for many years 
Instruments of this type located at high altitudes can be used for
$\gamma$-ray astronomy around a few TeV, as demonstrated with the
Tibet Air-Shower Array \citep{amenomori99} at 4300~m above sea level.
The ARGO-YBJ instrument is a 5800 m$^{2}$ complete ground coverage
instrument at the same site, utilizing resistive plate counters (RPCs)
and achieving a threshold of a few hundred GeV and a sensitivity 
similar to that of MILAGRO. However, this approach 
is likely prohibitively expensive for a much larger area 
next generation detector.

\section{VHE $\gamma$-ray sources}

\begin{figure}
\centerline{\psfig{figure=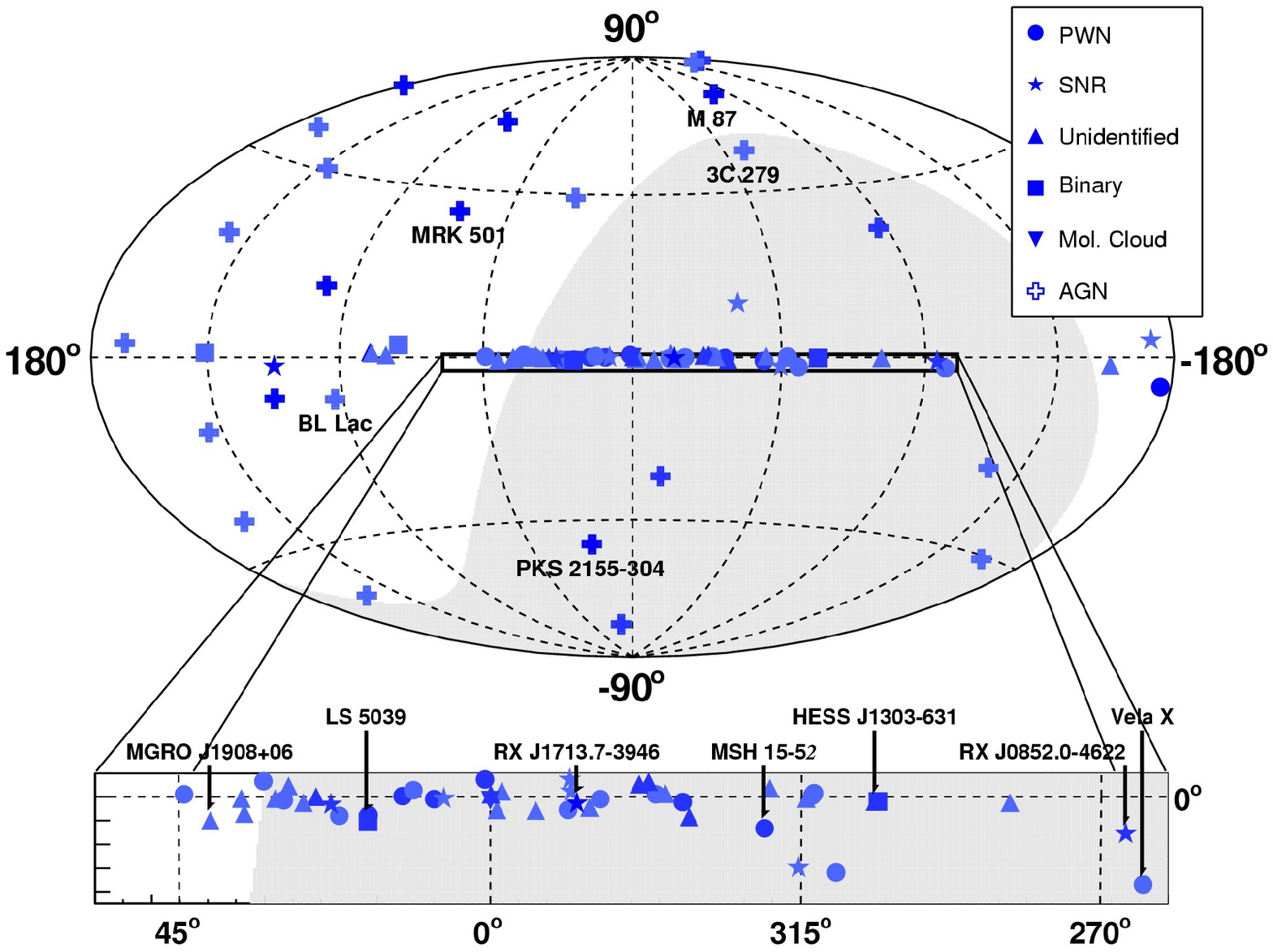,height=25pc}} 
\centerline{\psfig{figure=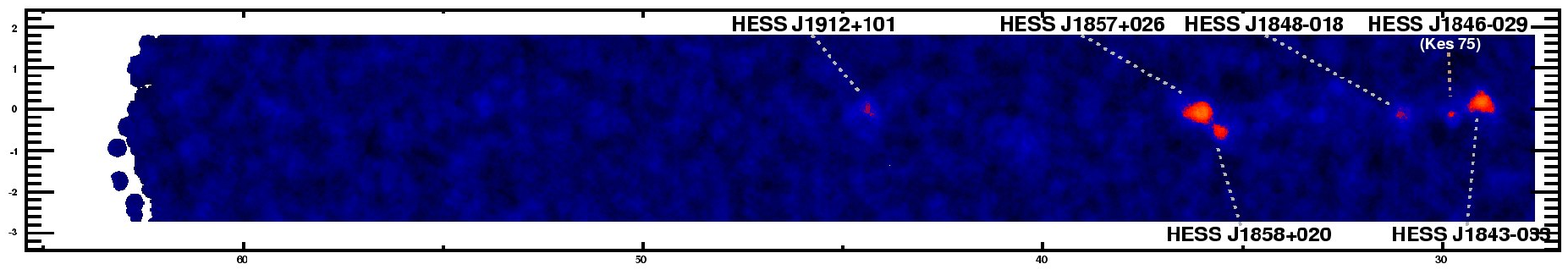,height=5pc}} 
\centerline{\psfig{figure=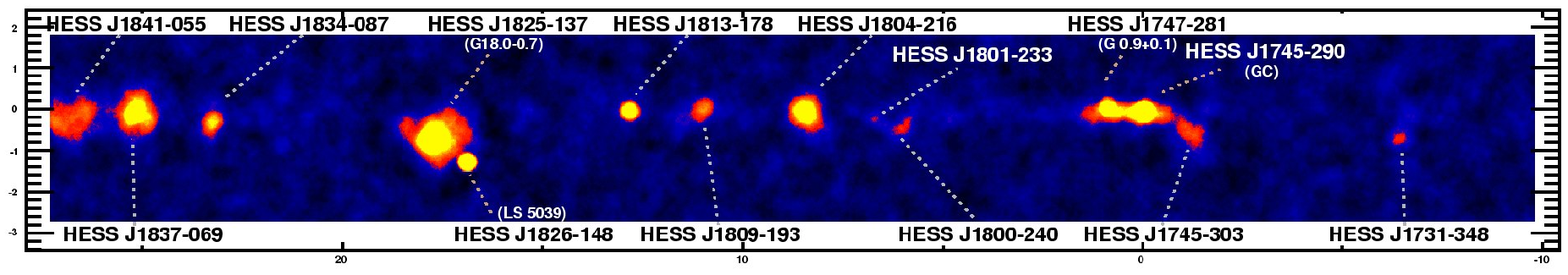,height=5pc}} 
\centerline{\psfig{figure=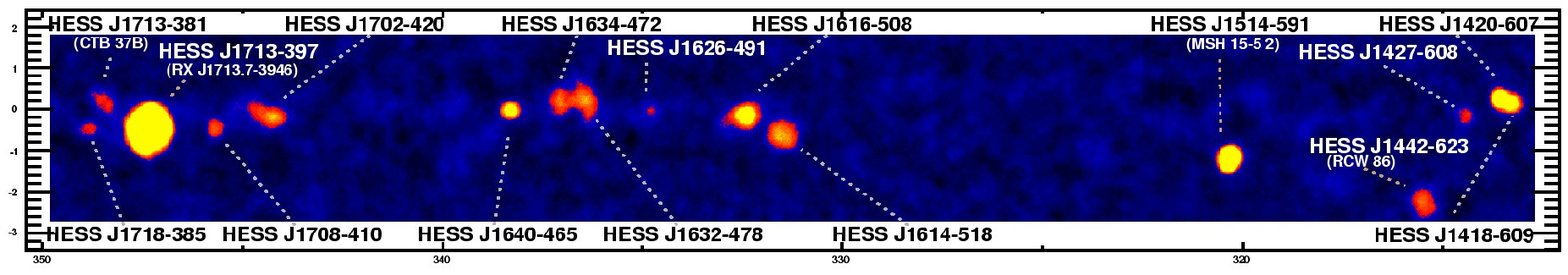,height=5pc}} 
\centerline{\psfig{figure=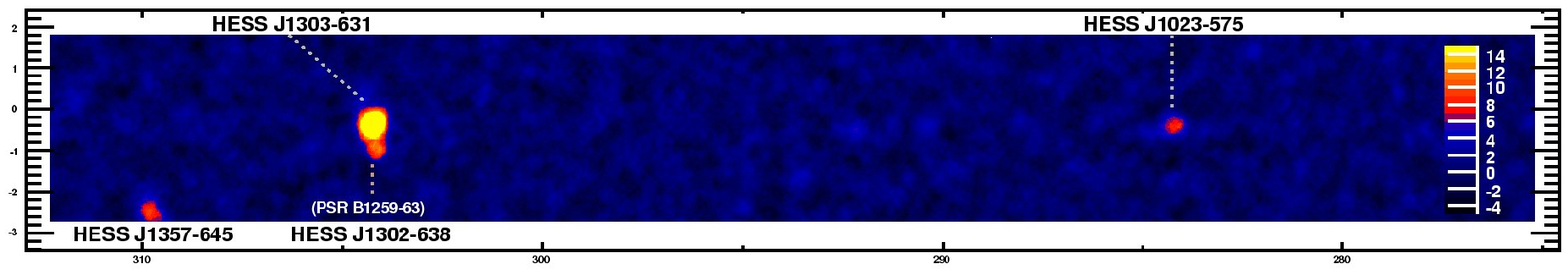,height=5pc}} 
\caption{
  {\it Top}: The current catalog of VHE $\gamma$-ray sources plotted 
  on the sky in Galactic coordinates. {\it Bottom}: the Milky Way viewed in 
  VHE $\gamma$-rays: the H.E.S.S. survey of the Galactic plane 
  \citep[reproduced from][]{HESS:scanicrc}.
}
\label{fig_tevsky}
\end{figure}

As can be seen from Figure~\ref{fig_tevsky}a, the TeV sky, despite a
modest number of known objects ($\sim$80), contains a diverse
collection of different object classes.  Numerically dominant are the
sources clustered along the plane of our galaxy (see 
Figure~\ref{fig_tevsky}b). In contrast to GeV energies, sources, rather
than diffuse emission, dominate our current view of the Galaxy. In the
following, we will attempt to summarize the important characteristics
of each class of TeV $\gamma$-ray sources 
in the context of the discussion
on particle acceleration, transport and radiation given in Section
\ref{sec:CB}. Table~\ref{tab:gal} gives a summary of prominent Galactic
VHE $\gamma$-ray emitters for which firm identifications exist.
There is insufficient space here for an adequate
discussion of diffuse TeV emission. We refer the reader to 
\citet{aharonian_buckley08} for a review of this topic and also
the complex situation at the Galactic center.

\begin{table}
    \begin{tabular}{|l|l|l|l|l|l|l|} \hline 
Object	&  Type	      & Method	& Flux  &Ref.  \\\hline
PSR B1259$-$63  &  Binary  &Pos/Var & 7$^{\star}$	& \cite{HESS:psrb1259}   \\
LS\,5039	&  Binary     &Pos/Per	& 3$^{\star}$		& \cite{HESS:ls5039p2}  \\
LS\,I\,+61\,303 &  Binary     &Pos/Var	& 16$^{\star}$	&  \cite{MAGIC:lsi61}   \\
RX\,J1713.7$-$39046 &  SNR   &	Mor	& 66	&  \cite{HESS:rxj1713p1} \\
Cassiopeia\,A	&  SNR	      &	Pos	& 3	& \cite{HEGRA:casA}    \\
Vela Junior	&  SNR  &	Mor	& 100	& \cite{CANGAROO:velajnr}  \\
RCW\,86		&  SNR  &	Mor	& 5-10?	& \cite{HESS:rcw86} \\
SN\,1006        & SNR   &  Mor & ? & \cite{HESS:sn1006} \\
Crab Nebula	&  PWN	      &	Pos	& 100		& \cite{weekes89} \\
G\,0.9+0.1	& PWN	      &	Pos     &  2     & \cite{HESS:g09}\\
MSH\,15-52	&  PWN	      &	Mor	& 15		& \cite{HESS:msh1552}  \\
HESS\,J1825$-$137
                &  PWN	      & EDMor	& 12    & \cite{HESS:1825p2}  \\
Vela\,X		&  PWN	      &	Mor	& 75	& \cite{HESS:velax}    \\
\hline
\end{tabular}
   \caption{ Selected galactic VHE $\gamma$-ray sources with well
     established multi-wavelength counterparts. Note that all these
     objects are X-ray sources.  Fluxes are approximate percentages of
     the TeV flux from the Crab Nebula, $^{\star}$ indicates variable
     emission. The final column lists the publications where a firm
     identification of the source was made. These associations were
     established through a range of methods, given here in abbreviated
     form: {\it Pos}: The centroid position of the VHE emission is
     established with sufficient precision that there is no ambiguity
     as to the counterpart.  {\it Mor}: There is a match between the
     $\gamma$-ray morphology and that seen at other wavelengths.  {\it
     EDMor}: Energy-dependent morphology is seen which approaches the
     morphology seen at other wavelengths at some limit, and is
     consistent with our physical understanding of the source. {\it
     Var}: $\gamma$-ray variability correlated with that in another
     waveband is observed. {\it Per}: periodicity in the $\gamma$-ray
     emission is seen, matching that seen at another wavelength. Table
     adapted from \cite{hinton08}.  }
   \label{tab:gal}
\end{table}

\subsection{Supernova remnants}

Ever since \citet{zwicky39}, supernova remnants have been viewed as
the most likely sources of galactic cosmic rays up to an energy of at
least that of the {\it knee} of the CR spectrum around $10^{15}$\,eV,
and possibly beyond $10^{17}$\,eV. This argument is based in part on
the energy input required to maintain the CR flux in the Galaxy,
$dE/dt \approx \rho V \tau \sim 5 \cdot 10^{40}$\,erg\,s$^{-1}$, with
the energy density in cosmic rays $\rho \sim 1$\,eV\,cm$^{-3}$, the
confinement volume of cosmic rays $V \sim 10^{67}$\,cm$^{3}$ and the
characteristic residence time $\tau \sim 10^7$\,yr. With a kinetic
energy output of $10^{51}$\,erg per explosion and a rate of a 2--3 per
100 years, an average of 10\% of the supernova kinetic energy needs to
be converted into CR energy. The first-order Fermi acceleration
process outlined in Section~\ref{sec:acc}, possibly enhanced by
magnetic field amplification, provides a means to plausibly reach this
efficiency and a maximum energy beyond $10^{15}$\,eV, and naturally
provides a power-law spectrum with an index around 2, which can
explain the observed CR index if energy-dependent diffusion and escape
is invoked~\citep[see e.g.][]{strong07}. While this scheme is
generally accepted, there are still open questions concerning the
consistency of required diffusion speeds and (the lack of a
significant) anisotropy of the CRs at the Earth. Equally important,
while the presence of high-energy electrons in SNRs can be inferred
from the non-thermal X-ray spectra measured in several objects, modulo
the a priori unknown, strength of $B$-fields, efficient acceleration
of nuclei still lacks undisputed evidence, and observational proof is
missing that acceleration in SNR can quantitatively account for the
observed CR spectrum.

\begin{figure}
  \centerline{\psfig{figure=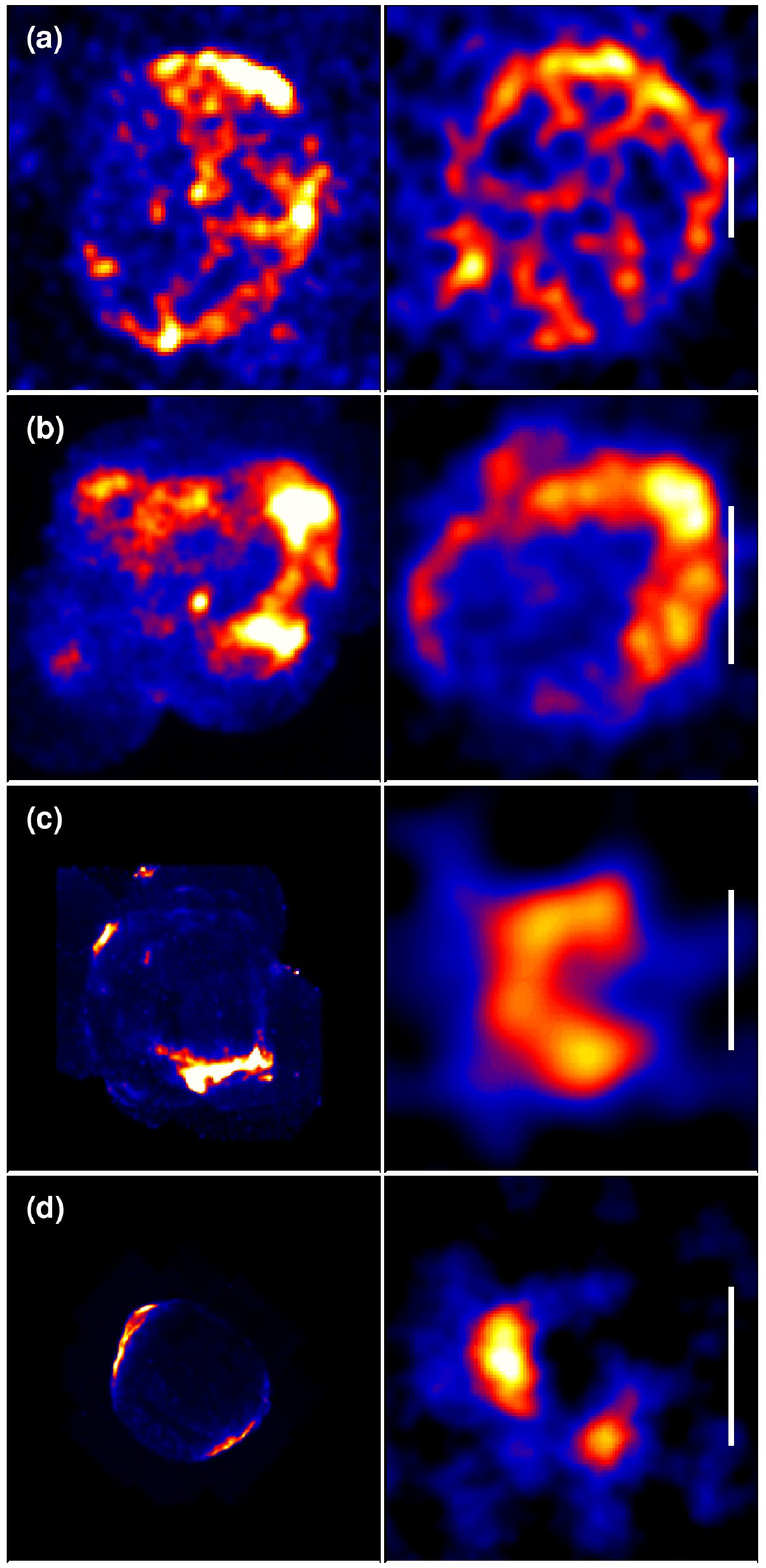,height=40pc}} 
  \caption{ Four SNRs imaged in (dominantly) non-thermal X-rays (left) and 
    resolved in VHE $\gamma$-rays with H.E.S.S. (right).
    a) RX\,J1713.7$-$3946 with 1--3 keV data from ASCA \citep{uchiyama02},
    b) RX\,J0852.0$-$4622  with ROSAT (1.3--2.4 keV) \citep{aschenbach98},
    c) RCW\,86 with 2--4 keV data from XMM-Newton \citep{vink06}
    d) SN\,1006 with Chandra archive data (0.5--10 keV).
    The H.E.S.S. data are taken from \citet{HESS:rxj1713p2, HESS:velajnr2, HESS:rcw86, HESS:sn1006}.
    The white scale bars are 0.5$^{\circ}$ long.
  }
  \label{fig_snr}
\end{figure}

\begin{figure}
  \centerline{\psfig{figure=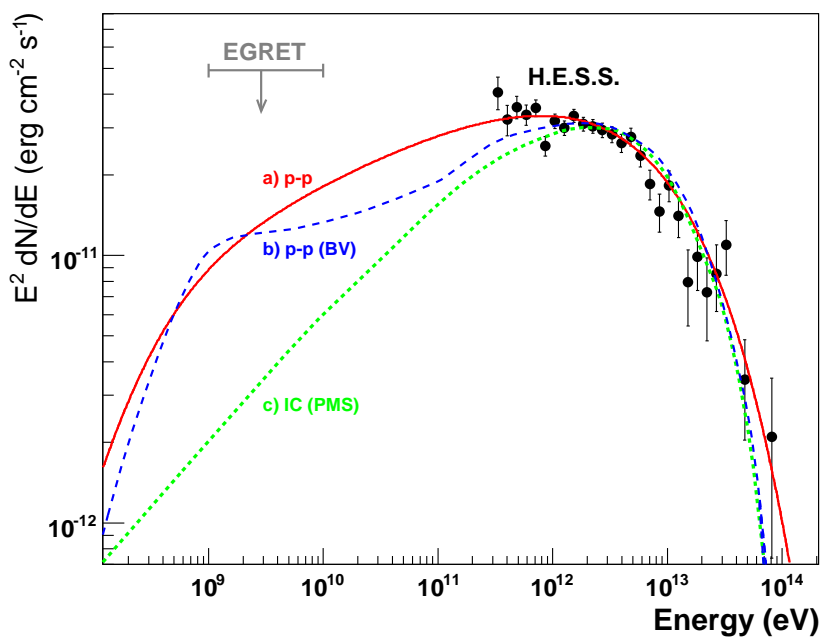,height=20pc}} 
  \caption{
    The SED of RX\,J1713.7$-$3946 at $\gamma$-ray energies.
    Three curves are shown are shown in comparison to the H.E.S.S. data
    \citep{HESS:rxj1713p3} and EGRET upper limit: a) the best-fit $\gamma$-ray 
    spectrum arising from interacting protons with an energy distribution 
    following a power-law with exponential cut-off, see \cite{kelner06},
    (b) hadronic emission as calculated by \cite{berezhko06}, and
    (c) IC emission as calculated by \cite{porter06}.
    Reproduced from \cite{hinton08}. 
  }
  \label{fig_rxj1713}
\end{figure}

VHE $\gamma$-rays trace the relevant populations of energetic
particles in SNRs. The flux of hadronic $\gamma$-rays is proportional
to the CR density times the density of target gas, and the electronic
IC $\gamma$-rays directly trace the electron density, given that
target photon fields are likely to be almost constant on the scale of
the SNR. The ratio of IC $\gamma$-rays and synchrotron X-rays is
determined by the strength of the magnetic field. A detailed study of
acceleration in an SNR shell, and distinction between the shell and
the nebula of a possible pulsar created in the explosion, requires the
shell to be resolved in VHE $\gamma$-rays. This has so far been
achieved for four SNRs: RX\,J1713.7$-$3946 \citep{HESS:rxj1713p2},
RX\,J0852.0$-$4622 (alias {\it Vela Junior}) \citep{HESS:velajnr2},
RCW\,86 \citep{HESS:rcw86} and, most recently, SN~1006
\citep{HESS:sn1006}. The $\gamma$-ray spectrum measured for
RX\,J1713.7$-$3946 extends to several tens of TeV and follows a power
law with index 2 up to about 20\,TeV where a cut-off sets in (Figure
\ref{fig_rxj1713}), demonstrating the presence of primaries in the
100s of TeV energy range \citep{HESS:rxj1713p3}.  In all cases, there
is a strong similarity between the morphology observed in VHE
$\gamma$-rays and in non-thermal X-rays, once the difference in PSF of
the instruments is taken into account (see Figure \ref{fig_snr}). The
correlation between X-rays and $\gamma$-rays seems to argue in favor
of a leptonic origin of $\gamma$-rays.  Correlation with gas density,
as traced by CO, is less pronounced. This may, however, be due to a
lack of resolution along the line of sight (i.e. in distance estimated
by line velocity), making it difficult to know if a given gas mass is
actually co-located with the accelerated particles.
Only in a few special cases can the
association between gas and remnant be proven by an observed high
velocity dispersion of the gas, resulting from the interaction with
the SNR shock, or by OH maser emission.

The complex morphology of the remnants, with non-uniform emission
along the rim, reflects their interaction with their environment and
makes interpretation difficult, in particular given the lack of
high-resolution, 3-dimensional information on the surrounding gas
density. In the case of SN\,1006, located off the Galactic plane
in a less complex environment, the observed bipolar emission pattern
can be modeled more easily: SN\,1006 is believed to be threaded by a
relatively uniform magnetic field, and non-thermal emission marks the
polar caps where the B-field vector is roughly parallel to the
expansion direction. In the equatorial regions, where field lines are
perpendicular to the direction of shock propagation, particle
injection into the Fermi process is presumably inefficient,
since particles spiral along field lines and are immediately overrun
by the shock. In the polar regions, injection is more effective,
resulting in significant CR current and most likely in
turbulent magnetic field amplification, rendering the acceleration process even
more efficient.

Explaining the VHE emission from RX\,J1713.7$-$3946 --- probably the
best studied case --- as IC emission of electrons requires magnetic
fields of around 10\,$\mu$G \citep{HESS:rxj1713p2}.  These magnetic
fields are at variance with the significantly higher fields of at
least 50\,$\mu$G determined from the width of X-ray emitting
filaments, translated into an electron cooling time
\citep[e.g. ][]{berezhko06}, and from the temporal variations observed
in X-ray structures \citep{2007Natur.449..576U}, which directly
measure the cooling time. Questions remain, though, whether these high
magnetic fields are characteristic of the entire SNR, and if the
observed structures are due to variations in electron density rather
than magnetic fields; only if this is the case, an IC origin of the
$\gamma$-rays is firmly ruled out. There are no significant variations
of $\gamma$-ray spectral shape across the TeV SNRs and the observed
spectra are well described by hadron acceleration models which
generate a power-law distribution with an index close to 2, with a
gradual high-energy cut-off (Figure \ref{fig_rxj1713}). Leptonic
models tend to generate spectra that are too hard at low energy,
reflecting the fact that the index of IC $\gamma$-rays is 1.5, for an
electron index of 2 (see Section~\ref{sec:elec}).  This problem can be
solved to a limited extent by adjusting the composition of target
radiation fields and by introducing multi-zone models with different
cut-offs, where different IC peaks are superimposed to mimic an
$E^{-2}$ spectrum \citep[e.g.][]{porter06}.  In hadronic models, on
the other hand, the strong X-ray/$\gamma$-ray correlation is
non-trivial to obtain. In the case of high B-fields, electron
lifetimes are comparable to acceleration timescales and the X-ray flux
is influenced both by the energy input in accelerated electrons and by
the strength of B-fields. Hadronic $\gamma$-rays, on the other hand,
reflect proton flux --- which should scale with the (injected)
electron flux --- multiplied by the gas density.  A strong correlation
between X-rays and $\gamma$-rays requires a link between gas density
and B-field strength; magnetic field amplification (see
Section~\ref{sec:acc}) may provide a mechanism for this.

A final demonstration of CR origin in SNRs may
be achieved by a combination of wider spectral $\gamma$-ray coverage,
improved resolution or morphology, and larger-scale measurements of
magnetic fields. 
Detection of neutrinos from SNRs would also
demonstrate hadronic origin, but is challenging even for the largest
instruments such as ICECUBE, 
and suffers similarly from the fact that,
for a quantitative analysis, the target gas density needs to be
known. Investigation of global CR energetics and spectra will
in any case certainly require the $\gamma$-ray detection and spectral analysis of a
representative sample of SNRs.

\begin{figure}
  \centerline{\psfig{figure=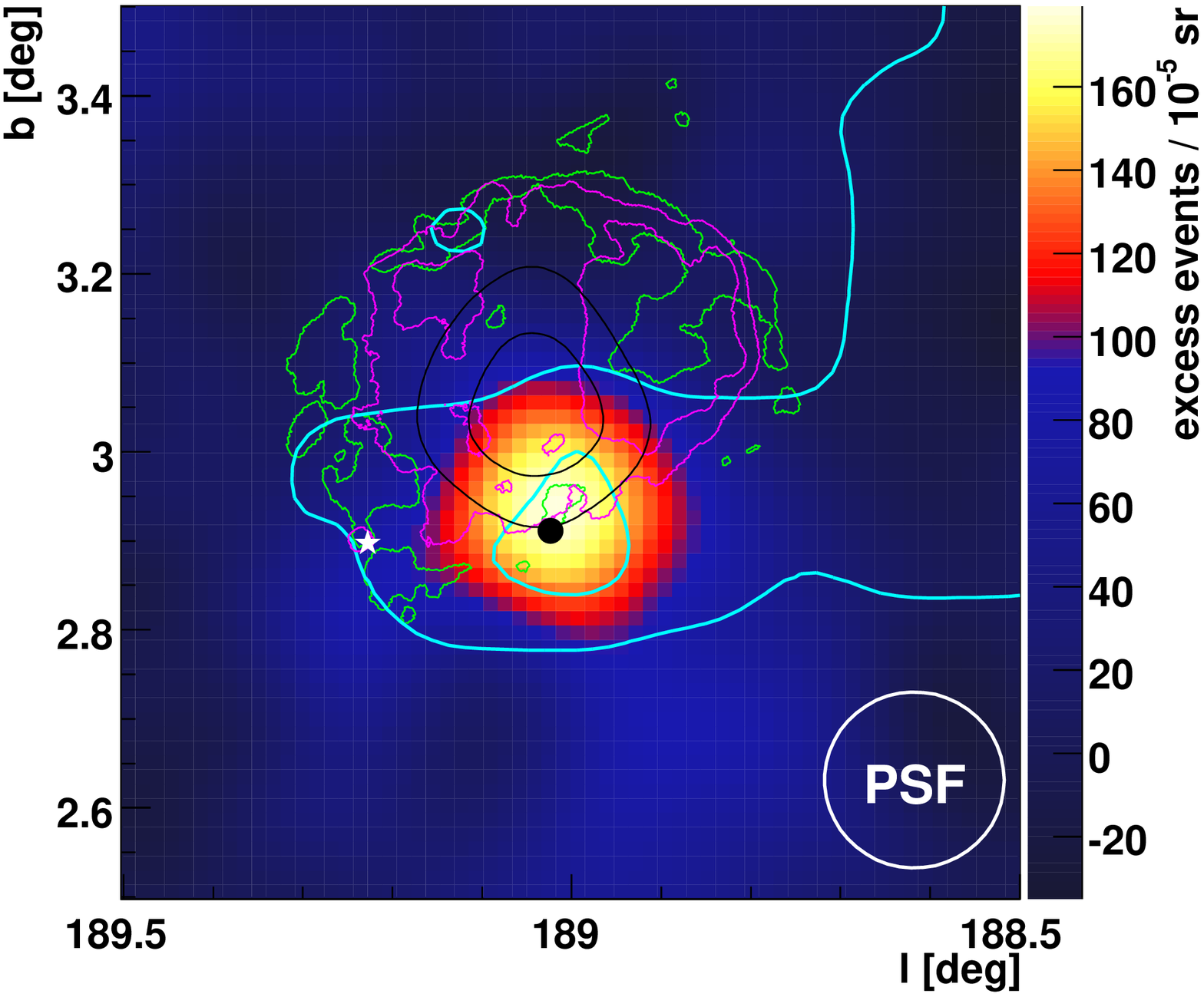,height=17pc} 
    \vspace{-2mm}\psfig{figure=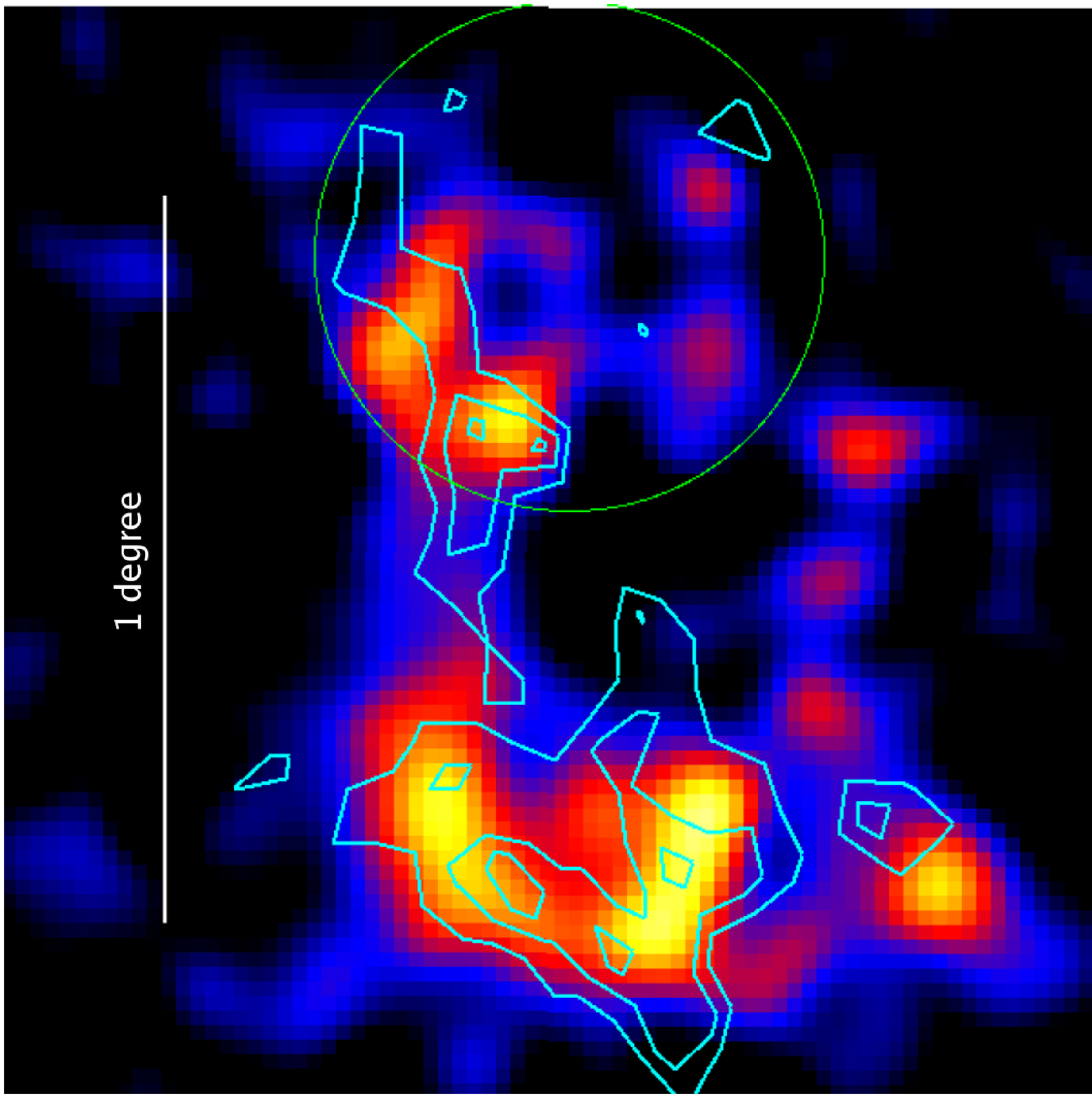,height=17pc}} 
  \caption{
    Multiwavelength views of IC\,443 (left) and W 28 (right). Molecular tracer $^{12}$CO (J=2$\to$1)
    is shown (cyan contours) in comparison to TeV data (color scale) from H.E.S.S. 
    \citep{HESS:w28} and MAGIC \citep{MAGIC:ic443}. 
  }
  \label{fig_w28}
\end{figure}

Another approach towards demonstrating CR acceleration in SNRs
is to look for dense molecular clouds adjacent to, or interacting with,
an SNR. In clouds, interactions of accelerated protons and nuclei will give rise to an enhanced
$\gamma$-ray flux proportional to the cloud's mass (Equation~\ref{eq:cloudflux}) whereas IC
radiation from electrons is not enhanced. 
Two candidate systems where
this might be occurring are W\,28 \citep{HESS:w28} and 
IC\,443 \citep{MAGIC:ic443} (Figure \ref{fig_w28}). W\,28 is an old remnant
(30--150 kyr) which has most likely released most of its CRs. VHE $\gamma$-ray data show four emission hot-spots coincident with
enhancements of gas density; if interpreted as proton interactions in
passive clouds, their masses imply a CR flux which is 10 to 30 times
the flux near Earth, a plausible value given the proximity (at least
in projection) of the remnant. A similar situation is seen in IC 443, 
where TeV emission coincides with a massive molecular cloud, with OH maser
emission indicating that the SNR shock wave is hitting the cloud. 


\subsection{Pulsars and pulsar wind nebulae}
\label{sec:pwn}

The first VHE $\gamma$-ray source to be detected, the Crab Nebula
\citep{weekes89}, is a pulsar wind nebula (PWN),
where populations of electrons and positions with energies up to 
PeV energies emit X-rays and IC $\gamma$-rays
\citep[e.g.][]{1996A&AS..120C.453A}. 
As PWNe have a well-defined central energy source and are typically close enough to
be spatially resolved, they allow relativistic flows, and the shocks 
which result when these winds collide with their surroundings, to be studied~\citep{gaensler06}.
PWNe are the most abundant
class amongst the sources discovered in the H.E.S.S. survey of the Galactic
Plane. Many of the fundamental concepts concerning PWNe are 
summarized in the seminal papers by \citet{1974MNRAS.167....1R} and \citet{1984ApJ...283..710K};
for a recent summaries see~\citet{gaensler06}. 

A supernova explosion may create a pulsar, a neutron star with a magnetic
field axis which is misaligned with the rotation axis. The rotating
magnetic dipole will emit electromagnetic radiation
at a luminosity $\dot{E} \sim 3 \cdot
10^{33} B_{12}^2 P^{-4}_{\rm ms}$\,erg s$^{-1}$ and will spin down; $B_{12}$ is the surface
magnetic field in units of $10^{12}$\,G and $P_{\rm ms}$ the period in ms. 
Pulsars with $\gamma$-ray PWN tend to have $\dot{E}$ around and above $10^{35}$ erg\,s$^{-1}$.
The time dependence of the spin-down energy loss is
given by 
\begin{equation}
\dot{E}(t) = \dot{E}_o/(1+t/\tau)^{p}
\end{equation}
with $p = (n+1)/(n-1)$, with the characteristic spin-down time
$\tau$. For pure dipole radiation one has a {\it breaking index} $n=2$
and $\tau \sim 10 P_{0,{\rm ms}}^2 /B_{12}^2$\,y where $P_{0,\rm{ms}}$
is the birth period of the pulsar; measured values for $n$ lie between
2 and 3. The rotating field creates a voltage drop of order $10^{17}
B_{12}/P_{\rm ms}^2$\,V which can be used to accelerate particles, fed
by electron-positron pair cascades in the giant electric and magnetic
fields near the pulsar surface. Accelerated electrons will emit
$\gamma$-radiation which appears pulsed to a stationary observer away
from the rotation axis.  Both the polar caps of pulsars and the
``outer gap'' at a couple of stellar radii have been considered as
emission regions \citep[see e.g.][for a review and
references]{2007arXiv0710.3517H}.  Radiation emitted near the polar
cap of the pulsar should cut off sharply at a few GeV since cascading
in the high fields in this region prevents escape of higher-energy
photons. In the region of the `outer gap', energies of tens of GeV may
be reached.

The relativistic electron-positron wind from the pulsar terminates in
a shock where the ram pressure of the wind is balanced by the pressure
of the surrounding nebula. At the shock, the kinetic energy of the
wind is transformed into random motion. Outside the shock, the
resulting relativistic electron-positron gas will convect outwards at
subsonic speeds, $v < c/\sqrt{3}$, $v$ decreasing initially as
$1/r^2$, forming an expanding PWN visible in sychrotron radiation and
IC $\gamma$-rays.  Assuming Bohm diffusion, convection of particles
will dominate over diffusive propagation in most of the nebula (see
e.g. \cite{ARxIV:0803.0116}).  The reverse shock created in the
expanding SNR may collide with the expanding PWN after some kyr and
may temporarily halt the expansion of the PWN.  The evolution of PWNe
is summarized for example in \citet{2001ApJ...563..806B}. 

In retrospect, it appears plausible that PWN are dominant among
galactic VHE $\gamma$-ray sources: the energy content of a PWN -- of
order $10^{49}$\,erg -- is small compared to the energy of $\sim
10^{51}$ erg dissipated in a supernova shock, but since a large
fraction of the energy is carried by relativistic electrons with
radiative lifetimes of 10$^{3}$--10$^{4}$ years
(Equation~\ref{eq:sync_cooltime}), kinetic energy is very efficiently
converted to radiation, compared to nuclei with interaction timescales
of ${\cal O}$(10$^7$ y) (see Section~\ref{sec:hadrons}).  In addition,
after ${\cal O}$(10 kyr), decelerating supernova shocks can no longer
confine the highest energy particles, cutting off emission at VHE
energies, whereas a powerful pulsar may drive a PWN significantly
longer.

Since electrons suffer energy-dependent energy losses as they
convect/diffuse away from the pulsar, both synchrotron X-ray and IC
$\gamma$-ray spectra should evolve with increasing distance from the
pulsar.  Indeed, this is observed in X-rays for a number of PWN, and
in a single object at TeV energies.

\begin{figure}
  \centerline{\psfig{figure=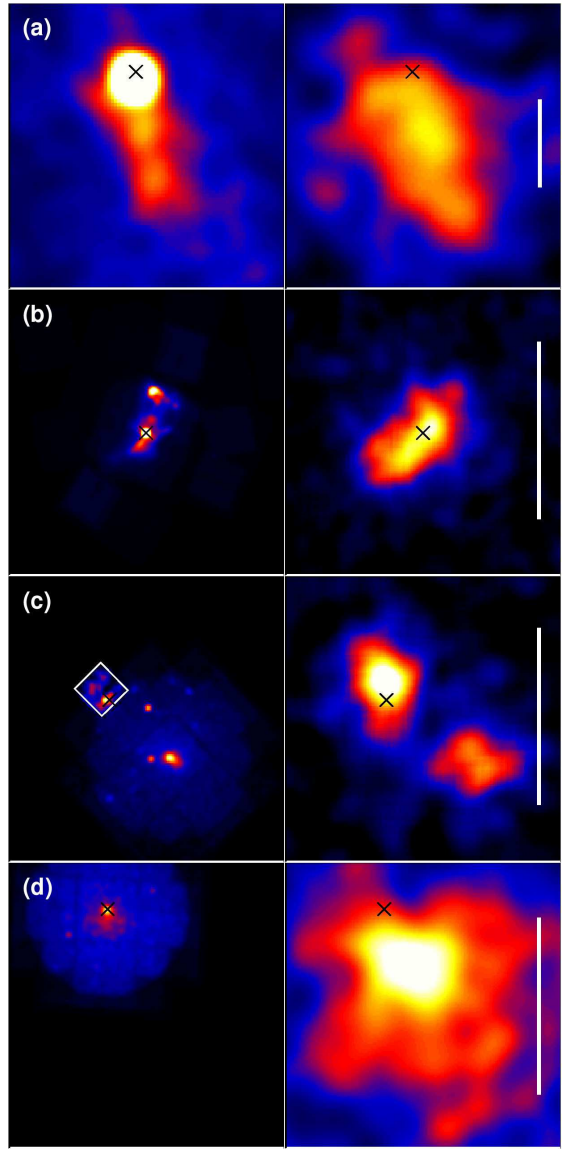,height=36pc}} 
  \caption{ Five $\gamma$-ray PWN candidates in X-rays (left) and TeV
    $\gamma$-rays (right).  a) Vela X, b) MSH 15$-$5{\it
    2}, c) the K3 and Rabbit PWNe in the Kookaburra Nebula, and
    d) G18.0$-$0.7 / HESS\,J1825$-$137.  The $\gamma$-ray images are all
    made using H.E.S.S., see
    \citet{HESS:velax,HESS:msh1552,HESS:kookaburra, HESS:1825p2}.
    Publicly available X-ray data have been reprocessed to produce the
    X-ray images: a) ROSAT, b) Chandra, c) XMM and Chandra (white inset), d) XMM.
    The positions of the associated radio pulsars are shown with
    crosses. The white scale bars are 0.5$^{\circ}$ long.  }
  \label{fig_pwn}
\end{figure}

Table \ref{tab:gal} lists $\gamma$-ray sources which are almost
certainly associated with PWNe.  Criteria for identification as a PWN
include positional coincidence with a known pulsar powerful enough to
potentially power the source, and the detection of an X-ray PWN.  In
some cases a radio SNR shell is also present providing a potential
contribution to the $\gamma$-ray emission.  Figure \ref{fig_pwn} shows
some of the best $\gamma$-ray PWN candidates, with characteristics
which were surprising at the time of discovery: (a) the large size of
the $\gamma$-ray sources --- typically $\sim 0.5^{\circ}$,
corresponding to a few tens of parsecs for typical few kpc distances,
sometimes an order of magnitude larger then the corresponding X-ray
PWNe and (b) the displacement of $\gamma$-ray sources from the pulsars
by as much as the source radius, frequently putting the pulsar at the
edge of the nebula.  That the association is nevertheless significant
can be demonstrated in two ways: a statistical study of sources in the
H.E.S.S. Galactic Plane Survey shows a significant $\gamma$-ray excess
near powerful pulsars \citep{2007arXiv0709.4094C} and in one case ---
HESS\,J1825$-$137 --- energy-dependent morphology is observed
\citep{HESS:1825p2}, with the source
shrinking towards the pulsar for higher-energy $\gamma$-rays, as expected
due to radiative cooling of electrons convecting away from the pulsar.
 
\begin{figure}
  \centerline{\psfig{figure=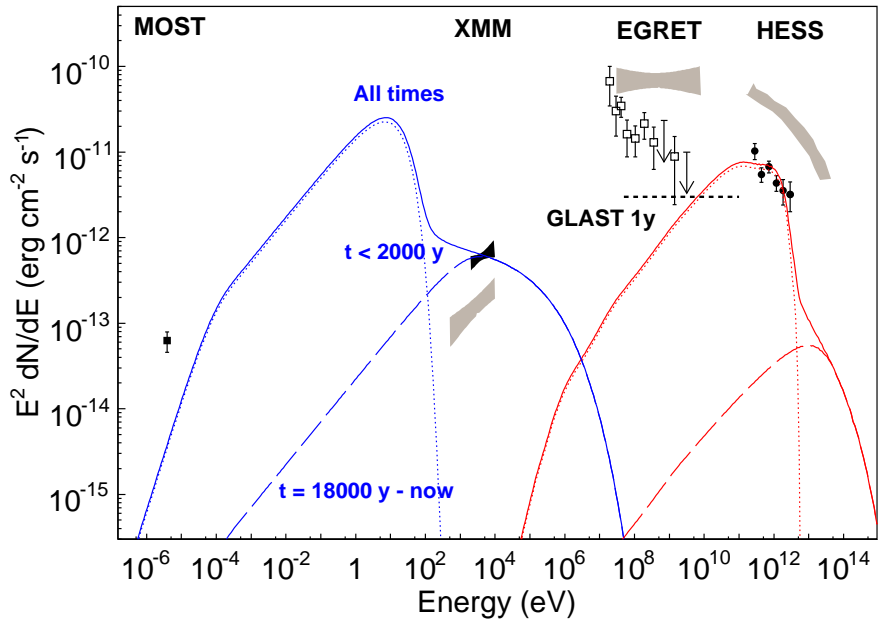,height=20pc}} 
  \caption{SED for HESS\,J1640$-$465,
    showing hypothetical contributions from young/old electrons.
    Data for HESS J1825$-$137 are shown for comparison (gray regions). 
    Reproduced from \citet{funk07a}.
  }
  \label{fig_1640}
\end{figure}

The difference in size between $\gamma$-ray PWNe and X-ray PWNe can be
attributed to the difference in energy of the electrons responsible
for the radiation. For inferred B-fields at the PWN core of some
$10\,\mu$G, electrons of many tens of TeV are required to produce
X-rays in the keV range (Equation~\ref{eq:esync}).  For these fields,
cooling timescales are of order kyr, resulting in a modest range of
particles.  Electrons emerging from this inner region of the PWN still
have energies sufficient to produce TeV $\gamma$-rays in interactions
with IR and optical photons in the KN regime
(Equation~\ref{eq:egammaIC_KN}), but in the assumed $\mu$G fields in
the outer regions of the PWN the synchrotron peak is shifted into the
optical, where it cannot be detected for all practical purposes. This
also implies that VHE $\gamma$-ray emitting electrons, with
characteristic cooling times of some 10~kyr, are accumulated over a
much longer history of the pulsar, for medium-aged pulsars reaching
back to the birth of the pulsar, where the energy output rate was one
to two orders of magnitude higher. As a result, the $\gamma$-ray
luminosity can reach and even exceed the current spin-down luminosity
of the pulsar; an equilibrium is only reached once the pulsar age is
large compared to cooling timescales, as is typically the case for
X-ray emitting electrons. An example for a resulting SED is shown in
Figure \ref{fig_1640} for the object HESS\,J1640$-$465
\citep{funk07a}. In the model shown here, electrons, up to 20\,kyr old
are responsible for the detected VHE $\gamma$-rays, with their
synchrotron radiation peaking (undetectably) in the visible.  Young
electrons, recently injected into the PWN, generate the compact X-ray
nebula seen with XMM-Newton.

Displacements between pulsars and their $\gamma$-ray PWNe appear to be
fairly common, but their origin is not fully understood; at least in
the two cases where the proper motion of the pulsar is known (Vela\,X
and HESS\,J1825$-$137), the displacement between nebula and pulsar is
almost orthogonal to the pulsar motion, eliminating the explanation
that the pulsar was created with a significant kick, leaving a 'relic'
PWN behind. The origin of the displacement is most likely a
consequence of the environment, with density gradients deforming the
evolution of the SNR shell and hence also the PWN
\citep{2001ApJ...563..806B}. Alternatively, target photon fields for
IC scattering may be enhanced locally due to stars or star clusters.

The sample of $\gamma$-ray PWNe is not large enough to systematically
assess how PWN properties depend on pulsar properties. However, among
the PWNe detected so far, two trends seem to emerge, with some caveats
concerning selection bias \citep[see e.g.][]{2008arXiv0811.0327M}: (a)
the ratio of $\gamma$-ray luminosity to spin-down loss tends to
decrease with increasing spin-down loss $\dot{E}$ (or decreasing
pulsar age), and the fraction of energy radiated in X-rays increases,
and (b) very energetic pulsars (such as the Crab) tend to have very
compact nebulae. A plausible explanation is that for rough
equipartition between particle energy and magnetic fields in the
nebula, the fields increase with $\dot{E}$, which implies that the
ratio of X-ray to IC luminosities increases with $\dot{E}$ and that
the lifetime and range of electrons decrease.  Furthermore, due to the
strong correlation between pulsar age and spin-down luminosity, high
spin-down pulsars tend to be young, such that the population of
slower-cooled $\gamma$-ray emitting electrons is still increasing
whereas X-ray emitting electrons have already reached their
equilibrium output. The size of evolved $\gamma$-ray nebulae tends to
saturate at some 10~pc (Figure \ref{fig_pwn}). This may be because the
nebulae themselves are confined by the ambient medium, or because
radiative losses and convection timescales are such that multi-TeV
electrons die out.

\begin{figure}
\centerline{\psfig{figure=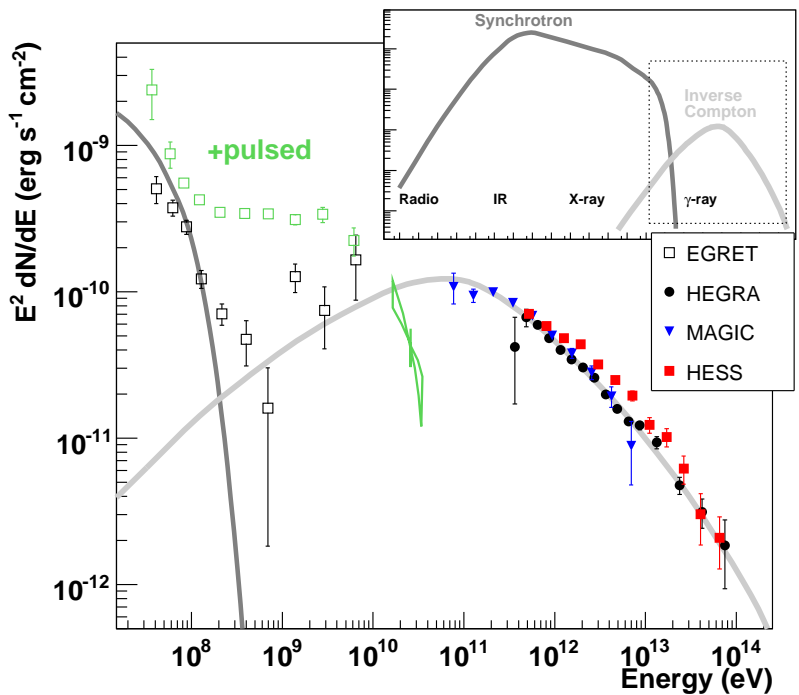,height=23pc}} 
\caption{The SED of the Crab nebula
  and pulsar, adapted from \citet{hinton08}.
}
\label{fig_crab}
\end{figure}

By far the best-studied PWN is the Crab Nebula
\citep{2008ARA&A..46..127H}. Its broad-band SED exhibits overlapping
synchrotron and IC spectra, the first extending from radio into the
EGRET energy range, the second from EGRET energies to $\sim$100 TeV
(Figure \ref{fig_crab}).  Recently, the MAGIC collaboration
succeeded in detecting, for the first time, a pulsed component in the
VHE energy range, above 25\,GeV, hence excluding a pure polar-cap
scenario as the origin of the pulsed emission~\citep{MAGIC:crabpulsed}.

\subsection{Compact object binary systems as $\gamma$-ray sources}

The physical environment inside a close binary system (or an
eccentric binary close to periastron) is radically different to that of
the diffuse ISM relevant to SNRs and PWN. This environment is
characterized by very high radiation densities ${\cal O}$(1\,erg\,cm$^{-3}$) 
of rather high frequency photons ${\cal O}$(1\,eV) and high magnetic
fields (mG--G). The consequence of this environment for relativistic
electrons is that rapid cooling is inevitable. In the case that the
radiation pressure dominates, the cooling of TeV electrons will occur
in the KN regime, with implications for the spectral shape 
discussed in Section~\ref{sec:elec}.

All relevant timescales in such a system are short in comparison to
the length of observation programs (typically years), for example the
acceleration and cooling time for relativistic electrons and the
orbital period. It is therefore expected that $\gamma$-ray emitting
binaries will be variable point-like objects if electrons dominate the
$\gamma$-ray production. Hadron accelerating binaries may produce
steady and extended $\gamma$-ray emission if the protons and nuclei
can escape the production region without significant energy losses.
If $\gamma$-rays are produced inside the binary system, then the
assumption of free escape is normally invalid. In the presence of
intense radiation fields, $\gamma$-$\gamma$ interactions produce
e$^{+}$/e$^{-}$ pairs which are likely to IC scatter, leading to
electromagnetic cascades. This effect, combined with KN IC cooling can
lead to $\gamma$-ray spectral shapes radically different from those
seen in diffuse sources \citep[see e.g.][]{khangulyan08}.

The most obvious energy source inside a binary system in which one
member is a compact object (neutron star or black hole) is
accretion. Particle acceleration in {\it jets} produced by accretion
onto a compact object is well established in extragalactic objects and
in the early 90s Galactic analogues to AGN were discovered and dubbed
{\it Micro-quasars} \citep{mirabel94}.  The internal and external
shocks associated with such jets provide potential sites for particle
acceleration. The primary alternative power-source is the collision of
stellar/neutron-star winds, with acceleration at the termination shock
of the wind, in a compressed version of a PWN. Of the three
well-established systems, one, PSR\,B1259$-$63/SS 2883, is a
unambiguously a binary PWN.  For the other two systems (LS\,5039 and
LS\,I\,+61\,303) both PWN and micro-quasar scenarios have been
extensively discussed. These objects are briefly described below. Note
that a much more complete historical account can be found in
\citet{aharonian_buckley08}.

\begin{figure}
  \centerline{\psfig{figure=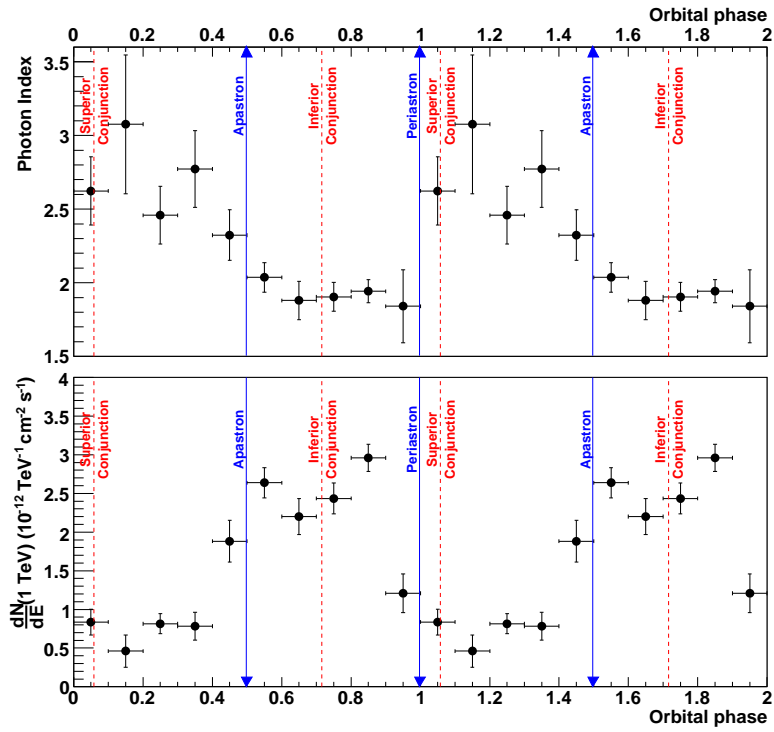,height=23pc}} 
  \caption{Phase-folded light curve and spectral index variations for the binary system\,LS 5039.
    Reproduced from \cite{HESS:ls5039p2}. }
  \label{fig_ls5039}
\end{figure}

The most detailed TeV measurements so far exist in the case of
LS\,5039, a 3.9\,day period system of a $\sim$20$M_{\sun}$ (O6.5V
type) star and a compact companion of mass $3.7^{+1.3}_{-1.0} M_{\sun}$
\citep{casares05}. Figure~\ref{fig_ls5039} shows the flux and spectral
index of the VHE emission of LS\,5039 as a function of orbital phase,
as measured using H.E.S.S. \citep{HESS:ls5039p2}.  As the distance
between the stars varies by a factor $\sim$2 around the orbit, the
observed modulation of flux and spectrum with period is not
unexpected. The maximum flux occurs at inferior conjunction, the
point where $\gamma$-$\gamma$ absorption is expected to be at a
minimum. However, the observed modulation is certainly not consistent
with the simple-minded expectation for such absorption (and/or
cascading). As discussed in Section~\ref{sec:abs} the peak absorption
should occur at $0.9/E_{T,eV} = 300$ GeV (as the thermal photons from
the companion star have $kT\approx3$ eV), whereas there is apparently
{\it no} modulation of flux at this energy. It seems likely that
effects such as the angular dependence of the IC cross-section,
adiabatic losses and the dependence of acceleration efficiency (and
also the efficiency of injection into the acceleration process) and
maximum energy, on distance between the massive star and compact
object (i.e. changing accretion rate, shock velocity...)  must be
taken into account, together with the geometry of the system, to reach
an understanding of the physical processes at work.  As LS\,5039
appears to host a bi-polar radio jet with a speed $\approx0.2c$
\citep{parades00} the accretion powered micro-quasar scenario has been
most extensively discussed, but despite extensive theoretical work
\citep[see][and references therein]{khangulyan08}
the acceleration site and nature of the compact object are still unclear in this system.

LS\,I\,+61\,303 is a longer period (26.5 day) system with a lower mass
companion ($\sim$10 $M_{\sun}$, type B0Ve) which has been extensively
observed by the MAGIC and VERITAS collaborations \citep[see][and
references therein]{MAGIC:lsi61b}.  As in the case of LS\,5039 VLBI
radio observations had revealed extended {\it jet-like} radio emission
in LS\,I\,+61\,303, leading to its classification as a
micro-quasar. However, more recent VLBI data from \citet{dhawan06}
suggest that the radio structure rotates with orbital phase, as might
be expected for the cometary emission of a shocked pulsar-wind
encountering a (higher momentum) stellar wind from the massive
companion.

In the remaining well-established $\gamma$-ray binary system the
pulsar nature of the compact object is well established by radio
pulsation measurements.  PSR B1259$-$63 and its B2e companion SS\,2883
form a highly eccentric binary with an orbital period of $\approx$3.4
years \citep{johnston92}. TeV emission from the system close to its
periastron passage was predicted by \citet{kirk99} and observed using
H.E.S.S.~\citep{HESS:psrb1259}.  PSR\,B1259$-$63 is powerful and close
enough ($\dot{E}\sim10^{36}$ erg s$^{-1}$, $d\approx$2.3 kpc) that TeV
emission might have been expected for a classical PWN.  The TeV
detection around periastron is usually attributed to the boost in IC
emission from the strong radiation field of SS\,2883.

The detection using the MAGIC telescope of a single flare from Cyg X-1
is extremely important as in this system there is no doubt about the
nature of the compact object - it is a $21\pm8 M_{\sun}$ black hole.
The $\approx$80 minute flare has a significance of approximately 4.1
$\sigma$ after accounting for statistical trials \citep{MAGIC:cygX1}.
At present there is therefore strong evidence for, rather than proof
of, VHE emission from Cyg X-1.

Given the relatively deep survey of most of the Galactic volume using
H.E.S.S., it is natural to ask if additional candidates exist for
$\gamma$-ray binary systems. Of the $\sim$20 unidentified VHE sources,
very few are point-like in nature, as would be expected both by
analogy with identified sources.  By far the best candidate for a new
$\gamma$-ray binary is HESS\,J0632+057 \citep{HESS:0632}, a point-like
TeV source coincident with a Be star and an EGRET source. Follow-up
observations of this object with XMM-Newton led to the discovery of a
variable X-ray source coincident with the star \citep{hinton08b}. If
all these objects are truly associated then the SED resembles that of
the known $\gamma$-ray binaries.  This discovery suggests that a
population of $\gamma$-ray binaries exists with somewhat lower X-ray,
radio and $\gamma$-ray fluxes then the 3 well established systems,
which have hence so far escaped detection.

\subsection{Stellar clusters and stellar winds}

All known Galactic sources of VHE $\gamma$-rays are associated
(directly or indirectly) to massive star formation. Both the
end-points of the massive stellar lifecycles, SNRs and pulsars, and
high mass stars with compact companions (HMXBs) are TeV emitters. It
is natural to consider whether massive stars can accelerate particles
to TeV energies in the absence of a compact object. The idea of
particle acceleration at the shock front formed by colliding stellar
winds in a binary system of two massive stars has been developed over
the last five years~\citep[see for
example][]{benaglia03,reimer06,pittard06}.  The discovery of
non-thermal hard X-ray emission from the massive binary Eta Carina
\citep{leyder08} has strengthened the case for high-energy particle
acceleration in these systems.

So far there are no unambiguous VHE detections of individual colliding
wind systems. However, extended TeV $\gamma$-ray emission has been
detected from in and around Westerlund\,2, the second largest
concentration of massive young stars in our galaxy
\citep{HESS:westerlund2}.  It seems plausible that this source is
powered by the collective effect of stellar winds within the
cluster. The association of massive stars Cyg\,OB2 has also been
suggested as the counterpart of the unidentified Galactic plane source
discovered using HEGRA~\citep{HEGRA:tevj2032}.  However, as
essentially all known types of Galactic $\gamma$-ray source are
associated (directly or indirectly) with high-mass star formation it
is plausible that these emission regions are associated with a single
object within the cluster, such as a PWN, which has not yet been
identified at other wavelengths, rather than the cluster as a whole.

\subsection{Unidentified VHE $\gamma$-ray sources}

Roughly one third of the $\sim$60 Galactic TeV sources have no
compelling counterpart at other wavelengths. In several cases
sensitive follow-up X-ray and radio observations have failed to
identify these sources. Due to the apparent lack of emission of these
TeV sources at lower frequencies, they are sometimes referred to as
``dark accelerators''. The main questions which arise for this
population are 1) do they represent a new class of objects, or are
they members of the known classes? and 2) is the emission hadronic or
leptonic in origin?  The identified VHE sources have low-frequency
counterparts with non-thermal emission usually attributed to the
synchrotron process. The lack of synchrotron counterparts to the
unidentified sources may be taken as a suggestion that the $\gamma$-ray
emission is produced by hadrons rather than leptons in most of these
unidentified sources. However, there are several complications to this
simple picture.

The distribution of sizes and spectra of the unidentified sources are
rather similar to those of the identified sources. However, these
properties may be common to most TeV sources on rather general
physical grounds.  For example, the relatively fast diffusion and slow
energy losses of $>$TeV particles make TeV sources, in general, rather
large. The expected size of an old (i.e. in equilibrium) source of
electrons cooled by synchrotron emission and with Bohm diffusion is
$r_{pc} \approx 30 B_{\mu{\rm G}}^{1.5}$ (from equations
\ref{eq:diffusion} and \ref{eq:sync_cooltime}).  Typical ISM magnetic
fields of a few $\mu$G therefore lead to sources with a scale of a few
parsecs, with apparent radii of $\sim$$0.1^{\circ}$ if located at
typical distances of a few kpc.  Unfortunately, the difficulty of
identifying a source increases rapidly with its angular extent (unless
closely matched in morphology to a bright object in another waveband
as in the case of the $\gamma$-ray SNR shells). The sensitivity of
existing X-ray and radio surveys for very extended $\sim$$0.5^{\circ}$
objects is limited, and along the Galactic plane several candidates
typically exist per degree of longitude. Source confusion can
therefore be a major difficulty.  Unidentified sources such as
HESS\,J1745$-$303 \citep{HESS:1745} lie in regions with a high density
of candidates and very likely have unresolvable contributions from
several objects. Even with better angular resolution the inherently
diffuse nature of CR sources may make such regions very difficult to
disentangle.

Leptonic scenarios for unidentified TeV sources are often dismissed as
requiring very low magnetic fields to avoid bright synchrotron
counterparts.  However, even for ${\cal O}$(10 $\mu$G) magnetic fields
it is plausible that the synchrotron emission accompanying the VHE IC
emission is largely confined to the FIR--UV band, and hence almost
impossible to detect for even the brightest sources (e.g. $\nu F\nu$
$\sim$ $10^{-10}$ erg cm$^{-2}$ s$^{-1}$) if they are extended on
typical $\sim$0.2$^{\circ}$ scales. X-ray synchrotron emission with
such B-fields requires higher electron energies than needed to produce
the $\sim1$ TeV IC $\gamma$-rays. Furthermore, it is also possible to
avoid {\it radio} counterparts if there is a low-energy cut-off in the
injection spectrum of electrons, quite plausible in the case of, for
example, pulsar wind nebulae.

The lack of a radio counterpart may be harder to explain in the
context of a hadronic scenario. Secondary electrons down to GeV
energies are inevitably produced in the same p-p collisions that
produce $\pi^{0}$ $\gamma$-rays.  The accumulation of these electrons
(assuming an $E^{-2}$ injection spectrum) will lead to a synchrotron
energy flux
\begin{equation}
\nu F_{sync} \sim 3 \times 10^{-4} (\nu_{\rm GHz})^{1/2} (B/10
\mu G)^{3/2} (t/ 10^{5} {\rm yr}) \nu F_{\rm TeV}
\end{equation}
at intermediate energies where the low energy turn-over in the
electron distribution can be ignored and cooling is not important. In
the saturated regime at high frequencies ($t_{\rm cool}\ll$ age) the
synchrotron energy flux is directly proportional to the $\gamma$-ray
flux and $F_{sync}\approx0.18 F_{\pi^{0}}$ (see
Figure~\ref{fig_sedcool}).  For a typical TeV source with an energy
flux of $\sim 10^{-12}$ erg\,cm$^{-2}$\,s$^{-1}$, the resulting
extended X-ray source would be difficult to detect with current
instruments.

In this context it is useful to consider some examples, both of
sources which remain unidentified and those which were initially
unidentified and now have likely counterparts.  HESS\,J1303$-$631
\citep{HESS:1303} is an example of a TeV source which originally
appeared to be without any compelling candidate at lower
frequencies. At the time of discovery a PWN associated with
PSR\,J1301$-$6305 was considered rather unlikely due to the required
extremely efficient conversion of spin-down power to TeV emission and
the large offset and rms size of the TeV source (both $\approx
0.16^{\circ}$). In the light of the TeV PWN discoveries discussed in
Section~\ref{sec:pwn}, PSR\,J1301$-$6305 is now considered a
compelling candidate. Given an age of $\sim10^{4}$ years, evolutionary
effects are likely to enhance the TeV signal (see
Figure~\ref{fig_1640}). Furthermore, the pulsar is quite plausibly
much closer than its nominal dispersion measure distance of 15~kpc,
making the offset and size very typical for a TeV PWN.

\begin{figure} 
  \centerline{\psfig{figure=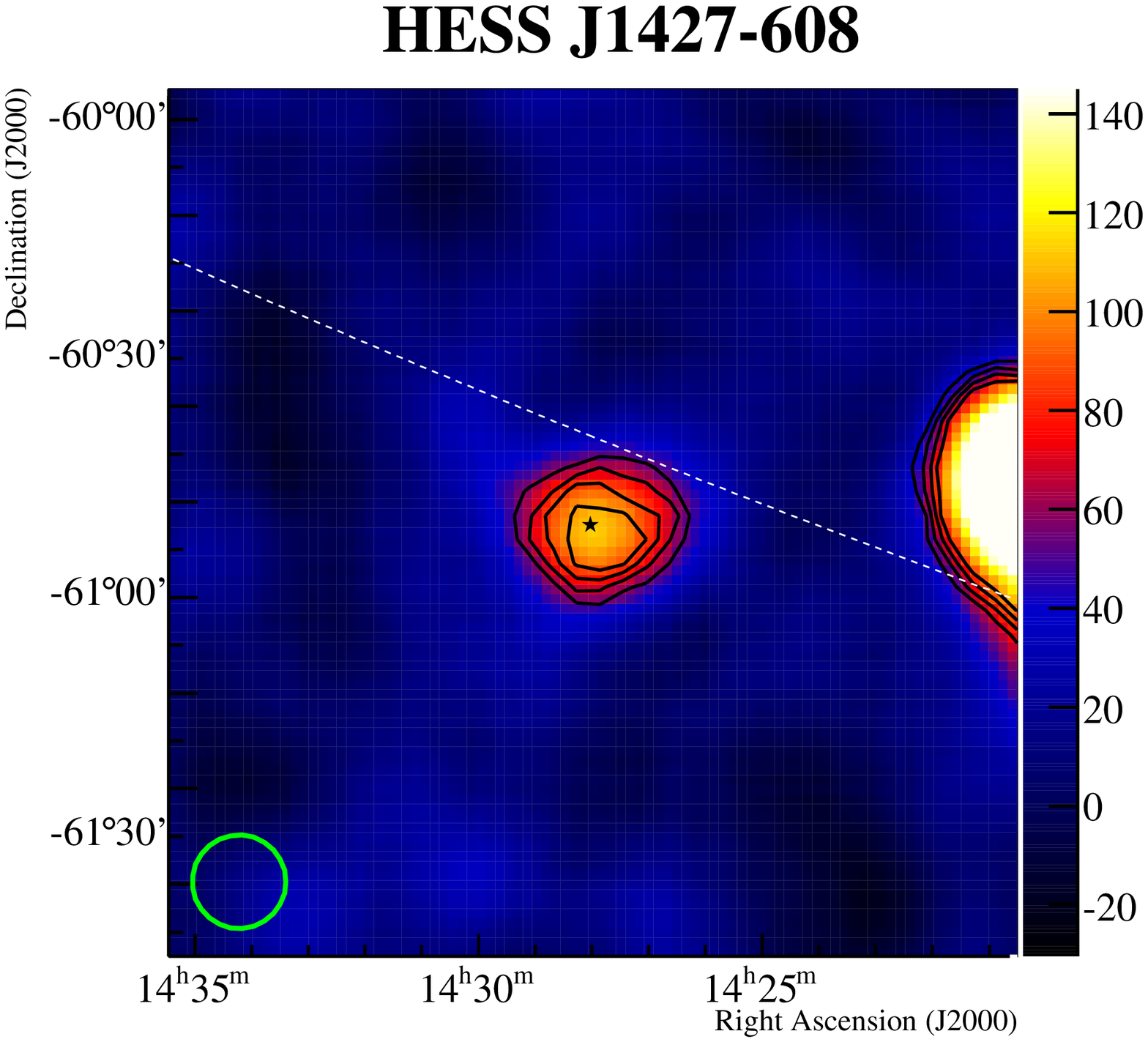,height=15pc}\psfig{figure=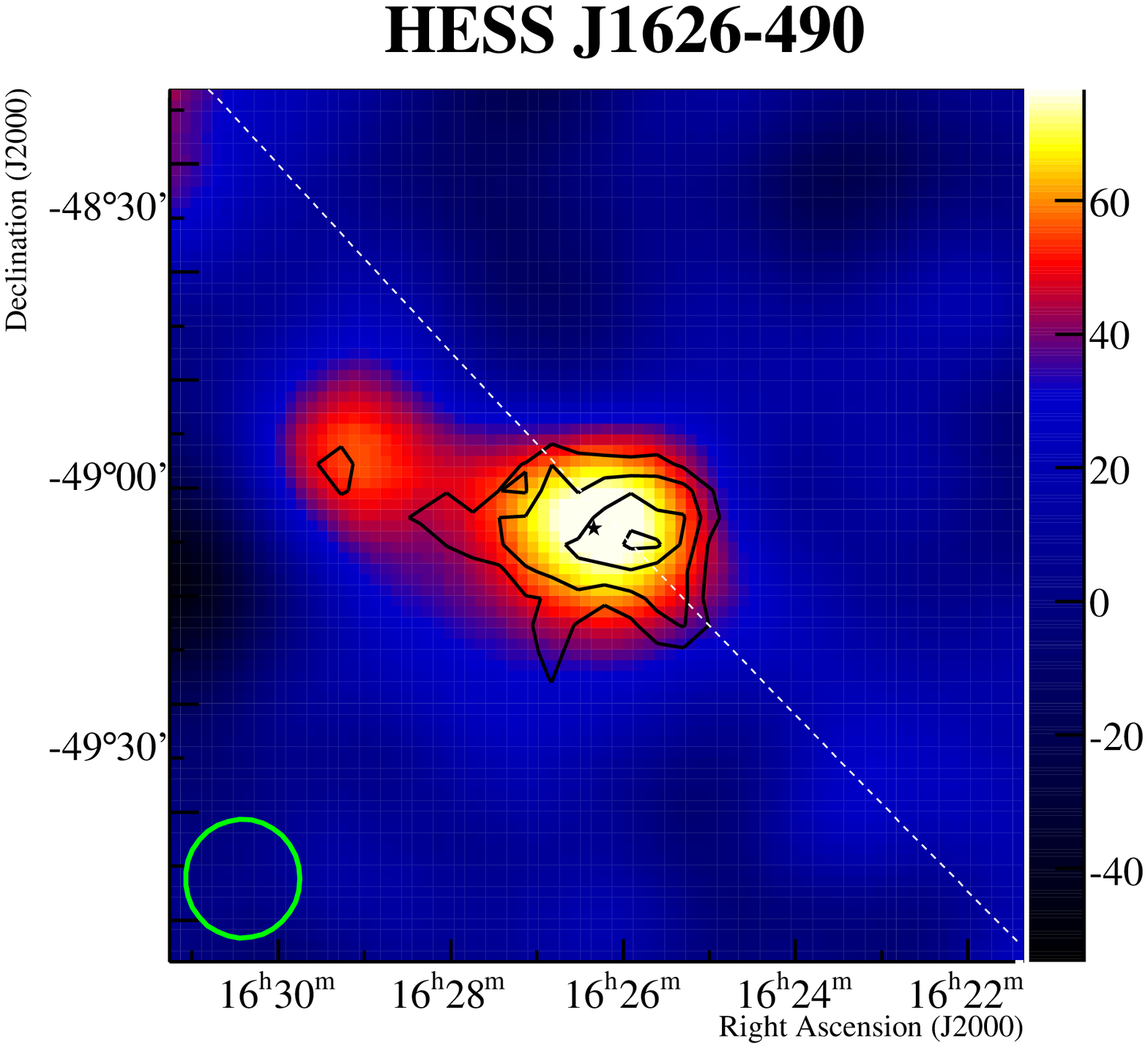,height=15pc}}
  \centerline{\psfig{figure=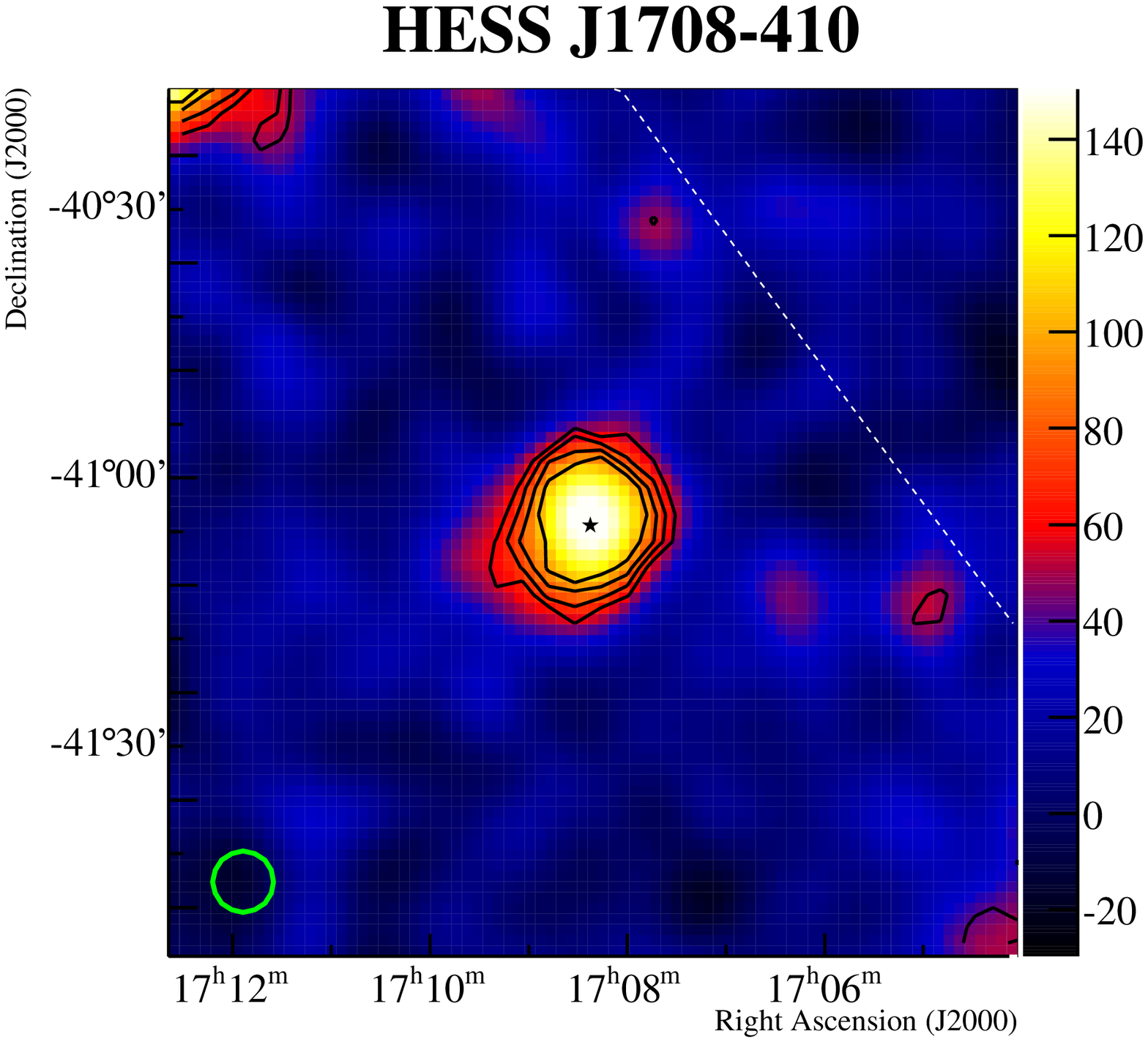,height=15pc}\psfig{figure=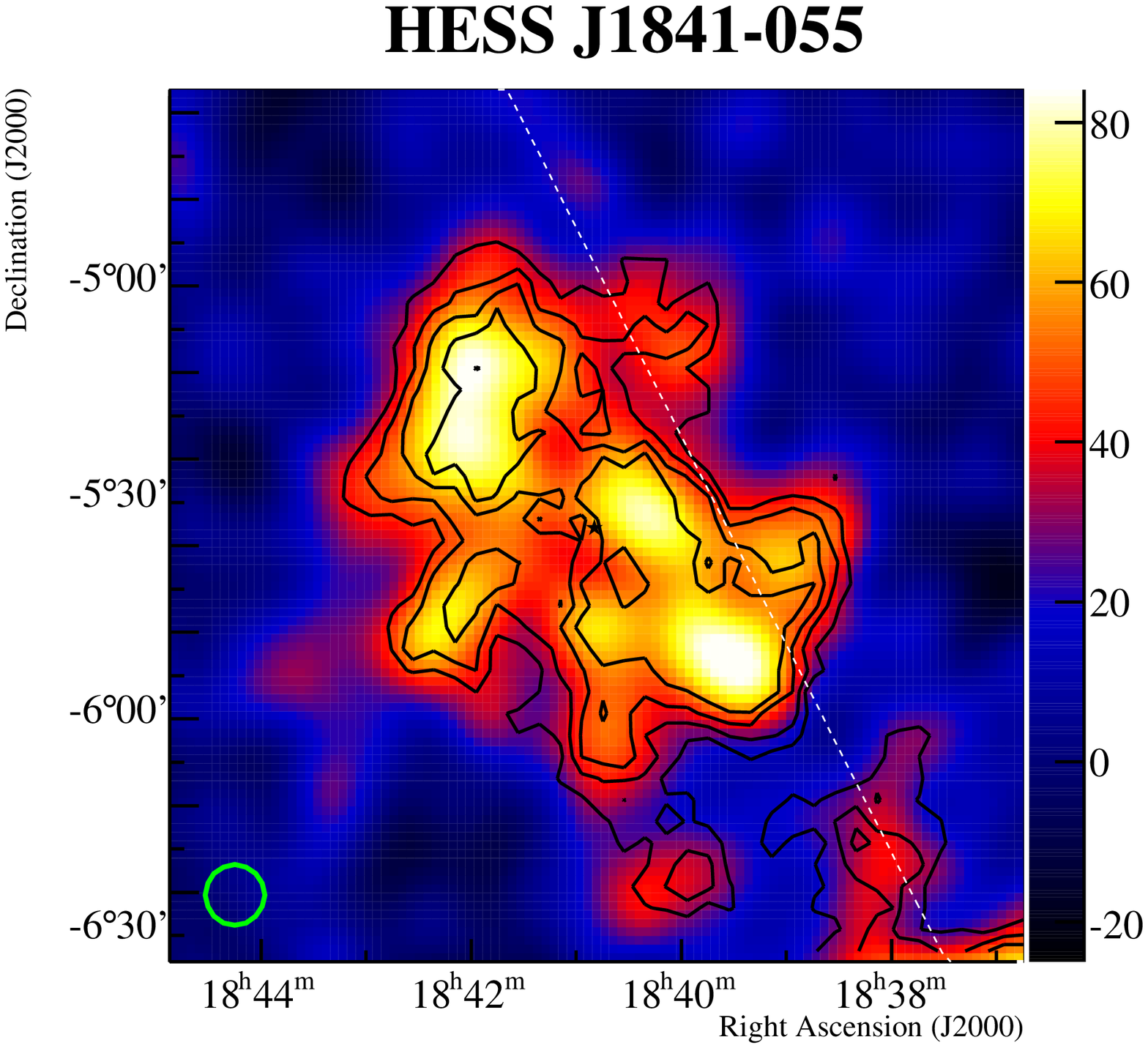,height=15pc}}
  \caption{Selected unidentified $\gamma$-ray sources as seen by
    H.E.S.S.  The Galactic plane is shown as a dashed line. The
    smoothed PSF is indicated by a circle at the bottom-left of each
    image. See \cite{HESS:dark} for details.  }
  \label{fig_unids}
\end{figure}

The second class of ``no-longer unidentified'' objects contains those
where follow-up observations have led to the discovery of new pulsars,
PWN and/or SNRs. HESS\,1813$-$178 is an example of such an object,
rapidly established as a new composite SNR following its discovery,
based on new and archival data from the VLA, INTEGRAL, XMM and Chandra
\citep[see ][ and references therein]{helfand07}.  HESS\,1813$-$178
is, however, an unusual case. It is a rather bright ($\approx 3\times
10^{-12}$ erg cm$^{-2}$ s$^{-1}$), compact ($\approx2'$) TeV source
with a bright X-ray counterpart (AX\,J1813$-$178). Such rapid
assignment to an existing source class seems unlikely for the
remaining unidentified sources. Figure~\ref{fig_unids} shows four
example VHE sources for which there are no plausible candidates at
lower frequencies. All lie within half a degree of the Galactic plane
and all are significantly extended beyond the instrumental PSF.

\subsection{Active galaxies}
\label{sec:AGN}

\begin{figure}
\centerline{\psfig{figure=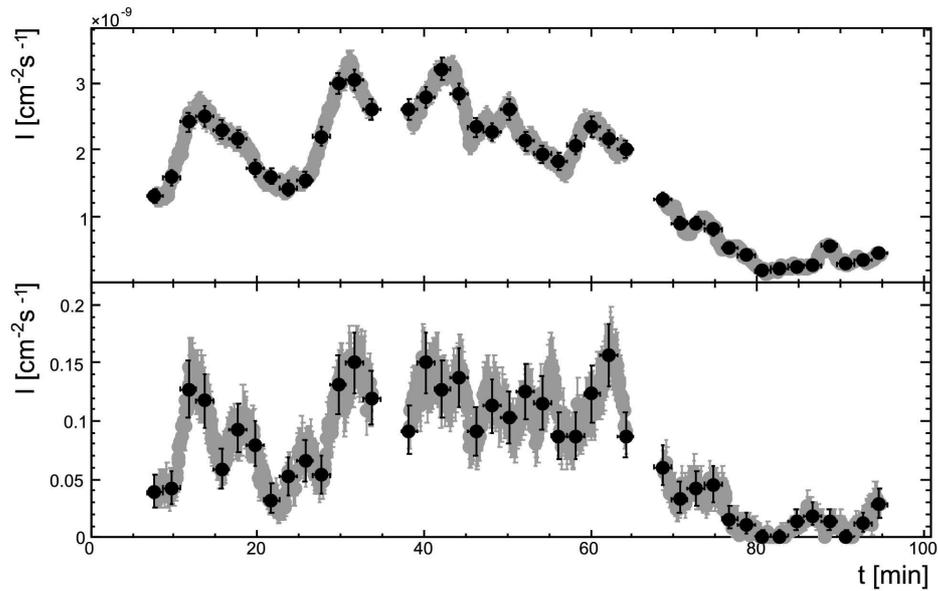,height=20pc}} 
  \caption{VHE light curve of PKS\,2155$-$304 during the July 
  2006 flare, in two energy bands: 200--800 GeV (bottom) and 
  above 800 GeV (top) \citep{2008PhRvL.101q0402A}. The light curve 
  is sampled in two-minute intervals around each point.}
  \label{fig_AGN_var}
\end{figure}

\begin{figure}
\centerline{\psfig{figure=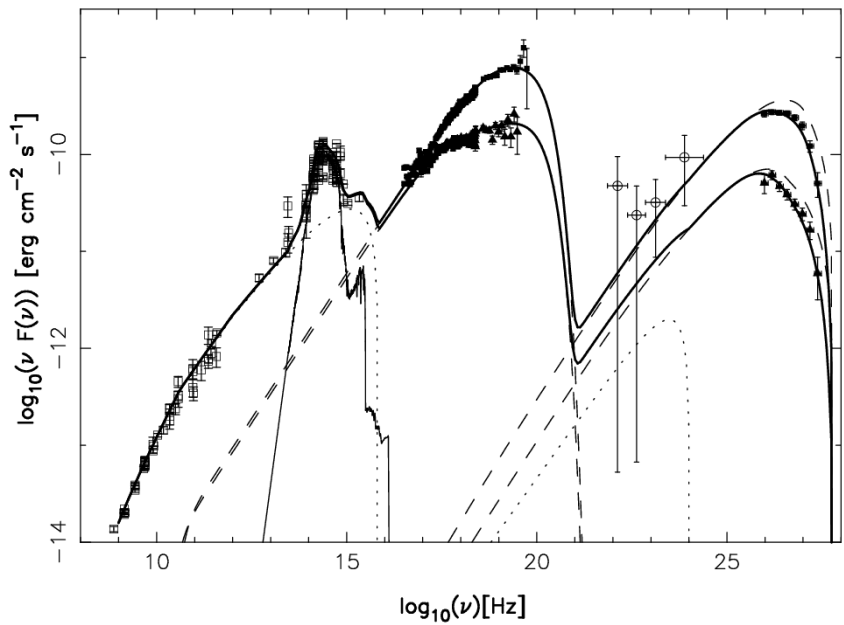,height=20pc}} 
  \caption{Spectral energy distribution of Mrk\,501 in two different
  states \citep{2001A&A...367..809K}. The two dominant peaks are
  interpreted as synchrotron and synchrotron self Compton (SSC)
  emission of electrons (dashed lines). Modeling of the SED at lower
  frequencies adds contributions where external photons are
  Compton-scattered (dotted lines) and emission by the host galaxy
  (thin full line).}  \label{fig_AGN_SED}
\end{figure}

The galaxy Mrk\,421 was the second  VHE $\gamma$-ray source
detected~\citep{1992Natur.358..477P}; 
the number of extragalactic VHE sources 
has now risen to well over 20, with
redshifts up to 0.536 (3C\,279, \citet{MAGIC:3c279}).  In order
of increasing redshift, the VHE $\gamma$-ray emitters include M\,87
($z=0.004$), Mrk\,421 ($z=0.030$), Mrk\,501 ($z=0.034$), 1ES\,2344+514
($z=0.044$), Mrk\,180 ($z=0.045$), 1ES\,1959+650 ($z=0.047$), 
PKS\,0548$-$322 ($z=0.069$), BL\,Lacertae ($z=0.069$), PKS\,2005$-$489
($z=0.071$), RGB\,J0152+017 ($z=0.08$) \citep{2008A&A...481L.103A}, W
Comae ($z=0.102$) \citep{2008ApJ...684L..73A}, PKS\,2155$-$304
($z=0.116$), H\,1426+428 ($z=0.129$), 1ES\,0806+524 ($z=0.138$)
\citep{2008arXiv0812.0978V}, 1ES\,0229+200 ($z=0.139$), H\,2356$-$309
($z=0.165$), 1ES\,1218+304 ($z=0.182$), 1ES\,0347$-$121 ($z=0.188$), 
1ES\,1101$-$232 ($z=0.186$), 1ES\,1011+469 ($z=0.212$), 3C\,279 ($z=0.536$), 
PG\,1553+113 ($z$ unknown).
A compilation of emission
characteristics and remaining references for these objects can be found in
\citet{2008MNRAS.385..119W}.  All these objects harbor Active Galactic
Nuclei (AGN), where a supermassive black hole with a mass from
millions to billions of solar masses accretes matter and powers jets ---
collimated highly relativistic outflows. Unlike Galactic VHE sources,
all objects appear point-like given the $\sim$5$'$ resolution of IACTs, 
and all VHE source positions are consistent with the nominal
location of the AGN.  With the exception of the radio galaxy M\,87
\citep{2006Sci...314.1424A}, all belong to the blazar
class where a jet points towards the observer.  
High-frequency peaked BL\,Lac objects (HBLs) dominate the sample. The 
few exceptions are the flat
spectrum radio quasar (FSRQ) 3C\,279, the intermediate-frequency peaked
BL\,Lac object (IBL) W Comae and the low-frequency peaked BL Lac object
(LBL) BL\,Lacertae itself. Some objects --- most notably Mrk\,421
\citep[e.g.][]{2008ApJ...677..906F}
and PKS\,2155$-$304
\citep[e.g.][]{2007ApJ...664L..71A}, see Figure~\ref{fig_AGN_var} --- exhibit
burst-like variability on timescales of a few minutes to a few tens of
minutes; emission from AGN such as M\,87 or 3C\,279 appears variable on
day timescales.  Despite the vastly larger distances, 
some AGN, at the peak of flares, outshine the strongest galactic 
sources by more than a factor of 10.  
While low-statistics measurements of
VHE blazar spectra are consistent with power-laws, well-measured
spectra are generally significantly curved, steepening with increasing
energy, for example Mrk\,421 \citep{2001ApJ...560L..45K}
and PKS\,2155$-$304 \citep{2007ApJ...664L..71A}.  Spectral indices tend to increase with
source distance, at least partly due to absorption of
high-energy $\gamma$-rays on infrared intergalactic photon fields, see
Section~\ref{sec:EBL}, but perhaps also related to the fact that distant
AGN must be intrinsically brighter to be detectable.  Spectral
indices vary with flux for some sources, with a tendency for
spectra to harden with increasing activity
\citep[see for example][]{2002ApJ...575L...9K}.
All VHE-detected blazars are also relatively strong radio and X-ray 
sources; in fact, observation targets are typically selected on the
basis of their output in these bands \citep{2002A&A...384...56C}.
Optical measurements have also proven useful in tagging high activity
states of AGN for targeted observations by VHE instruments
\citep[see e.g.][]{2006ApJ...648L.105A}. The broad-band spectral energy distributions
of AGN exhibit a double-humped shape, with a high-frequency
peak in the GeV to TeV regime, and a low-frequency peak in the
optical to X-ray regime (Figure \ref{fig_AGN_SED}). Simultaneous measurements in
the X-ray and VHE $\gamma$-rays bands reveal --- with a few exceptions
\citep{2005ApJ...621..181D} --- strong correlations between the X-ray
and $\gamma$-ray fluxes \citep[e.g.][]{2008ApJ...677..906F}, 
suggesting a common electron population as
the origin of the radiation, with synchrotron X-ray emission and inverse
Compton $\gamma$-ray emission.  Similarly, average X-ray and $\gamma$-ray
luminosities show a pronounced correlation
\citep{2008MNRAS.385..119W}.

In these blazars, VHE $\gamma$-ray emission is thought to arise inside
the jets; models typically assume a spherical `blob' of high-energy particles moving
along the jet axis, with flares created when high-energy electrons are
freshly injected into the blob \citep[e.g.][]{1998ApJ...509..608T}.
The relativistic motion of the blob beams and
boosts the emission and reduces power requirements by a factor
$\delta^4$, with $\delta$ denoting the Doppler factor $\delta =
(\Gamma (1 - \beta cos \theta))^{-1}$ where $\Gamma$ is the Lorentz
factor describing bulk motion of the jet and $\theta$ is the angle
between jet and observer. The blob size is usually estimated by
causality arguments from the minimum timescale $t_{var}$ of
variability, $R \approx c t_{var} \delta$ . Minimum Lorentz factors
are then obtained from the requirement that the energy density in the
blob rest-frame must be low enough to make the blob reasonably
transparent for $\gamma$-rays \citep[e.g.][]{1995MNRAS.273..583D}, 
and are in the range of 10 to almost 100. Magnetic fields in the blob, as
well as parameters of the electron spectrum, can then be determined from
the relative spacing and height of the two peaks of the SED and
from the requirement that the cooling time of electrons matches
the flare timescale, and are typically in the range 0.01\,G to 1\,G. 
For the shape of the electron spectrum, a broken power-law is frequently
employed. Parameter estimates should, however, be taken with a grain
of salt; well-sampled AGN light curves are composed of a range of
Fourier components with a red-noise spectral distribution
\citep[e.g.][]{2007ApJ...664L..71A}, without a highest frequency which
could be unambiguously associated with the size of a blob. In
Synchrotron-Self-Compton (SSC) models \citep[e.g.][]{1998ApJ...509..608T}
synchrotron photons
generated in the blob provide the dominant target for IC
up-scattering by electrons, resulting to first approximation in a
quadratic increase of $\gamma$-ray flux with electron number and X-ray
flux. Detailed modelling of broad-band
spectra, however, also indicates the relevance of external target
radiation components from other regions of the jet and from the
high-temperature gas surrounding the central engine
\citep[e.g.][]{2001A&A...367..809K} (Figure \ref{fig_AGN_SED}).

Phenomenological models frequently leave open how particles in the
blob are accelerated; possibilities include shock-wave acceleration in
MHD turbulence in the jet, centrifugal acceleration of particles along
rotating magnetic field lines near the base of the jet, or shear
\citep{2002A&A...396..833R}.  However, X-ray observations show that
acceleration of high-energy particles must occur along the jet, since
travel times from the base of the jet to X-ray emitting knots exceed
cooling times. Many fundamental aspects of AGN jets and of particle
acceleration in these jets are poorly understood, including the
mechanisms which launch the jets and their composition.  Not even the
character of the emitting particle populations is firmly established:
cascades induced by high-energy protons \citep[see e.g.][]{1993A&A...269...67M}
can successfully reproduce most features of AGN SEDs, with the
possible exception of very fast variability, which is hard to model
due to the long energy loss timescales of protons in comparison to
electrons.  VHE measurements are a promising method for probing jet
properties and ultimately studying the environment of supermassive
black holes.

The discovery of VHE $\gamma$-ray emission from M\,87 established radio
galaxies as a new class of source where $\gamma$-rays are
emitted at significant angles to the jet, not relying on Doppler
boosting (and hence only observable for nearby objects).
The observation of fast - day scale - variability of the
emission \citep{2006Sci...314.1424A,2008ApJ...685L..23A} excludes the
extended jet or the radio lobes as sources and imply a compact
emission region with a size comparable to the radius of the SMBH at
the center of M\,87, most likely the nucleus itself or possibly the knot
HST-1 in the inner jet.

\subsection{Probing background radiation fields with $\gamma$-rays}
\label{sec:EBL}

\begin{figure}
\centerline{\psfig{figure=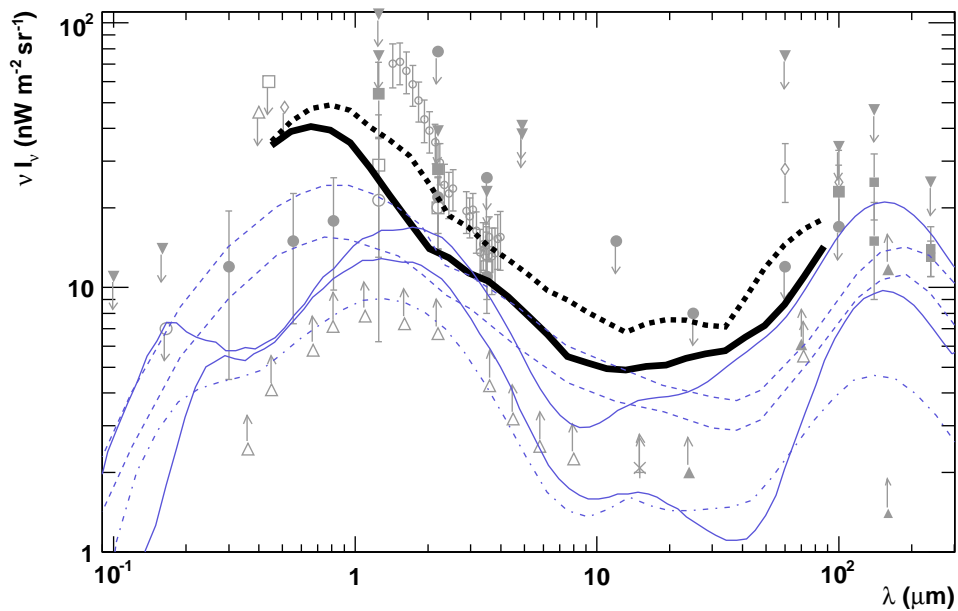,height=20pc}} 
  \caption{EBL limits obtained from VHE $\gamma$-ray spectra using
  plausible assumptions for intrinsic spectra (solid black line), 
  compared to lower limits from direct observations --- potentially 
  hampered by incomplete subtraction
  of foreground emission --- and including recent EBL models. From
  \citet{2007A&A...471..439M}; see there for details and references.}
  \label{fig_EBLn}
\end{figure}

The energy spectra of extragalactic $\gamma$-ray sources are modified by
interactions with the diffuse extragalactic background light (EBL, see
\citet{2001ARA&A..39..249H} for a review), see Section~\ref{sec:abs}
and Figure~\ref{fig_abs}. Since the absorption cross section peaks
near threshold, there is an approximate mapping between the wavelength
of an absorbing EBL photon and the energy of an interacting
$\gamma$-ray, $E_{\rm TeV} \approx 0.7\lambda_{\mu {\rm m}}$.  The
level of the EBL, in particular at the relevant IR wavelengths, is
very difficult to determine by direct observation, due to overwhelming
foreground light sources, hindering the interpretation of AGN spectra
due to the uncertain correction for EBL absorption.  On the other
hand, the EBL is important in its own right and the information
provided by this TeV absorption is potentially very useful.  The EBL
represents the integrated, red-shifted, emission from all epochs of
the evolution of the Universe. The EBL, with its peaks around $1 \mu$m
from starlight and around $100 \mu$m due to starlight reprocessed by
dust (Figure~\ref{fig_EBLn}), has embedded within it information on
the history of galaxy and star formation in the Universe.  Assuming
`plausible' shapes for blazar spectra, one can use absorption features
in $\gamma$-ray spectra to constrain the level and spectral
distribution of the EBL~\citep{1992ApJ...390L..49S}.  The `plausible'
assumptions usually include a $\gamma$-ray spectral index
$\Gamma_{\gamma} \ge 1.5$ at the source, and emitted spectra which are
power-laws, possibly curved with the index increasing --- but not
decreasing --- with increasing energy.  These assumptions are
supported by nearby sources, which are less affected by EBL
absorption. However, it can not be firmly excluded that the more
distant sources used to `measure' the EBL are subject to a selection
bias and differ in their intrinsic spectra
\citep[e.g.][]{2007ApJ...667L..29S}.  EBL de-absorption with an optical
depth $\tau(E)$ transforms a measured $\gamma$-ray flux $\Phi_{o}(E)$ into an
intrinsic flux at the source $\Phi_{i}(E) = \Phi_{o}(E) e^{\tau(E)}$.  A
rapid increase of $\tau(E)$ at higher energies, as predicted in some
EBL scenarios, will decrease the intrinsic spectral index and may
conflict with the assumptions concerning intrinsic spectra.
Sources with higher redshifts and correspondingly large
$\tau(E)$ and with hard measured spectra and wide energy coverage will
provide the most stringent constraints. On the basis of spectra
measured for 1ES\,1101$-$232 ($z=0.186$) and H\,2356$-$309 ($z=0.165$), upper
limits for the EBL density at optical/near-IR wavelengths were derived
which were within a factor 1.5 of the lower limits provided by the
integrated light of resolved galaxies, and below the EBL level previously
assumed \citep{2006Natur.440.1018A}.  Strong constraints on the shape
of the EBL spectrum in the 5--10 $\mu$m region were obtained based on
the spectrum of 1ES\,0229+200 ($z=0.139$) \citep{2007A&A...475L...9A}.  The
dection of VHE $\gamma$-rays from 3C\,279 ($z=0.536$)
\citep{MAGIC:3c279} provided another demonstration of low EBL
levels and hence the relatively high transparency of the Universe to $\gamma$-rays.
Limits derived for the EBL density often include --- beyond the
assumptions on intrinsic source spectra --- certain minimal assumptions
about the shape of the EBL spectrum itself.
\citet{2007A&A...471..439M} have developed an approach where arbitrary EBL 
shapes are tested and constrained, confirming earlier EBL limits
(Figure~\ref{fig_EBLn}). EBL determinations using absorption of VHE
$\gamma$-rays have driven development of models for EBL formation, and
the latest models \citep[see e.g.][]{2008A&A...487..837F} and calculations,
based on hierarchical structure formation and employing detailed models
for galaxy formation and evolution and the reprocessing of starlight
by dust, reproduce the low EBL levels measured \citep[see e.g.][]{2008arXiv0811.3230P}.

\subsection{Other extragalactic source classes}

A range of extragalactic objects beyond AGN is expected to emit VHE
$\gamma$-rays at some level, such as normal galaxies, starburst
galaxies, galaxy clusters, GRBs, and the sources of ultra-high-energy
cosmic rays (UHECR); most of these are predicted to emit at a level
not far below the sensitivity of current instruments, and many of these
classes raise interesting new science topics.

{\bf Nearby galaxies.} Diffuse $\gamma$-rays from nearby normal galaxies are
produced in CR interactions with the interstellar
medium; their detection - combined with estimates of supernova rates
in these galaxies - would allow concepts of CR
acceleration and propagation in galaxies to be tested. Expected
fluxes are discussed in \citet{2001ApJ...558...63P}, applied to the
EGRET energy range. However, detection of local group Galaxies
at VHE energies will be difficult even for next-generation
instruments.

{\bf Starburst galaxies and ultraluminous infrared galaxies.} Starburst
galaxies with strongly enhanced supernova rates and enhanced gas
density, and their more extreme cousins the ultralumious infrared galaxies, 
should provide VHE $\gamma$-ray fluxes not far below current sensitivities,
providing another test of concepts of CR acceleration.  Amongst the best
candidates are NGC\,253 \citep{2005A&A...444..403D} and M\,82
\citep{2008A&A...486..143P}; in fact, NGC\,253 was 
claimed as a TeV emitter by the CANGAROO collaboration \citep{2003A&A...402..443I}, 
but this detection was not confirmed by later, more sensitive, observations
with H.E.S.S. and CANGAROO~\citep{2007A&A...462...67I}. Bounds on emission 
from the ultralumious infrared galaxy Arp~220 have been reported by 
the MAGIC collaboration~\citep{2007ApJ...658..245A}.

{\bf Galaxy clusters.} Clusters of galaxies are the largest gravitationally
bound systems in the Universe, and should contain a significant
non-thermal particle population, fed by particle acceleration in
accretion shocks during the assembly of the cluster, by supernova
activity in cluster galaxies, and by particle acceleration by cluster 
AGN \citep[e.g.][]{1996SSRv...75..279V,2007astro.ph..1545B}.
The time required for CRs to diffuse out of a cluster is generally 
supposed to be larger than the age of the Universe.
As a consequence the CR abundance in clusters
provides a measure of the time-integrated CR
production rate, and the CR spectrum is not softened by escape
as it is, for example, inside our own galaxy. VHE $\gamma$-ray flux limits 
have been derived using the Whipple telescope for the Perseus and Abell 2029 
clusters \citep{2006ApJ...644..148P}, limiting the CR energy density to
as little as 8\% of the thermal energy density in the case of the Perseus
cluster, depending on assumptions concerning the distribution of
CRs relative to gas in the cluster. Such a value is well
within the range estimated for
different acceleration mechanisms, indicating that detections
may be possible in the near future.

{\bf GRBs.} Gamma ray bursts - GRBs - are usually explained by fireball
models, with emission produced by relativistic shocks
\citep[e.g.][]{2006RPPh...69.2259M}. Both prompt and delayed $\gamma$-ray
emission at GeV and TeV energies has been predicted
\citep[e.g.][]{2005ApJ...633.1018P,2007ApJ...671..645A}, 
due to both leptonic and hadronic mechanisms. No VHE $\gamma$-rays have been
detected from GRBs so far. Truly simultaneous observations are
primarily possible with wide-field instruments such as MILAGRO
\citep[see e.g.][for limits on 17 GRBs]{2007ApJ...666..361A}, at the
expense of sensitivity and energy threshold in comparison to current
Cherenkov telescopes. 
In one case --- GRB 060602B ---
an (unusually soft) GRB occurred in the field of view of ongoing
observations of the H.E.S.S. telescope, however, without giving rise to
a detectable signal \citep{HESS:promptGRB}. Upper limits 
were reported, for example, from the MAGIC collaboration
for 9 GRBs \citep{2007ApJ...667..358A},
with observations starting as early as 40\,s after the burst, due to the
fast slewing capabilities of the MAGIC telescope. A strong and 
nearby GRB, observed rather promptly, is needed to challenge GRB
models. None of the limits obtained so far fall in this class
and it looks as if patience will be required to obtain a VHE
$\gamma$-ray detection of a GRB, even if a TeV emission component
is present, due to the limited duty cycle and redshift reach of the
sensitive instruments. An in-depth discussion and detailed references concerning
high-energy $\gamma$-rays from GRBs is given in
\citet{2008arXiv0810.0444B}.

{\bf Sources of UHECR.} Nearby accelerators of ultra high energy cosmic
rays (UHECRs) - within the $\approx$100\,Mpc GZK radius - should also be
sources of VHE $\gamma$-rays. Firstly, interactions during the 
propagation of UHECR beyond the GZK cut-off give rise to cascades feeding 
energy down to the TeV range \citep[see e.g.][]{ferrigno05}. Secondly,
VHE photon production during acceleration of UHECRs is expected in some
scenarios \citep[see e.g.][]{levinson00}.
Again, fluxes for sources from which a few UHECR are detected
are typically expected to lie at the lower edge of the sensitivity of current
instruments.

\section{VHE $\gamma$-rays and astroparticle physics}

The areas of VHE $\gamma$-ray astronomy discussed so far concern 
primarily issues in astronomy and astrophysics. The development of VHE $\gamma$-ray
instrumentation has also been driven to a significant extent by
questions in the field of astroparticle physics, above all the indirect search for
dark matter. Results obtained by VHE instruments also have relevance to
fundamental physics, for example in tests for an energy dependence of the
speed of light, as predicted in some models of quantum gravity. The former
aspect will be addressed in more detail, the latter rather briefly.

\subsection{Searching for dark matter}

In the standard cosmological model \citep[e.g.][]{FriemanAARA2008},
which has emerged over the last decade, only about 4\% of the energy
density in the Universe is in the form of baryons, about 25\% is cold
dark matter (DM) and the remainder is dark energy. Evidence for the
gravitational effects of DM is observed on all scales ranging from
galaxies, where flat rotation curves require that visible matter is
embedded in extended DM halos, to clusters of galaxies. DM is
essential in explaining structure formation in the Universe, driving
the formation of potential wells in which baryonic matter accumulates,
forming galaxies. Structure formation favors cold dark matter ---
i.e., DM in the form of non-relativistic particles --- since rapidly
streaming DM would have smoothed inhomogeneities in the matter
distribution on small scales. 
Some form of weakly interacting massive particle (WIMPs
\citep{1985NuPhB.253..375S}) is required 
to explain the existing observations. 

Creation and annihilation of DM particles was in equilibrium in the
early Universe, but under the rapid expansion of the Universe the WIMP
component was rapidly frozen out, the small interaction cross-section
no longer sustaining sufficient annihilation rates. The current DM
abundance emerges naturally assuming typical weak interaction cross
sections and WIMPs in the mass range of some tens of GeV to
TeV. Candidates include supersymmetric particles, where the lightest
supersymmetric particle is stable due to (assumed) R-parity
conservation, and Kaluza-Klein particles arising in theories with
TeV-scale extra dimensions. As a result of gravitationally-driven
structure formation, DM particles form halos in which galaxies are
embedded, with a pronounced peaks at galactic centers. Simulations of
structure formation predict a $1/r$ (NFW) density profile near the
center of galaxies, with a significant number of sub-halos scattered
elsewhere within galaxies. However, details of the merger history of
halos/galaxies as well as the influence of central black-hole
formation may significantly influence the distribution of DM,
resulting in large uncertainties in the predicted densities.

\begin{figure}
\centerline{\psfig{figure=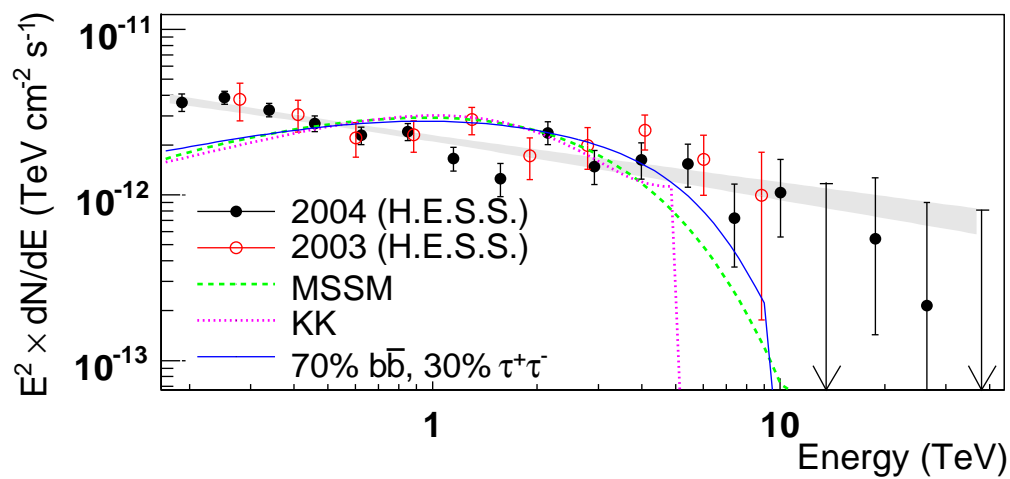,height=15pc}} 
  \caption{Spectral energy distribution from DM annihilation compared to the spectrum
  of $\gamma$-rays from the Galactic center as measured using H.E.S.S. Dashed line: 
  Annihilation of ``typical'' 14~TeV WIMPs, 
  solid line: mix of 70\% $b \bar{b}$ and 30\% $\tau^+\tau^-$ modes of 10~TeV WIMPs,
  dotted: annihilation of 5~TeV KK particles \citep{HESS:gcprl}.
  }
  \label{fig_DMspectra1}
\end{figure}

\begin{figure}
  \centerline{\psfig{figure=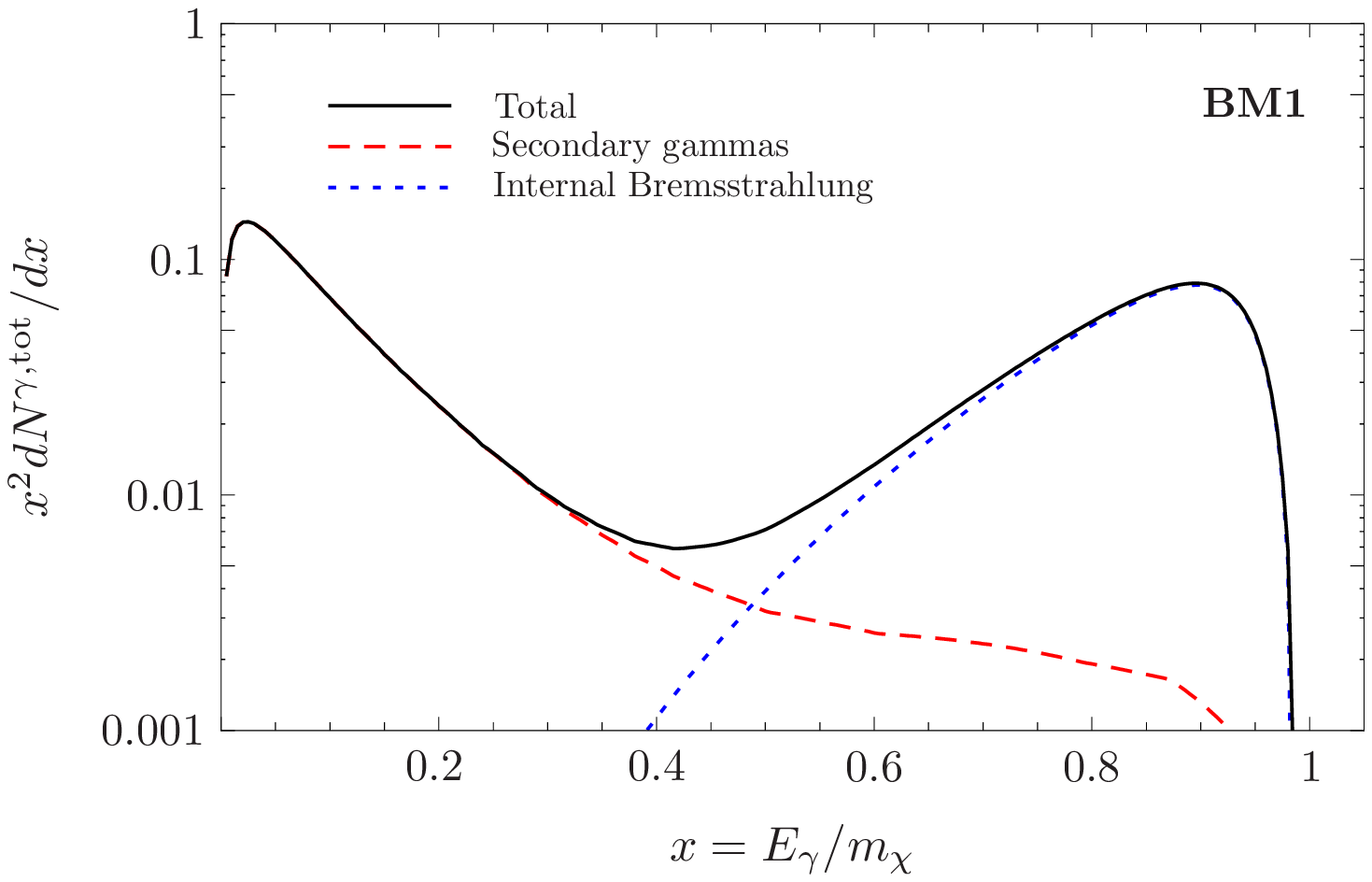,height=20pc}} 
  \caption{$\gamma$-ray spectrum resulting from DM annihilation in one of the benchmark scenarios of 
  \citep{Bringmann2008}, with strong contributions from internal bremsstrahlung. 
  }
  \label{fig_DMspectra2}
\end{figure}

Identification of the nature of DM constituents
\citep[e.g.][]{HooperARNP2008} has been a major theme of particle and
astroparticle physics in the last one to two decades. Possibilities
for the detection of cosmological DM include challenging
direct-detection experiments, where DM particles --- typically assumed
to have a local density of around 0.3~GeV\,cm$^{-3}$ --- scatter off
nuclei, resulting in keV-scale recoil energies
\citep{GaitskellARNP2004}. Annihilation of DM particles in the
galactic vicinity of the Sun will enhance CR electron and positron
fluxes and in particular the positron/electron ratio, as well as the
antiproton/proton ratio, depending on the specific annihilation
channels. Finally --- and most relevant for VHE $\gamma$-ray astronomy
--- annihilation of DM particles will generate an enhanced
$\gamma$-ray flux from regions of high DM density, the annihilation
rate reflecting two-particle interactions, proportional to the square
of the density \citep[see][and references therein]{whitepDM}.
$\gamma$-rays may be produced directly in the primary annihilation
process, by the decay of hadrons created via annihilation into
fermion-antifermion pairs or vector bosons, or by IC scattering of
electrons created in the annihilation or in the decay chains of
annihilation particles. $\gamma$-rays from DM annihilation are
identified by their spectral and directional signature. The expected
flux is generally written as a product of a ``particle physics''
factor and an ``astrophysics'' factor,
\begin{equation}
\Phi_\gamma = \left( {<\sigma v> \over M^2} D_\gamma(E) \right) \left( {1 \over 4\pi} \int_{LOS} \rho^2 dl \right)
\end{equation}
 where $<\sigma v>$ is the velocity-weighted annihilation cross
section of WIMP particles of mass $M$, $D_\gamma(E)$ the spectral
yield per annihilation, and $\rho$ the density of DM particles, with
the annihilation flux being proportional to the line-of-sight (LOS)
integrated squared density.  For all decay modes the
spectrum $D_\gamma(E)$ cuts off at the mass of the DM particles, which
annihilate effectively at rest. The spectral details depend very
much on the decay modes. Two-body decays into $\gamma$-rays, such as
$\gamma\gamma$ or $Z^0\gamma$ or $H\gamma$, resulting in a
monoenergetic line signature, are possible via loop processes, but are
strongly suppressed in most scenarios. Dominant decays into
quark-antiquark pairs or boson pairs, followed by hadronization and
hadron decays, generate a broad featureless spectrum with an SED hump
at about 10\% of the WIMP mass (see Figure~\ref{fig_DMspectra1}). Decays
into tau leptons result in a harder $\gamma$-ray continuum. In one class of
models, two-body decays are helicity suppressed, boosting diagrams
with internal bremsstrahlung such as $W^+W^-\gamma$ and a $\gamma$-ray
enhancement near the kinematic limit \citep{Bringmann2008}, almost
equivalent to line emission given the finite energy resolution of VHE
$\gamma$-ray detectors (Figure~\ref{fig_DMspectra2}).

The expected flux of annihilation $\gamma$-rays is proportional to the
LOS integrated squared DM density.  Searches for signatures of DM
annihilation in the VHE regime therefore concentrate on objects with
spikes in DM concentration. Favorable objects include the Galactic
center, dwarf spheriodal galaxies such as Draco, Ursa Minor and Ursa
Major, or Willman 1 \citep{2008ApJ...678..614S} as well as microhalos
and DM enhancements predicted around intermediate-mass black holes (IMBHs) in
the Galactic vicinity \citep{Bertone2005_0509565}, globular clusters,
and supermassive black holes in external galaxies. 
A DM signal would
have to be identified by a combination of spectral and directional
signatures, with nearby sources exhibiting a narrow spike towards the
center of the source combined with an extended tail.

In the VHE domain, despite extensive searches, no DM candidate source
has been identified so far. In fact, predicted fluxes tend to be well
below the sensitivity of current instruments, requiring ``boost
factors'' of several orders of magnitude to make signals detectable,
arising from enhancements in the DM density.  Detection of a DM signal
from the most obvious candidate source, the Galactic Center, is
hampered by the presence of a strong astrophysical source of
$\gamma$-rays, responsible for a power-law spectrum of $\gamma$-rays
extending beyond 10 TeV, which may hide a faint low-energy
annihilation signal. Limits on annihilation fluxes have been obtained
e.g for the Galactic Center \citep{HESS:gcprl}, the dwarf galaxies
Draco \citep{Wood2008ApJ...678..594W,Albert2008ApJ...679..428A}, Ursa
Minor \citep{Wood2008ApJ...678..594W} and Sagittarius
\citep{Aharonian2008APh....29...55A}, for local group galaxies
\citep{Wood2008ApJ...678..594W} and for the globular cluster M15
\citep{Wood2008ApJ...678..594W}, DM annihilation around IMBHs in the
Galactic halo predicts detectable fluxes even in the absence of large
additional boost factors \citep{Bertone2005_0509565} and limits
restricting the model parameter space have been obtained
\citep{Aharonian2008_0806.2981}, but the process relies on a specific
astrophysical scenario which may not be realized in
nature. Interestingly, DM annihilation fluxes from IMBHs depend only
weakly on the annihilation cross-section, since close to an IMBH the
DM density is so high that annihilation limits the peak density, so
that the resulting $\rho^2 \sigma$ varies only as $\sigma^{2/7}$.

\subsection{Violation of Lorentz invariance and quantum gravity}

In the framework of quantum gravity models, Lorentz invariance may be
violated such that the propagation speed of radiation depends on the
photon energy and possibly on its helicity 
\citep[see e.g.][]{1998Natur.395Q.525A}. For energies small compared to the
scale of quantum gravity, the photon speed $c'$ can be parametrized
as
\begin{equation}
c' = c\left( 1+\xi {E \over E_P} + \zeta {E^2 \over E_P^2} \right)
\end{equation}
where $E_P = 1.22 \cdot 10^{19}$ GeV is the Planck energy and $\xi$
and $\zeta$ are free parameters which are expected to be of order
unity, unless specific symmetries forbid the corresponding terms; for
example, there are classes of models with $\xi = 0$. GRBs and flares
from AGN can be used to perform ``time of flight'' studies, searching
for an energy dependence of the peak time of arrival of the
emission. The best sensitivity in terms of limits on $\xi$ and $\zeta$
is obtained for short bursts with well-located peak emission, for
distant objects and for large energy differences. A positive detection
of energy-dependent peak emission could of course also be attributed
to mechanisms inside the source, where such effects can arise rather
naturally, governed by acceleration and/or cooling time scales. To
confirm that a detection implies violation of Lorentz invariance, one
would need to demonstrate that the effect increases with source
distance, and that the distance dependence cannot be attributed to
evolution of the source population. At VHE energies, the technique was
first applied by the Whipple group to a flare of Mrk\,421 ($z \approx
0.03$) \citep{1999PhRvL..83.2108B}, resulting in a limit of $|\xi| <
200$. More recently, MAGIC reported an energy dependence in the peak
position of a flare of Mrk\,501 ($z \approx 0.03$), finally quoting a
limit $|\xi| < 60$ \citep{AlbertPLB2008}. In the giant flare of
PKS\,2155$-$306 ($z \approx 0.12$) detected using H.E.S.S., no energy
dispersion was visible, resulting in $|\xi| < 17$
\citep{2008PhRvL.101q0402A}. VHE $\gamma$-ray observations are hence
beginning to probe energy-scales up to a few \% of the Planck
energy. While such time-of-flight measurements provide the most
model-independent tests of a possible energy dependence of the speed
of light, it must be mentioned that alternative methods --- which make
some additional assumptions --- are significantly more sensitive,
providing limits from $|\xi| < 10^{-7}$ for helicity-dependent speed
modifications (e.g. \citet{2007MNRAS.376.1857F} down to $|\xi| <
10^{-14}$ (e.g. \citet{2008PhRvD..78f3003G} based on decay kinematics.

\section{Conclusions}

VHE $\gamma$-ray astronomy can now legitimately claim to be a mature
astronomical discipline, with resolved source morphologies and
well-sampled light curves and energy spectra. Furthermore, as some objects
have their peak energy output in the TeV range, VHE observations are
clearly critical to our understanding of some astronomical objects.
The primary success of the field so far is in the unambiguous
identification of sites of cosmic particle acceleration. TeV photons
are currently the {\it only} effective probe of relativistic hadrons
in astrophysical environments. Two complementary techniques exist for ground-based $\gamma$-ray 
astronomy: IACTs and ground-based particle detectors. Both have 
significant further potential, with order of magnitude more sensitive
instruments such as CTA and HAWC being planned. The IACT technique
has the potential to cover the energy range from 10 GeV to hundreds 
of TeV, with an angular resolution of better than 1$'$ achievable
at a few TeV. Together, the Fermi satellite and IACTs will 
soon provide unbroken sensitive coverage over 7 orders of magnitude in
$\gamma$-ray energy. 

The number of VHE $\gamma$-ray sources is now over 80.  More importantly,
many different source types are represented. Supernova remnants and pulsar
wind nebulae appear to dominate the Galactic population, but pulsed 
magnetospheric emission from the Crab pulsar has now been detected from the
ground and a number of binary systems have been established as TeV 
emitters. There is also TeV emission apparently associated with 
clusters of massive stars, perhaps as a consequence of stellar wind interactions.
Variable emission from active galactic nuclei dominates the extragalactic TeV sky.
A complex mixture of different mechanisms appear to be at work in
these TeV sources.  Bulk flow appears to be the main energy source;
relativistic jets/winds in AGN/pulsars and 
non-relativistic shocks driven by supernovae ejecta. 
Acceleration processes transfer energy from these bulk motions into
populations of relativistic particles with power-law energy distributions. 
The convection, diffusion and cooling of these particles then determine 
the properties of the high-energy radiation. For accelerated hadrons cooling does not
usually modify the injection spectrum and the observed $\gamma$-ray
spectral shape is close to that of the injected hadrons -- possibly
with modifications due to energy dependent propagation. The situation
for electrons is more complex, cooling can in general not be neglected
and there is a balance between synchrotron and IC losses -- complex
spectral shapes can emerge, particularly in the case of IC dominated 
cooling.

Galactic TeV sources are typically extended on 10~pc scales.  
Compact $\gamma$-ray sources require both compact accelerators and dense compact target
material/radiation fields/magnetic fields, and such objects appear to
be rare --- with binary systems as the only established example. The
cooling time for 10~TeV electrons in $10^{-5}$ G
magnetic fields is $\sim10^{4}$ years -- comparable to the active lifetime of the
dominant sources, energetic pulsars and young SNR. 
Both bulk motion at 1000~km\,s$^{-1}$ over this timescale, 
and Bohm diffusion of electrons in older (cooling limited) objects, 
naturally produce 10~pc scale TeV sources.
For sources
on this scale the detection sensitivity of IACTs is very competitive
to that of X-ray telescopes for detection of synchrotron counterparts.

The current extragalactic TeV sources are characterized by
relativistic bulk motion and compact regions of acceleration and
cooling, leading to variability on timescales down to minutes. For the
blazar class, the boosting associated with this bulk motion allows us
to see distant objects --- with a current redshift record of 0.536.
The combination of TeV data with observations of optical to X-ray
synchrotron emission promises to be a powerful probe of the
inner jets of these AGN. Extragalactic TeV sources also provide a
useful tool to study the cosmic infra-red background and fundamental
physics such as the potential violation of Lorentz invariance at high
energies.

The next decade of TeV astronomy should bring the source count to
${\cal O}$(1000) and increase the diversity of known Galactic
and extragalactic sources. Predicted VHE fluxes for several classes of
astrophysical object are close to detectability with current
instruments, for example clusters of galaxies, starbursts and GRBs.
The indirect detection of Dark Matter is a major objective of the
field and considerable phase-space is available for discovery via this
channel. For known source classes more sensitive and precise observations
will bring improved understanding, in particular a quantitative 
understanding of Galactic CR origin seems within reach of the next 
generation of IACT detectors.
To conclude, TeV astronomy has emerged as a powerful tool for
high-energy astrophysics and looks set to become our primary window on the 
non-thermal universe in the years to come.

\clearpage

\section*{Acknowledgements}

We would like to thank everyone who helped with the preparation of
this review, including Konrad Bernol\"ohr, Julia Brucker, Rolf
B\"uhler, Stefan Funk, Mathieu de Naurois, Ullrich Schwanke, Joanna
Skilton, Melitta Naumann-Godo, Gavin Rowell and Jacco Vink. JAH is
supported by an STFC Advanced Fellowship.


\bibliographystyle{Astronomy}
\bibliography{ARAA}

\begin{thebibliography}{}
\expandafter\ifx\csname natexlab\endcsname\relax\def\natexlab#1{#1}\fi

\bibitem[{{Abdo} et~al.(2008){Abdo}, {Allen}, {Aune}, {Berley}, {Blaufuss}
  et~al.}]{abdo08}
{Abdo} AA, {Allen} B, {Aune} T, {Berley} D, {Blaufuss} E, et~al. 2008.
\newblock \textit{ArXiv e-prints} 805.0417

\bibitem[{{Abdo} et~al.(2007{\natexlab{a}}){Abdo}, {Allen}, {Berley},
  {Casanova}, {Chen} et~al.}]{abdo07}
{Abdo} AA, {Allen} B, {Berley} D, {Casanova} S, {Chen} C, et~al.
  2007{\natexlab{a}}.
\newblock \textit{\apjl} 664:L91--L94

\bibitem[{{Abdo} et~al.(2007{\natexlab{b}}){Abdo}, {Allen}, {Berley},
  {Blaufuss}, {Casanova} et~al.}]{2007ApJ...666..361A}
{Abdo} AA, {Allen} BT, {Berley} D, {Blaufuss} E, {Casanova} S, et~al.
  2007{\natexlab{b}}.
\newblock \textit{\apj} 666:361--367

\bibitem[{{Acciari} et~al.(2008){Acciari}, {Aliu}, {Beilicke}, {Benbow},
  {B{\"o}ttcher} et~al.}]{2008ApJ...684L..73A}
{Acciari} VA, {Aliu} E, {Beilicke} M, {Benbow} W, {B{\"o}ttcher} M, et~al.
  2008.
\newblock \textit{\apjl} 684:L73--L77

\bibitem[{{Aharonian} et~al.(2001){Aharonian}, {Akhperjanian}, {Barrio},
  {Bernl{\"o}hr}, {B{\"o}rst} et~al.}]{HEGRA:casA}
{Aharonian} F, {Akhperjanian} A, {Barrio} J, {Bernl{\"o}hr} K, {B{\"o}rst} H,
  et~al. 2001.
\newblock \textit{\aap} 370:112--120

\bibitem[{{Aharonian} et~al.(2002){Aharonian}, {Akhperjanian}, {Beilicke},
  {Bernl{\"o}hr}, {B{\"o}rst} et~al.}]{HEGRA:tevj2032}
{Aharonian} F, {Akhperjanian} A, {Beilicke} M, {Bernl{\"o}hr} K, {B{\"o}rst} H,
  et~al. 2002.
\newblock \textit{\aap} 393:L37--L40

\bibitem[{{Aharonian} et~al.(2005{\natexlab{a}}){Aharonian}, {Akhperjanian},
  {Aye}, {Bazer-Bachi}, {Beilicke} et~al.}]{HESS:msh1552}
{Aharonian} F, {Akhperjanian} AG, {Aye} KM, {Bazer-Bachi} AR, {Beilicke} M,
  et~al. 2005{\natexlab{a}}.
\newblock \textit{\aap} 435:L17--L20

\bibitem[{{Aharonian} et~al.(2005{\natexlab{b}}){Aharonian}, {Akhperjanian},
  {Aye}, {Bazer-Bachi}, {Beilicke} et~al.}]{HESS:psrb1259}
{Aharonian} F, {Akhperjanian} AG, {Aye} KM, {Bazer-Bachi} AR, {Beilicke} M,
  et~al. 2005{\natexlab{b}}.
\newblock \textit{\aap} 442:1--10

\bibitem[{{Aharonian} et~al.(2005{\natexlab{c}}){Aharonian}, {Akhperjanian},
  {Aye}, {Bazer-Bachi}, {Beilicke} et~al.}]{HESS:1303}
{Aharonian} F, {Akhperjanian} AG, {Aye} KM, {Bazer-Bachi} AR, {Beilicke} M,
  et~al. 2005{\natexlab{c}}.
\newblock \textit{\aap} 439:1013--1021

\bibitem[{{Aharonian} et~al.(2005{\natexlab{d}}){Aharonian}, {Akhperjanian},
  {Aye}, {Bazer-Bachi}, {Beilicke} et~al.}]{HESS:g09}
{Aharonian} F, {Akhperjanian} AG, {Aye} KM, {Bazer-Bachi} AR, {Beilicke} M,
  et~al. 2005{\natexlab{d}}.
\newblock \textit{\aap} 432:L25--L29

\bibitem[{{Aharonian} et~al.(2008{\natexlab{a}}){Aharonian}, {Akhperjanian},
  {Barres de Almeida}, {Bazer-Bachi}, {Becherini} et~al.}]{2008PhRvL.101q0402A}
{Aharonian} F, {Akhperjanian} AG, {Barres de Almeida} U, {Bazer-Bachi} AR,
  {Becherini} Y, et~al. 2008{\natexlab{a}}.
\newblock \textit{Physical Review Letters} 101:170402--+

\bibitem[{{Aharonian} et~al.(2007{\natexlab{a}}){Aharonian}, {Akhperjanian},
  {Barres de Almeida}, {Bazer-Bachi}, {Behera} et~al.}]{2007A&A...475L...9A}
{Aharonian} F, {Akhperjanian} AG, {Barres de Almeida} U, {Bazer-Bachi} AR,
  {Behera} B, et~al. 2007{\natexlab{a}}.
\newblock \textit{\aap} 475:L9--L13

\bibitem[{{Aharonian} et~al.(2008{\natexlab{b}}){Aharonian}, {Akhperjanian},
  {Barres de Almeida}, {Bazer-Bachi}, {Behera} et~al.}]{2008A&A...481L.103A}
{Aharonian} F, {Akhperjanian} AG, {Barres de Almeida} U, {Bazer-Bachi} AR,
  {Behera} B, et~al. 2008{\natexlab{b}}.
\newblock \textit{\aap} 481:L103--L107

\bibitem[{{Aharonian} et~al.(2008{\natexlab{c}}){Aharonian}, {Akhperjanian},
  {Barres de Almeida}, {Bazer-Bachi}, {Behera} et~al.}]{HESS:1745}
{Aharonian} F, {Akhperjanian} AG, {Barres de Almeida} U, {Bazer-Bachi} AR,
  {Behera} B, et~al. 2008{\natexlab{c}}.
\newblock \textit{\aap} 483:509--517

\bibitem[{{Aharonian} et~al.(2008{\natexlab{d}}){Aharonian}, {Akhperjanian},
  {Barres de Almeida}, {Bazer-Bachi}, {Behera} et~al.}]{HESS:dark}
{Aharonian} F, {Akhperjanian} AG, {Barres de Almeida} U, {Bazer-Bachi} AR,
  {Behera} B, et~al. 2008{\natexlab{d}}.
\newblock \textit{\aap} 477:353--363

\bibitem[{{Aharonian} et~al.(2009){Aharonian}, {Akhperjanian}, {Barres
  DeAlmeida}, {Bazer-Bachi}, {Behera} et~al.}]{HESS:promptGRB}
{Aharonian} F, {Akhperjanian} AG, {Barres DeAlmeida} U, {Bazer-Bachi} AR,
  {Behera} B, et~al. 2009.
\newblock \textit{\apj} 690:1068--1073

\bibitem[{{Aharonian} et~al.(2007{\natexlab{b}}){Aharonian}, {Akhperjanian},
  {Bazer-Bachi}, {Behera}, {Beilicke} et~al.}]{2007ApJ...664L..71A}
{Aharonian} F, {Akhperjanian} AG, {Bazer-Bachi} AR, {Behera} B, {Beilicke} M,
  et~al. 2007{\natexlab{b}}.
\newblock \textit{\apjl} 664:L71--L74

\bibitem[{{Aharonian} et~al.(2008{\natexlab{e}}){Aharonian}, {Akhperjanian},
  {Bazer-Bachi}, {Behera}, {Beilicke} et~al.}]{HESS:w28}
{Aharonian} F, {Akhperjanian} AG, {Bazer-Bachi} AR, {Behera} B, {Beilicke} M,
  et~al. 2008{\natexlab{e}}.
\newblock \textit{\aap} 481:401--410

\bibitem[{{Aharonian} et~al.(2006{\natexlab{a}}){Aharonian}, {Akhperjanian},
  {Bazer-Bachi}, {Beilicke}, {Benbow} et~al.}]{HESS:ls5039p2}
{Aharonian} F, {Akhperjanian} AG, {Bazer-Bachi} AR, {Beilicke} M, {Benbow} W,
  et~al. 2006{\natexlab{a}}.
\newblock \textit{\aap} 460:743--749

\bibitem[{{Aharonian} et~al.(2006{\natexlab{b}}){Aharonian}, {Akhperjanian},
  {Bazer-Bachi}, {Beilicke}, {Benbow} et~al.}]{HESS:rxj1713p2}
{Aharonian} F, {Akhperjanian} AG, {Bazer-Bachi} AR, {Beilicke} M, {Benbow} W,
  et~al. 2006{\natexlab{b}}.
\newblock \textit{\aap} 449:223--242

\bibitem[{{Aharonian} et~al.(2006{\natexlab{c}}){Aharonian}, {Akhperjanian},
  {Bazer-Bachi}, {Beilicke}, {Benbow} et~al.}]{2006Natur.440.1018A}
{Aharonian} F, {Akhperjanian} AG, {Bazer-Bachi} AR, {Beilicke} M, {Benbow} W,
  et~al. 2006{\natexlab{c}}.
\newblock \textit{\nat} 440:1018--1021

\bibitem[{{Aharonian} et~al.(2006{\natexlab{d}}){Aharonian}, {Akhperjanian},
  {Bazer-Bachi}, {Beilicke}, {Benbow} et~al.}]{HESS:kookaburra}
{Aharonian} F, {Akhperjanian} AG, {Bazer-Bachi} AR, {Beilicke} M, {Benbow} W,
  et~al. 2006{\natexlab{d}}.
\newblock \textit{\aap} 456:245--251

\bibitem[{{Aharonian} et~al.(2006{\natexlab{e}}){Aharonian}, {Akhperjanian},
  {Bazer-Bachi}, {Beilicke}, {Benbow} et~al.}]{HESS:1825p2}
{Aharonian} F, {Akhperjanian} AG, {Bazer-Bachi} AR, {Beilicke} M, {Benbow} W,
  et~al. 2006{\natexlab{e}}.
\newblock \textit{\aap} 460:365--374

\bibitem[{{Aharonian} et~al.(2006{\natexlab{f}}){Aharonian}, {Akhperjanian},
  {Bazer-Bachi}, {Beilicke}, {Benbow} et~al.}]{2006Sci...314.1424A}
{Aharonian} F, {Akhperjanian} AG, {Bazer-Bachi} AR, {Beilicke} M, {Benbow} W,
  et~al. 2006{\natexlab{f}}.
\newblock \textit{Science} 314:1424--1427

\bibitem[{{Aharonian} et~al.(2006{\natexlab{g}}){Aharonian}, {Akhperjanian},
  {Bazer-Bachi}, {Beilicke}, {Benbow} et~al.}]{HESS:velax}
{Aharonian} F, {Akhperjanian} AG, {Bazer-Bachi} AR, {Beilicke} M, {Benbow} W,
  et~al. 2006{\natexlab{g}}.
\newblock \textit{\aap} 448:L43--L47

\bibitem[{{Aharonian} et~al.(2006{\natexlab{h}}){Aharonian}, {Akhperjanian},
  {Bazer-Bachi}, {Beilicke}, {Benbow} et~al.}]{HESS:gcprl}
{Aharonian} F, {Akhperjanian} AG, {Bazer-Bachi} AR, {Beilicke} M, {Benbow} W,
  et~al. 2006{\natexlab{h}}.
\newblock \textit{Physical Review Letters} 97:221102--+

\bibitem[{{Aharonian} et~al.(2007{\natexlab{c}}){Aharonian}, {Akhperjanian},
  {Bazer-Bachi}, {Beilicke}, {Benbow} et~al.}]{HESS:westerlund2}
{Aharonian} F, {Akhperjanian} AG, {Bazer-Bachi} AR, {Beilicke} M, {Benbow} W,
  et~al. 2007{\natexlab{c}}.
\newblock \textit{\aap} 467:1075--1080

\bibitem[{{Aharonian} et~al.(2007{\natexlab{d}}){Aharonian}, {Akhperjanian},
  {Bazer-Bachi}, {Beilicke}, {Benbow} et~al.}]{HESS:velajnr2}
{Aharonian} F, {Akhperjanian} AG, {Bazer-Bachi} AR, {Beilicke} M, {Benbow} W,
  et~al. 2007{\natexlab{d}}.
\newblock \textit{\apj} 661:236--249

\bibitem[{{Aharonian} et~al.(2007{\natexlab{e}}){Aharonian}, {Akhperjanian},
  {Bazer-Bachi}, {Beilicke}, {Benbow} et~al.}]{HESS:rxj1713p3}
{Aharonian} F, {Akhperjanian} AG, {Bazer-Bachi} AR, {Beilicke} M, {Benbow} W,
  et~al. 2007{\natexlab{e}}.
\newblock \textit{\aap} 464:235--243

\bibitem[{{Aharonian} et~al.(2008{\natexlab{f}}){Aharonian}, {Akhperjanian},
  {Bazer-Bachi}, {Beilicke}, {Benbow} et~al.}]{Aharonian2008APh....29...55A}
{Aharonian} F, {Akhperjanian} AG, {Bazer-Bachi} AR, {Beilicke} M, {Benbow} W,
  et~al. 2008{\natexlab{f}}.
\newblock \textit{Astroparticle Physics} 29:55--62

\bibitem[{{Aharonian} et~al.(2008{\natexlab{g}}){Aharonian}, {Akhperjanian},
  {de Almeida}, {Bazer-Bachi}, {Behera} et~al.}]{Aharonian2008_0806.2981}
{Aharonian} F, {Akhperjanian} AG, {de Almeida} UB, {Bazer-Bachi} AR, {Behera}
  B, et~al. 2008{\natexlab{g}}.
\newblock \textit{\prd} 78:072008--+

\bibitem[{{Aharonian} et~al.(2008{\natexlab{h}}){Aharonian}, {Buckley},
  {Kifune} \& {Sinnis}}]{aharonian_buckley08}
{Aharonian} F, {Buckley} J, {Kifune} T, {Sinnis} G. 2008{\natexlab{h}}.
\newblock \textit{Reports on Progress in Physics} 71:096901--+

\bibitem[{{Aharonian} \& {{et al.}}(2008)}]{HESS:rcw86}
{Aharonian} F, {{et al.}} 2008.
\newblock \textit{ArXiv e-prints} 0810.2689

\bibitem[{{Aharonian}(1991)}]{aharonian91}
{Aharonian} FA. 1991.
\newblock \textit{\apss} 180:305--320

\bibitem[{{Aharonian} et~al.(2004){Aharonian}, {Akhperjanian}, {Aye},
  {Bazer-Bachi}, {Beilicke} et~al.}]{HESS:rxj1713p1}
{Aharonian} FA, {Akhperjanian} AG, {Aye} KM, {Bazer-Bachi} AR, {Beilicke} M,
  et~al. 2004.
\newblock \textit{\nat} 432:75--77

\bibitem[{{Aharonian} et~al.(2007{\natexlab{f}}){Aharonian}, {Akhperjanian},
  {Bazer-Bachi}, {Behera}, {Beilicke} et~al.}]{HESS:0632}
{Aharonian} FA, {Akhperjanian} AG, {Bazer-Bachi} AR, {Behera} B, {Beilicke} M,
  et~al. 2007{\natexlab{f}}.
\newblock \textit{\aap} 469:L1--L4

\bibitem[{{Aharonian}, {Khangulyan} \& {Costamante}(2008)}]{aharonian08}
{Aharonian} FA, {Khangulyan} D, {Costamante} L. 2008.
\newblock \textit{\mnras} 387:1206--1214

\bibitem[{{Albert} et~al.(2008{\natexlab{a}}){Albert}, {Aliu}, {Anderhub},
  {Antonelli}, {Antoranz} et~al.}]{2008ApJ...685L..23A}
{Albert} J, {Aliu} E, {Anderhub} H, {Antonelli} LA, {Antoranz} P, et~al.
  2008{\natexlab{a}}.
\newblock \textit{\apjl} 685:L23--L26

\bibitem[{{Albert} et~al.(2008{\natexlab{b}}){Albert}, {Aliu}, {Anderhub},
  {Antonelli}, {Antoranz} et~al.}]{MAGIC:3c279}
{Albert} J, {Aliu} E, {Anderhub} H, {Antonelli} LA, {Antoranz} P, et~al.
  2008{\natexlab{b}}.
\newblock \textit{Science} 320:1752--

\bibitem[{{Albert} et~al.(2006{\natexlab{a}}){Albert}, {Aliu}, {Anderhub},
  {Antoranz}, {Armada} et~al.}]{2006ApJ...648L.105A}
{Albert} J, {Aliu} E, {Anderhub} H, {Antoranz} P, {Armada} A, et~al.
  2006{\natexlab{a}}.
\newblock \textit{\apjl} 648:L105--L108

\bibitem[{{Albert} et~al.(2006{\natexlab{b}}){Albert}, {Aliu}, {Anderhub},
  {Antoranz}, {Armada} et~al.}]{MAGIC:lsi61}
{Albert} J, {Aliu} E, {Anderhub} H, {Antoranz} P, {Armada} A, et~al.
  2006{\natexlab{b}}.
\newblock \textit{Science} 312:1771--1773

\bibitem[{{Albert} et~al.(2007{\natexlab{a}}){Albert}, {Aliu}, {Anderhub},
  {Antoranz}, {Armada} et~al.}]{MAGIC:ic443}
{Albert} J, {Aliu} E, {Anderhub} H, {Antoranz} P, {Armada} A, et~al.
  2007{\natexlab{a}}.
\newblock \textit{\apjl} 664:L87--L90

\bibitem[{{Albert} et~al.(2007{\natexlab{b}}){Albert}, {Aliu}, {Anderhub},
  {Antoranz}, {Armada} et~al.}]{2007ApJ...658..245A}
{Albert} J, {Aliu} E, {Anderhub} H, {Antoranz} P, {Armada} A, et~al.
  2007{\natexlab{b}}.
\newblock \textit{\apj} 658:245--248

\bibitem[{{Albert} et~al.(2007{\natexlab{c}}){Albert}, {Aliu}, {Anderhub},
  {Antoranz}, {Armada} et~al.}]{2007ApJ...667..358A}
{Albert} J, {Aliu} E, {Anderhub} H, {Antoranz} P, {Armada} A, et~al.
  2007{\natexlab{c}}.
\newblock \textit{\apj} 667:358--366

\bibitem[{{Albert} et~al.(2007{\natexlab{d}}){Albert}, {Aliu}, {Anderhub},
  {Antoranz}, {Armada} et~al.}]{MAGIC:cygX1}
{Albert} J, {Aliu} E, {Anderhub} H, {Antoranz} P, {Armada} A, et~al.
  2007{\natexlab{d}}.
\newblock \textit{\apjl} 665:L51--L54

\bibitem[{{Albert} et~al.(2008{\natexlab{c}}){Albert}, {Aliu}, {Anderhub},
  {Antoranz}, {Backes} et~al.}]{MAGIC:lsi61b}
{Albert} J, {Aliu} E, {Anderhub} H, {Antoranz} P, {Backes} M, et~al.
  2008{\natexlab{c}}.
\newblock \textit{\apj} 684:1351--1358

\bibitem[{{Albert} et~al.(2008{\natexlab{d}}){Albert}, {Aliu}, {Anderhub},
  {Antoranz}, {Backes} et~al.}]{Albert2008ApJ...679..428A}
{Albert} J, {Aliu} E, {Anderhub} H, {Antoranz} P, {Backes} M, et~al.
  2008{\natexlab{d}}.
\newblock \textit{\apj} 679:428--431

\bibitem[{Albert et~al.(2008)}]{AlbertPLB2008}
Albert J, et~al. 2008.
\newblock \textit{Phys. Lett.} B668:253--257

\bibitem[{{Aliu} et~al.(2008){Aliu}, {Anderhub}, {Antonelli} \& {et
  al.}}]{MAGIC:crabpulsed}
{Aliu} E, {Anderhub} H, {Antonelli} LA, {et al.} 2008.
\newblock \textit{Science} 322:1221--1224

\bibitem[{{Amelino-Camelia} et~al.(1998){Amelino-Camelia}, {Ellis},
  {Mavromatos}, {Nanopoulos} \& {Sarkar}}]{1998Natur.395Q.525A}
{Amelino-Camelia} G, {Ellis} J, {Mavromatos} NE, {Nanopoulos} DV, {Sarkar} S.
  1998.
\newblock \textit{\nat} 395:525--+

\bibitem[{{Amenomori} et~al.(1999){Amenomori}, {Ayabe}, {Cao}, {Danzengluobu},
  {Ding} et~al.}]{amenomori99}
{Amenomori} M, {Ayabe} S, {Cao} PY, {Danzengluobu}, {Ding} LK, et~al. 1999.
\newblock \textit{\apjl} 525:L93--L96

\bibitem[{{Asano} \& {Inoue}(2007)}]{2007ApJ...671..645A}
{Asano} K, {Inoue} S. 2007.
\newblock \textit{\apj} 671:645--655

\bibitem[{{Aschenbach}(1998)}]{aschenbach98}
{Aschenbach} B. 1998.
\newblock \textit{\nat} 396:141--142

\bibitem[{{Atkins} et~al.(2003){Atkins}, {Benbow}, {Berley}, {Blaufuss},
  {Bussons} et~al.}]{atkins03}
{Atkins} R, {Benbow} W, {Berley} D, {Blaufuss} E, {Bussons} J, et~al. 2003.
\newblock \textit{\apj} 595:803--811

\bibitem[{{Atoyan} \& {Aharonian}(1996)}]{1996A&AS..120C.453A}
{Atoyan} AM, {Aharonian} FA. 1996.
\newblock \textit{\aaps} 120:C453+

\bibitem[{{Benaglia} \& {Romero}(2003)}]{benaglia03}
{Benaglia} P, {Romero} GE. 2003.
\newblock \textit{\aap} 399:1121--1134

\bibitem[{{Berezhko}, {Ksenofontov} \& {V{\"o}lk}(2003)}]{berezhko03}
{Berezhko} EG, {Ksenofontov} LT, {V{\"o}lk} HJ. 2003.
\newblock \textit{\aap} 412:L11--L14

\bibitem[{{Berezhko} \& {V{\"o}lk}(2006)}]{berezhko06}
{Berezhko} EG, {V{\"o}lk} HJ. 2006.
\newblock \textit{\aap} 451:981--990

\bibitem[{{Bernlohr}(2000)}]{2000APh....12..255B}
{Bernlohr} K. 2000.
\newblock \textit{Astroparticle Physics} 12:255--268

\bibitem[{{Bertone}, {Zentner} \& {Silk}(2005)}]{Bertone2005_0509565}
{Bertone} G, {Zentner} AR, {Silk} J. 2005.
\newblock \textit{\prd} 72:103517--+

\bibitem[{{Biller} et~al.(1999){Biller}, {Breslin}, {Buckley}, {Catanese},
  {Carson} et~al.}]{1999PhRvL..83.2108B}
{Biller} SD, {Breslin} AC, {Buckley} J, {Catanese} M, {Carson} M, et~al. 1999.
\newblock \textit{Physical Review Letters} 83:2108--2111

\bibitem[{{Blasi}, {Gabici} \& {Brunetti}(2007)}]{2007astro.ph..1545B}
{Blasi} P, {Gabici} S, {Brunetti} G. 2007.
\newblock \textit{ArXiv Astrophysics e-prints} astro-ph/0701545

\bibitem[{{Blondin}, {Chevalier} \& {Frierson}(2001)}]{2001ApJ...563..806B}
{Blondin} JM, {Chevalier} RA, {Frierson} DM. 2001.
\newblock \textit{\apj} 563:806--815

\bibitem[{{Blumenthal} \& {Gould}(1970)}]{blumenthal70}
{Blumenthal} GR, {Gould} RJ. 1970.
\newblock \textit{Reviews of Modern Physics} 42:237--271

\bibitem[{{Bringmann}, {Bergstr{\"o}m} \& {Edsj{\"o}}(2008)}]{Bringmann2008}
{Bringmann} T, {Bergstr{\"o}m} L, {Edsj{\"o}} J. 2008.
\newblock \textit{Journal of High Energy Physics} 1:49--+

\bibitem[{{Buckley} et~al.(2008{\natexlab{a}}){Buckley}, {Baltz}, {Bertone},
  {Byrum}, {Fegan} et~al.}]{whitepDM}
{Buckley} J, {Baltz} EA, {Bertone} G, {Byrum} K, {Fegan} S, et~al.
  2008{\natexlab{a}}.
\newblock \textit{ArXiv e-prints} 0812.0795

\bibitem[{{Buckley} et~al.(2008{\natexlab{b}}){Buckley}, {Byrum}, {Dingus},
  {Falcone}, {Kaaret} et~al.}]{2008arXiv0810.0444B}
{Buckley} J, {Byrum} K, {Dingus} B, {Falcone} A, {Kaaret} P, et~al.
  2008{\natexlab{b}}.
\newblock \textit{ArXiv e-prints} 0810.0444

\bibitem[{{Caprioli} et~al.(2008){Caprioli}, {Blasi}, {Amato} \&
  {Vietri}}]{caprioli08}
{Caprioli} D, {Blasi} P, {Amato} E, {Vietri} M. 2008.
\newblock \textit{\apjl} 679:L139--L142

\bibitem[{{Carrigan} et~al.(2007){Carrigan}, {Hinton}, {Hofmann}, {Kosack},
  {Lohse} et~al.}]{2007arXiv0709.4094C}
{Carrigan} S, {Hinton} JA, {Hofmann} W, {Kosack} K, {Lohse} T, et~al. 2007.
\newblock \textit{ArXiv e-prints} 0709.4094

\bibitem[{{Casares} et~al.(2005){Casares}, {Rib{\'o}}, {Ribas}, {Paredes},
  {Mart{\'{\i}}} \& {Herrero}}]{casares05}
{Casares} J, {Rib{\'o}} M, {Ribas} I, {Paredes} JM, {Mart{\'{\i}}} J, {Herrero}
  A. 2005.
\newblock \textit{\mnras} 364:899--908

\bibitem[{{Costamante} \& {Ghisellini}(2002)}]{2002A&A...384...56C}
{Costamante} L, {Ghisellini} G. 2002.
\newblock \textit{\aap} 384:56--71

\bibitem[{{Daniel} et~al.(2005){Daniel}, {Badran}, {Bond}, {Boyle}, {Bradbury}
  et~al.}]{2005ApJ...621..181D}
{Daniel} MK, {Badran} HM, {Bond} IH, {Boyle} PJ, {Bradbury} SM, et~al. 2005.
\newblock \textit{\apj} 621:181--187

\bibitem[{{de Jager} \& {Djannati-Ata{\"i}}(2008)}]{ARxIV:0803.0116}
{de Jager} OC, {Djannati-Ata{\"i}} A. 2008.
\newblock \textit{ArXiv e-prints} 0803.0116

\bibitem[{{Dhawan}, {Mioduszewski} \& {Rupen}(2006)}]{dhawan06}
{Dhawan} V, {Mioduszewski} A, {Rupen} M. 2006.
\newblock In \textit{VI Microquasar Workshop: Microquasars and Beyond}

\bibitem[{{Domingo-Santamar{\'{\i}}a} \& {Torres}(2005)}]{2005A&A...444..403D}
{Domingo-Santamar{\'{\i}}a} E, {Torres} DF. 2005.
\newblock \textit{\aap} 444:403--415

\bibitem[{{Dondi} \& {Ghisellini}(1995)}]{1995MNRAS.273..583D}
{Dondi} L, {Ghisellini} G. 1995.
\newblock \textit{\mnras} 273:583--595

\bibitem[{{Fan}, {Wei} \& {Xu}(2007)}]{2007MNRAS.376.1857F}
{Fan} YZ, {Wei} DM, {Xu} D. 2007.
\newblock \textit{\mnras} 376:1857--1860

\bibitem[{{Ferrigno}, {Blasi} \& {de Marco}(2005)}]{ferrigno05}
{Ferrigno} C, {Blasi} P, {de Marco} D. 2005.
\newblock \textit{Astroparticle Physics} 23:211--226

\bibitem[{{Fossati} et~al.(2008){Fossati}, {Buckley}, {Bond}, {Bradbury},
  {Carter-Lewis} et~al.}]{2008ApJ...677..906F}
{Fossati} G, {Buckley} JH, {Bond} IH, {Bradbury} SM, {Carter-Lewis} DA, et~al.
  2008.
\newblock \textit{\apj} 677:906--925

\bibitem[{{Franceschini}, {Rodighiero} \&
  {Vaccari}(2008)}]{2008A&A...487..837F}
{Franceschini} A, {Rodighiero} G, {Vaccari} M. 2008.
\newblock \textit{\aap} 487:837--852

\bibitem[{{Frieman}, {Turner} \& {Huterer}(2008)}]{FriemanAARA2008}
{Frieman} JA, {Turner} MS, {Huterer} D. 2008.
\newblock \textit{\araa} 46:385--432

\bibitem[{{Funk} et~al.(2007){Funk}, {Hinton}, {P{\"u}hlhofer}, {Aharonian},
  {Hofmann} et~al.}]{funk07a}
{Funk} S, {Hinton} JA, {P{\"u}hlhofer} G, {Aharonian} FA, {Hofmann} W, et~al.
  2007.
\newblock \textit{\apj} 662:517--524

\bibitem[{{Gabici}, {Aharonian} \& {Blasi}(2007)}]{gabici07a}
{Gabici} S, {Aharonian} FA, {Blasi} P. 2007.
\newblock \textit{\apss} 309:365--371

\bibitem[{{Gaensler} \& {Slane}(2006)}]{gaensler06}
{Gaensler} BM, {Slane} PO. 2006.
\newblock \textit{\araa} 44:17--47

\bibitem[{{Gaitskell}(2004)}]{GaitskellARNP2004}
{Gaitskell} RJ. 2004.
\newblock \textit{Annual Review of Nuclear and Particle Science} 54:315--359

\bibitem[{{Galaverni} \& {Sigl}(2008)}]{2008PhRvD..78f3003G}
{Galaverni} M, {Sigl} G. 2008.
\newblock \textit{\prd} 78:063003--+

\bibitem[{{Harding}(2007)}]{2007arXiv0710.3517H}
{Harding} AK. 2007.
\newblock \textit{ArXiv e-prints} 0710.3517

\bibitem[{{Hartman} et~al.(1999){Hartman}, {Bertsch}, {Bloom}, {Chen},
  {Deines-Jones} et~al.}]{hartman99}
{Hartman} RC, {Bertsch} DL, {Bloom} SD, {Chen} AW, {Deines-Jones} P, et~al.
  1999.
\newblock \textit{\apjs} 123:79--202

\bibitem[{{Hauser} \& {Dwek}(2001)}]{2001ARA&A..39..249H}
{Hauser} MG, {Dwek} E. 2001.
\newblock \textit{\araa} 39:249--307

\bibitem[{{Helfand} et~al.(2007){Helfand}, {Gotthelf}, {Halpern}, {Camilo},
  {Semler} et~al.}]{helfand07}
{Helfand} DJ, {Gotthelf} EV, {Halpern} JP, {Camilo} F, {Semler} DR, et~al.
  2007.
\newblock \textit{\apj} 665:1297--1303

\bibitem[{{Hester}(2008)}]{2008ARA&A..46..127H}
{Hester} JJ. 2008.
\newblock \textit{\araa} 46:127--155

\bibitem[{{Hillas}(2005)}]{hillas05}
{Hillas} AM. 2005.
\newblock \textit{Journal of Physics G Nuclear Physics} 31:95--+

\bibitem[{{Hinton}(2008)}]{hinton08}
{Hinton} J. 2008.
\newblock \textit{ArXiv e-prints} 803.1609

\bibitem[{{Hinton} \& {Aharonian}(2007)}]{hinton07}
{Hinton} JA, {Aharonian} FA. 2007.
\newblock \textit{\apj} 657:302--307

\bibitem[{{Hinton} et~al.(2009){Hinton}, {Skilton}, {Funk}, {Brucker},
  {Aharonian} et~al.}]{hinton08b}
{Hinton} JA, {Skilton} JL, {Funk} S, {Brucker} J, {Aharonian} FA, et~al. 2009.
\newblock \textit{\apjl} 690:L101--L104

\bibitem[{{Hofmann}(2006)}]{2006astro.ph..3076H}
{Hofmann} W. 2006.
\newblock \textit{ArXiv Astrophysics e-prints} astro-ph/0603076

\bibitem[{{Hooper} \& {Baltz}(2008)}]{HooperARNP2008}
{Hooper} D, {Baltz} EA. 2008.
\newblock \textit{Annual Review of Nuclear and Particle Science} 58:293--314

\bibitem[{{Hoppe} \& {{et al.}}(2007)}]{HESS:scanicrc}
{Hoppe} S, {{et al.}} 2007.
\newblock \textit{ArXiv e-prints} 0710.3528

\bibitem[{{Hunter} et~al.(1997){Hunter}, {Bertsch}, {Catelli}, {Dame}, {Digel}
  et~al.}]{hunter97}
{Hunter} SD, {Bertsch} DL, {Catelli} JR, {Dame} TM, {Digel} SW, et~al. 1997.
\newblock \textit{\apj} 481:205--+

\bibitem[{{Itoh} et~al.(2007){Itoh}, {Enomoto}, {Yanagita}, {Yoshida},
  {Asahara} et~al.}]{2007A&A...462...67I}
{Itoh} C, {Enomoto} R, {Yanagita} S, {Yoshida} T, {Asahara} A, et~al. 2007.
\newblock \textit{\aap} 462:67--71

\bibitem[{{Itoh} et~al.(2003){Itoh}, {Enomoto}, {Yanagita}, {Yoshida},
  {Tanimori} et~al.}]{2003A&A...402..443I}
{Itoh} C, {Enomoto} R, {Yanagita} S, {Yoshida} T, {Tanimori} T, et~al. 2003.
\newblock \textit{\aap} 402:443--455

\bibitem[{{Johnston} et~al.(1992){Johnston}, {Manchester}, {Lyne}, {Bailes},
  {Kaspi} et~al.}]{johnston92}
{Johnston} S, {Manchester} RN, {Lyne} AG, {Bailes} M, {Kaspi} VM, et~al. 1992.
\newblock \textit{\apjl} 387:L37--L41

\bibitem[{{Kardashev}(1962)}]{kardashev62}
{Kardashev} NS. 1962.
\newblock \textit{Soviet Astronomy} 6:317--+

\bibitem[{{Katagiri} et~al.(2005){Katagiri}, {Enomoto}, {Ksenofontov}, {Mori},
  {Adachi} et~al.}]{CANGAROO:velajnr}
{Katagiri} H, {Enomoto} R, {Ksenofontov} LT, {Mori} M, {Adachi} Y, et~al. 2005.
\newblock \textit{\apjl} 619:L163--L166

\bibitem[{{Katarzy{\'n}ski}, {Sol} \& {Kus}(2001)}]{2001A&A...367..809K}
{Katarzy{\'n}ski} K, {Sol} H, {Kus} A. 2001.
\newblock \textit{\aap} 367:809--825

\bibitem[{{Kelner}, {Aharonian} \& {Bugayov}(2006)}]{kelner06}
{Kelner} SR, {Aharonian} FA, {Bugayov} VV. 2006.
\newblock \textit{\prd} 74:034018

\bibitem[{{Kennel} \& {Coroniti}(1984)}]{1984ApJ...283..710K}
{Kennel} CF, {Coroniti} FV. 1984.
\newblock \textit{\apj} 283:710--730

\bibitem[{{Khangulyan}, {Aharonian} \& {Bosch-Ramon}(2008)}]{khangulyan08}
{Khangulyan} D, {Aharonian} F, {Bosch-Ramon} V. 2008.
\newblock \textit{\mnras} 383:467--478

\bibitem[{{Kirk}, {Ball} \& {Skjaeraasen}(1999)}]{kirk99}
{Kirk} JG, {Ball} L, {Skjaeraasen} O. 1999.
\newblock \textit{Astroparticle Physics} 10:31--45

\bibitem[{{Krennrich} et~al.(2001){Krennrich}, {Badran}, {Bond}, {Bradbury},
  {Buckley} et~al.}]{2001ApJ...560L..45K}
{Krennrich} F, {Badran} HM, {Bond} IH, {Bradbury} SM, {Buckley} JH, et~al.
  2001.
\newblock \textit{\apjl} 560:L45--L48

\bibitem[{{Krennrich} et~al.(2002){Krennrich}, {Bond}, {Bradbury}, {Buckley},
  {Carter-Lewis} et~al.}]{2002ApJ...575L...9K}
{Krennrich} F, {Bond} IH, {Bradbury} SM, {Buckley} JH, {Carter-Lewis} DA,
  et~al. 2002.
\newblock \textit{\apjl} 575:L9--L13

\bibitem[{{Levinson}(2000)}]{levinson00}
{Levinson} A. 2000.
\newblock \textit{Physical Review Letters} 85:912--915

\bibitem[{{Leyder}, {Walter} \& {Rauw}(2008)}]{leyder08}
{Leyder} JC, {Walter} R, {Rauw} G. 2008.
\newblock \textit{\aap} 477:L29--L32

\bibitem[{{Lucek} \& {Bell}(2000)}]{lucek00}
{Lucek} SG, {Bell} AR. 2000.
\newblock \textit{\mnras} 314:65--74

\bibitem[{{Mannheim}(1993)}]{1993A&A...269...67M}
{Mannheim} K. 1993.
\newblock \textit{\aap} 269:67--76

\bibitem[{{Mattana} et~al.(2009){Mattana}, {Falanga}, {G{\"o}tz}, {Terrier},
  {Esposito} et~al.}]{2008arXiv0811.0327M}
{Mattana} F, {Falanga} M, {G{\"o}tz} D, {Terrier} R, {Esposito} P, et~al. 2009.
\newblock \textit{\apj} 694:12--17

\bibitem[{{Mazin} \& {Raue}(2007)}]{2007A&A...471..439M}
{Mazin} D, {Raue} M. 2007.
\newblock \textit{\aap} 471:439--452

\bibitem[{{Meszaros}(2006)}]{2006RPPh...69.2259M}
{Meszaros} P. 2006.
\newblock \textit{Reports on Progress in Physics} 69:2259--2322

\bibitem[{{Mirabel} \& {Rodriguez}(1994)}]{mirabel94}
{Mirabel} IF, {Rodriguez} LF. 1994.
\newblock \textit{\nat} 371:46--+

\bibitem[{{Moderski} et~al.(2005){Moderski}, {Sikora}, {Coppi} \&
  {Aharonian}}]{moderski05a}
{Moderski} R, {Sikora} M, {Coppi} PS, {Aharonian} F. 2005.
\newblock \textit{\mnras} 363:954--966

\bibitem[{{Moskalenko}, {Porter} \& {Strong}(2006)}]{moskalenko06}
{Moskalenko} IV, {Porter} TA, {Strong} AW. 2006.
\newblock \textit{\apjl} 640:L155--L158

\bibitem[{{Naumann-Godo} \& {{et al.}}(2006)}]{HESS:sn1006}
{Naumann-Godo} M, {{et al.}} 2006.
\newblock In \textit{Heidelberg International Symposium on Gamma-Ray Astronomy}

\bibitem[{{Paredes} et~al.(2000){Paredes}, {Mart{\'{\i}}}, {Rib{\'o}} \&
  {Massi}}]{parades00}
{Paredes} JM, {Mart{\'{\i}}} J, {Rib{\'o}} M, {Massi} M. 2000.
\newblock \textit{Science} 288:2340--2342

\bibitem[{{Pavlidou} \& {Fields}(2001)}]{2001ApJ...558...63P}
{Pavlidou} V, {Fields} BD. 2001.
\newblock \textit{\apj} 558:63--71

\bibitem[{{Pe'er} \& {Waxman}(2005)}]{2005ApJ...633.1018P}
{Pe'er} A, {Waxman} E. 2005.
\newblock \textit{\apj} 633:1018--1026

\bibitem[{{Perkins} et~al.(2006){Perkins}, {Badran}, {Blaylock}, {Bradbury},
  {Cogan} et~al.}]{2006ApJ...644..148P}
{Perkins} JS, {Badran} HM, {Blaylock} G, {Bradbury} SM, {Cogan} P, et~al. 2006.
\newblock \textit{\apj} 644:148--154

\bibitem[{{Persic}, {Rephaeli} \& {Arieli}(2008)}]{2008A&A...486..143P}
{Persic} M, {Rephaeli} Y, {Arieli} Y. 2008.
\newblock \textit{\aap} 486:143--149

\bibitem[{{Pittard} \& {Dougherty}(2006)}]{pittard06}
{Pittard} JM, {Dougherty} SM. 2006.
\newblock \textit{\mnras} 372:801--826

\bibitem[{{Porter}, {Moskalenko} \& {Strong}(2006)}]{porter06}
{Porter} TA, {Moskalenko} IV, {Strong} AW. 2006.
\newblock \textit{\apjl} 648:L29--L32

\bibitem[{{Primack}, {Gilmore} \& {Somerville}(2008)}]{2008arXiv0811.3230P}
{Primack} JR, {Gilmore} RC, {Somerville} RS. 2008.
\newblock \textit{ArXiv e-prints} 0811.3230

\bibitem[{{Protheroe} \& {Stanev}(1993)}]{protheroe93}
{Protheroe} RJ, {Stanev} T. 1993.
\newblock \textit{\mnras} 264:191--+

\bibitem[{{Punch} et~al.(1992){Punch}, {Akerlof}, {Cawley}, {Chantell}, {Fegan}
  et~al.}]{1992Natur.358..477P}
{Punch} M, {Akerlof} CW, {Cawley} MF, {Chantell} M, {Fegan} DJ, et~al. 1992.
\newblock \textit{\nat} 358:477--+

\bibitem[{{Rees} \& {Gunn}(1974)}]{1974MNRAS.167....1R}
{Rees} MJ, {Gunn} JE. 1974.
\newblock \textit{\mnras} 167:1--12

\bibitem[{{Reimer}, {Pohl} \& {Reimer}(2006)}]{reimer06}
{Reimer} A, {Pohl} M, {Reimer} O. 2006.
\newblock \textit{\apj} 644:1118--1144

\bibitem[{{Reynolds}(2008)}]{reynolds08}
{Reynolds} SP. 2008.
\newblock \textit{\araa} 46:89--126

\bibitem[{{Rieger} \& {Mannheim}(2002)}]{2002A&A...396..833R}
{Rieger} FM, {Mannheim} K. 2002.
\newblock \textit{\aap} 396:833--846

\bibitem[{{Stecker}, {Baring} \& {Summerlin}(2007)}]{2007ApJ...667L..29S}
{Stecker} FW, {Baring} MG, {Summerlin} EJ. 2007.
\newblock \textit{\apjl} 667:L29--L32

\bibitem[{{Stecker}, {de Jager} \& {Salamon}(1992)}]{1992ApJ...390L..49S}
{Stecker} FW, {de Jager} OC, {Salamon} MH. 1992.
\newblock \textit{\apjl} 390:L49--L52

\bibitem[{{Steigman} \& {Turner}(1985)}]{1985NuPhB.253..375S}
{Steigman} G, {Turner} MS. 1985.
\newblock \textit{Nuclear Physics B} 253:375--386

\bibitem[{{Strigari} et~al.(2008){Strigari}, {Koushiappas}, {Bullock},
  {Kaplinghat}, {Simon} et~al.}]{2008ApJ...678..614S}
{Strigari} LE, {Koushiappas} SM, {Bullock} JS, {Kaplinghat} M, {Simon} JD,
  et~al. 2008.
\newblock \textit{\apj} 678:614--620

\bibitem[{{Strong}, {Moskalenko} \& {Ptuskin}(2007)}]{strong07}
{Strong} AW, {Moskalenko} IV, {Ptuskin} VS. 2007.
\newblock \textit{Annual Review of Nuclear and Particle Science} 57:285--327

\bibitem[{{Tavecchio}, {Maraschi} \& {Ghisellini}(1998)}]{1998ApJ...509..608T}
{Tavecchio} F, {Maraschi} L, {Ghisellini} G. 1998.
\newblock \textit{\apj} 509:608--619

\bibitem[{{Thompson}(2004)}]{thompson04}
{Thompson} DJ. 2004.
\newblock \textit{New Astronomy Review} 48:543--549

\bibitem[{{Uchiyama} et~al.(2007){Uchiyama}, {Aharonian}, {Tanaka}, {Takahashi}
  \& {Maeda}}]{2007Natur.449..576U}
{Uchiyama} Y, {Aharonian} FA, {Tanaka} T, {Takahashi} T, {Maeda} Y. 2007.
\newblock \textit{\nat} 449:576--578

\bibitem[{{Uchiyama}, {Takahashi} \& {Aharonian}(2002)}]{uchiyama02}
{Uchiyama} Y, {Takahashi} T, {Aharonian} FA. 2002.
\newblock \textit{\pasj} 54:L73--L77

\bibitem[{{V.~Acciari} et~al.(2009){V.~Acciari}, {Aliu}, {Arlen}, {Bautista},
  {Beilicke} et~al.}]{2008arXiv0812.0978V}
{V.~Acciari}, {Aliu} E, {Arlen} T, {Bautista} M, {Beilicke} M, et~al. 2009.
\newblock \textit{\apjl} 690:126--129 

\bibitem[{{Vink} et~al.(2006){Vink}, {Bleeker}, {van der Heyden}, {Bykov},
  {Bamba} \& {Yamazaki}}]{vink06}
{Vink} J, {Bleeker} J, {van der Heyden} K, {Bykov} A, {Bamba} A, {Yamazaki} R.
  2006.
\newblock \textit{\apjl} 648:L33--L37

\bibitem[{{Vink} \& {Laming}(2003)}]{vink03}
{Vink} J, {Laming} JM. 2003.
\newblock \textit{\apj} 584:758--769

\bibitem[{{V{\"o}lk}, {Aharonian} \&
  {Breitschwerdt}(1996)}]{1996SSRv...75..279V}
{V{\"o}lk} HJ, {Aharonian} FA, {Breitschwerdt} D. 1996.
\newblock \textit{Space Science Reviews} 75:279--297

\bibitem[{{Wagner}(2008)}]{2008MNRAS.385..119W}
{Wagner} RM. 2008.
\newblock \textit{\mnras} 385:119--135

\bibitem[{{Weekes} et~al.(1989){Weekes}, {Cawley}, {Fegan}, {Gibbs}, {Hillas}
  et~al.}]{weekes89}
{Weekes} TC, {Cawley} MF, {Fegan} DJ, {Gibbs} KG, {Hillas} AM, et~al. 1989.
\newblock \textit{\apj} 342:379--395

\bibitem[{{Wood} et~al.(2008){Wood}, {Blaylock}, {Bradbury}, {Buckley}, {Byrum}
  et~al.}]{Wood2008ApJ...678..594W}
{Wood} M, {Blaylock} G, {Bradbury} SM, {Buckley} JH, {Byrum} KL, et~al. 2008.
\newblock \textit{\apj} 678:594--605

\bibitem[{{Zirakashvili} \& {Aharonian}(2007)}]{zirakashvili07}
{Zirakashvili} VN, {Aharonian} F. 2007.
\newblock \textit{\aap} 465:695--702

\bibitem[{{Zwicky}(1939)}]{zwicky39}
{Zwicky} F. 1939.
\newblock \textit{Proceedings of the National Academy of Science} 25:338--344

\end{thebibliography}
\end{document}